\documentclass[preprint]{elsarticle} % preprint|review

\usepackage[dvipsnames]{xcolor}

\usepackage{adjustbox}
\usepackage{varioref}

\usepackage{caption}
\usepackage{draftwatermark}
\SetWatermarkText{DRAFT}
\SetWatermarkScale{3}
\SetWatermarkLightness{0.9}

\usepackage[utf8]{inputenc}
\usepackage[pagebackref=false,hyperindex=true]{hyperref}

\usepackage[pdf]{graphviz}

\usepackage{amssymb}
\usepackage{amsmath}

\usepackage{algorithm}
\usepackage{algpseudocode}

\usepackage{rotating}

\usepackage{tikz}
\usetikzlibrary{matrix,shapes,arrows,positioning,chains,fit}

\newcounter{nodecount}
\newcommand\tn[1]{\addtocounter{nodecount}{1} \tikz \node (\arabic{nodecount}) {#1};}

\newcounter{nodecounta}

\usepackage{pgf}

\usepackage{listings}

\newcommand*\rot{\rotatebox{60}}

\usepackage{ifthen}
\newboolean{internaledf}
\setboolean{internaledf}{false}   

\ifthenelse{\boolean{internaledf}}{\newcommand{\opensource}[1]{}}{\newcommand{\opensource}[1]{#1}}

\makeatletter
\newcommand*{\doubleequation}[4]{%
	\par\vskip\abovedisplayskip\noindent%
	\noindent\begin{minipage}{.5\linewidth}\begin{equation}\label{#1}#3\end{equation}\end{minipage}%
	\begin{minipage}{.5\linewidth}\begin{equation}\label{#2}#4\end{equation}\end{minipage}%
	\par\vskip\belowdisplayskip%
}
\makeatother

\begin{document}

% Define block styles
\tikzset{
desicion/.style={
    diamond,
    draw,
    text width=4em,
    text badly centered,
    inner sep=0pt
},
block/.style={
    rectangle,
    draw,
    text width=10em,
    text centered,
    rounded corners
},
data/.style={
    ellipse,
    draw,
    text width=10em
},
cloud/.style={
    draw,
    ellipse,
    minimum height=2em
},
descr/.style={
    fill=white,
    inner sep=2.5pt
},
connector/.style={
    -latex,
    font=\scriptsize
},
rectangle connector/.style={
    connector,
    to path={(\tikztostart) -- ++(#1,0pt) \tikztonodes |- (\tikztotarget) },
    pos=0.5
},
rectangle connector/.default=-2cm,
straight connector/.style={
    connector,
    to path=--(\tikztotarget) \tikztonodes
}
}

% declare images to be used later
% @see https://tex.stackexchange.com/questions/2152/tikz-using-external-images-as-building-blocks

% from https://www.flaticon.com/packs/humans-3
\pgfdeclareimage[interpolate=true,height=0.6cm]{household1}{icons/1person}
\pgfdeclareimage[interpolate=true,height=0.7cm]{household2}{icons/2persons}
\pgfdeclareimage[interpolate=true,height=0.7cm]{household3}{icons/3persons}
\pgfdeclareimage[interpolate=true,height=1cm]{household4}{icons/4persons}

\pgfdeclareimage[interpolate=true,height=0.6cm]{home1}{icons/home_small}
\pgfdeclareimage[interpolate=true,height=0.8cm]{home2}{icons/home_medium}
\pgfdeclareimage[interpolate=true,height=0.9cm]{home3}{icons/home_big}

% Direct
% Probabilisitic
% Pairing

% (Matching)

% of
% Entities 
% depending on 
% Attributes

\title{Pairing for Generation of Synthetic Populations: the Direct Probabilistic Pairing method}

% designed the pairing algorithm
% implemented the algorithm
% complexity analysis
\author[edf,eifer]{Samuel Thiriot}\corref{cor1}\ead{samuel.thiriot@res-ear.ch}
\cortext[cor1]{Corresponding author}
\address[edf]{EDF Lab Paris-Saclay, 7 Boulevard Gaspard Monge, 91120 Palaiseau, France}

% proposed the application case
% identified data 
\author[eifer]{Marie Sevenet}\ead{marie.sevenet@eifer.org}
\address[eifer]{EIFER, Emmy-Noether-Straße 11, 76131 Karlsruhe, Germany}

% state of the art
% validation criteria
% implemented the data model
%\author[kc]{Kevin Chapuis}\ead{kevin.chapuis@gmail.com}
%\address[kc]{IRD UMMISCO, Vietnam}

\bibliographystyle{elsarticle-num}

\begin{abstract}
\ifthenelse{\boolean{internaledf}}{\textbf{This draft is confidential. For EDF group employees only.}\newline}{}
Methods for the Generation of Synthetic Populations do generate the entities required for micro models or multi-agent models, such as they match field observations or hypothesis on the population under study. 
% problem?
We tackle here the specific question of creating synthetic populations made of two types of entities linked together by 0, 1 or more links. 
% it is useful
Potential applications include the creation of dwellings inhabited by households, households owning cars, dwellings equipped with appliances, worker employed by firms, etc. 
% what we do
We propose a theoretical framework to tackle this problem. We then highlight how this problem is over-constrained and requires relaxation of some constraints to be solved. We propose a method to solve the problem analytically which lets the user select which input data should be preserved and adapts the others in order to make the data consistent.
% illustrate
We illustrate this method by synthesizing a population made of dwellings containing 0, 1 or 2 households in the city of Lille (France). In this population, the distributions of the dwellings' and households' characteristics are preserved, and both are linked according to statistical pairing statistics. 
%
% the question of creating composition links between two samples given pairing probabilities. 
%In this type of problem, a pairing algorithm has to deal with inconsistency of data. 
%We propose a pairing algorithm which includes a preprocessing step which defines a consistent target
%given the characteristics of the two samples and the user parameters. 
%Thanks to this step, the pairing algorithm does not has to deal with difficult problems anymore, and becomes straightforward, leading to the creation of links in one unique pass.  
%The overall algorithm enables a one-pass creation of links with a clear quantification of the biases due to the original inconsistency of data and the errors imputable to the pairing algorithm. 
%This algorithm might be suitable for large populations.
%We demonstrate the application of this algorithm on the creation of 1 to n composition links between dwellings and households. 
\end{abstract}


\maketitle

% tikz stuff

% circle one unique cell
% takes as parameters: the name of the tikz cell to circle, and the color to use
\newcommand\tikzaround[2]{\draw [#2](#1.north west) -- (#1.north east) -- (#1.south east) -- (#1.south west) -- cycle;}

% circles an entire line
% takes as parameters: the name of the first and last tikz cell, and the color to use
\newcommand\tikzline[3]{\draw [#3](#1.north west) -- (#2.north east) -- (#2.south east) -- (#1.south west) -- cycle;}

\newcommand\tikzcol[3]{\draw [#3](#1.north west) -- (#1.north east) -- (#2.south east) -- (#2.south west) -- cycle;}

\section{Introduction}

\subsection{Generation of Synthetic Populations}

The study, design and operation of sociotechnical systems rely nowadays on the construction and usage of \textit{disaggregate models} in which the entities of interest (households, persons, cars, buildings, etc.) are explicitly represented and simulated. 
Disaggregate models are the core of several modeling approaches including microsimulation \cite{orcutt1957,Merz1991,baroni_2007,Lovelace2016}, 
agent-based models of geographical systems \cite{heppenstall2011agent}, social sciences \cite{samuel_thiriot:bib_sma_simulation:gilbert_1999_1} or socio technical systems \cite{phan_2007,Smajgl_20014}. 

In order to simulate the evolution of the sociotechnical system of interest, such a  model obviously requires the population of the entities to simulate as an input of each simulation experiment. The actual population can rarely be used, either because data collection would be intractable or illegal for privacy reasons, or because the population to simulate is a future or a past one. As a consequence, this population has to be \textit{synthesized}. Generation of Synthetic Populations (GoSP) refers to the \textit{methods and tools} to \textit{generate populations of entities} which \textit{fulfill the model's and experiment's requirements}, and fit \textit{the data or hypothesis available on the population} of interest for a given \textit{study area}. 
GoSP can thus deal with diverse application fields such as:
the generation of spatialized populations of households and persons for activity-based modeling of transportation \cite{Waddell2002a,Balmer2006,Salvini2005};
the generation of dwellings inhabited by households and associated with appliances for the simulation of residential consumption \cite{Amouroux2014};
% TODO BEBOP, copains?
the generation of workers and firms for economical studies \cite{thiriot2011referral};
planning support systems \cite{Batty2007}.

%The characteristics of this population depend on the model it is generated for. The population might be \textit{spatialised}, as are typically households and activities in activity-based simulations. The population is often made of several \textit{distinct types of entities} having different characteristics such as households, firms and cars in transport models. 
%The population might be \textit{structured vertically by composition or aggregation relationships} between distinct types of entities. Activity-based models typically simulate households made of persons; other models require populations of workers composed into firms, etc. 
%Some type of models also might require horizontally structured populations, for instance to represent contacts in epidemiology, social influence for the diffusion of innovations, etc. 

\subsection{The Pairing problem}

\begin{figure}[th]
	\centering
	\vspace{0.5em}
	\tikzstyle{every picture}+=[remember picture,baseline]
	\tikzstyle{every node}+=[inner sep=1pt,anchor=base,
	minimum width=1.4cm,minimum height=1cm,align=right,text depth=0.5ex,outer sep=1pt]
	\tikzstyle{every path}+=[thick, rounded corners]
	\begin{tikzpicture}
	\matrix(a1)
	{
		\node (home1){\pgfbox[center,bottom]{\pgfuseimage{home1}}}; & 
		\node (home2){\pgfbox[center,bottom]{\pgfuseimage{home3}}}; & 
		\node (home3){\pgfbox[center,bottom]{\pgfuseimage{home2}}}; & 
		\node (home4){\pgfbox[center,bottom]{\pgfuseimage{home1}}}; & 
		\node (home5){\pgfbox[center,bottom]{\pgfuseimage{home3}}}; & 
		\node (home6){\pgfbox[center,bottom]{\pgfuseimage{home3}}}; &
		\\
		\tikzstyle{every node}+=[minimum height=1.9cm]
		\node (household1){\pgfbox[center,top]{\pgfuseimage{household1}}}; & 
		\node (household2){\pgfbox[center,top]{\pgfuseimage{household3}}}; &
		\node (household3){\pgfbox[center,top]{\pgfuseimage{household2}}}; &
		\node (household4){\pgfbox[center,top]{\pgfuseimage{household4}}}; & 
		\node (household5){\pgfbox[center,top]{\pgfuseimage{household4}}}; & 
		\node (household6){\pgfbox[center,top]{\pgfuseimage{household2}}}; &
		\node (household7){\pgfbox[center,top]{\pgfuseimage{household1}}}; 
		\\
	};
	
	\draw [black,line width=2pt]([yshift=-3pt]home1.center) -- ([yshift=5pt]household1.center);
	\draw [black,line width=2pt]([yshift=-3pt]home2.center) -- ([yshift=5pt]household2.center);
	\draw [black,line width=2pt]([yshift=-3pt]home3.center) -- ([yshift=5pt]household3.center);
	\draw [black,line width=2pt]([yshift=-3pt]home4.center) -- ([yshift=5pt]household4.center);
	\draw [black,line width=2pt]([yshift=-3pt]home5.center) -- ([yshift=5pt]household5.center);
	\draw [black,line width=2pt]([yshift=-3pt]home6.center) -- ([yshift=5pt]household6.center);
	\draw [black,line width=2pt]([yshift=-3pt]home6.center) -- ([yshift=5pt]household7.center);
	\end{tikzpicture}
	\vspace{0.5em}
	\caption{Illustration of the pairing problem tackled in this paper, illustrated on a case of composition of dwellings (top) of different surfaces and households of various sizes. The two types of entities are linked together by pair and might have several links (with a degree depending to their characteristics). 
%	The links in this example represent composition, but might also represent composition. 
	This figure depicts the example of dwellings composing households, but any type of entity might be substituted to these examples.}\label{fig_expected_result}
\end{figure}
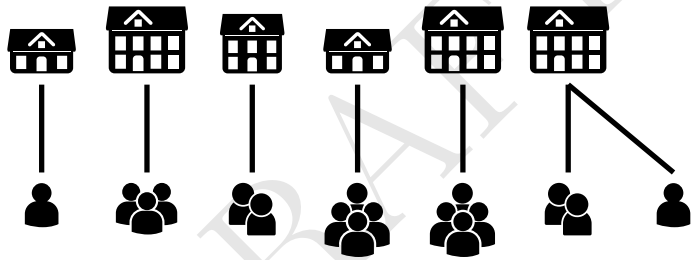

% what is it ?
Among the numerous questions open in GoSP, we tackle in this study the problem of the \textit{generation of synthetic populations made of two types of entities A and B linked together with $n:n$ relationships}, that is in these populations, each entity $a \in A$ might be associated with 0, 1 or $n$ entities of B, and each entity $b \in B$ might be associated with 0, 1 or $n$ entities of A. 
Practical applications of this problem include the creation of buildings made of several dwellings; dwellings composing one or more households; households and cars; households having main residence and secondary residence; firms and workers; etc. 
Figure~\vref{fig_expected_result} illustrates an example of expected result
with a population of dwellings characterized by a surface (small, medium or large), and households characterized by a size (1, 2, 3 or 4 persons). In such an example, each dwelling might host 0, 1 or several households, depending to its characteristics (larger dwellings are more likely to host several households). Each household would be housed by exactly one dwelling. 
We name this type of generation problem the \textit{pairing problem}, in order to emphasis how the main objective is not to generate the populations A and B but to create links between them.

%
%% justifier PIJ
%\paragraph*{}
%In the target synthetic population, the existence of a link depends on the properties of the two entities linked by that link. More precisely, the existence of a link might depend on several of the characteristics of each entity. The possibility to be linked does not relies on systematic rules but rather probabilities. For instance, if bigger households tend to live in bigger dwellings and smaller families in smaller dwellings, this rule is not absolute. In the same way, if men and women tend to mary more in similar social groups and similar sociodemographic (homophily), they also might TODO. As a consequence, a pairing algorithm should accept probabilities densities which describe the statitical dependancy between individuals properties and the probability for these indivduals to be paired together.
%
%% justifier degre
%\paragraph*{}
%In some cases, an entity would only be expected to have one link connected to it; that would be the case of a dwelling to its owner, TODO. Some entities might have no link (homeless people have no link, etc). 
%In some cases, one might have several links of the same type. 
%For instance dwellings might welcome several households. 
%Human beings might have several children or friends.
%% TODO notion degree
%Experiments on network generators have demonstrated that randomly creating links woul 
%
%% TODO discussion: extrapoler vers l'accumulation successive de liens de différents types

\subsection{State of the art on the generation of structured populations}

Several methods in the literature already deal with variations of the pairing problem. 
In order to generate populations of households made of persons, 
the prominent methods in the state of the art rely on samples of households and persons and either reweigh (section \ref{indoc_stateart_reweighting}) or recombine (\ref{indoc_stateart_co}) them according to summary data. 
Because both these approaches rely on samples, they only can be applied in specific cases (\ref{indoc_stateart_pbsample}).
The sample-free methods (\ref{indoc_stateart_samplefree}) constitute alternatives to generate households composing persons (or rarely, any type of entity) from summary data. 

%
%We review here only the methods related to the generation of structured populations.
%Detailed reviews include \cite{samuel_thiriot:bib_sma_simulation:harland_2012_1}, TODO.
%

%
%% grille de lecture
%Note that by purpose, any generation algorithm can be analyzed as: 
%the \textit{type of population} it aims to generate (structured or not, spatial or not), 
%the \textit{kind of input data} it requires (summary data, micro samples, probabilities...),
%\textit{the constrainsts it enforces }(e.g. fit summary data on the distribution of households' characteristics),
%and its \textit{general approach} (reweighting, based on mathematical theory, etc.).
% 
%

% TODO good reveiw \cite{Sun2015a}

\subsubsection{Reweigthing of samples of persons composed in households\label{indoc_stateart_reweighting}}

The stream of methods known as Synthetic Reconstruction (SR) sample-based methods was developed to feed transportation models with populations of \textit{households made of persons} spatialized in local subdivisions of space denoted "small areas". 
These methods take as inputs samples of entities A (households) and B (persons) \textit{which should already include the relationship of composition between A and B}. This relationships takes the form of a unique identifier being associated to every household, and every person being referring to one household identifier; so in the original dataset, each person belongs a household and a household is made of 1 or more persons. Such micro samples known as PUMS (Public Use Micro Samples) in the U.S.A are weighted to be statistically representative at the national level; 
for every small area, statistical institutes also publish summary data which describes the proportions of various control variables of households and persons for each small area. 
These methods propose to \textit{reweigh micro samples of households made of persons} in order to \textit{generate a spatialized population statistically representative at the local scale}.

In his pioneering study \cite{beckman1996creating}, Beckman proposed a fit-and-generate scheme \cite{muller2011hierarchical}. For \textit{fitting}, one first sums the weights of households for the various combinations of control variables in the form of a k-way table (for instance the proportion of households of various sizes, income and ethnicity for each small area). The cells of this table describe the joint distribution present in the sample which is statistically representative at a national scale. The marginals of this table describe the distributions of each  variables, such as the distribution of the ages of the households' head; these marginals should match summary data for each small area for the population to be representative. After proposing this vision of the problem, Beckman propose to \textit{solve this inconsistency between the original and target distributions of attributes by reweighing the cells of the k-way table so the marginals sum up to known summary data}. He proposes to use the Iterated Proportional Fitting (IPF) procedure \cite{samuel_thiriot:bib_sma_simulation:deming_1940_1} which iteratively adapts weights of each dimension of the k-way table so it fits the marginals, and converges to a table which matches all the marginals.
Once the fitting is done, Beckman changes these probabilities to integers through an \textit{integerization step} (multiplication of the float values by a constant and rounding). He then selects households from the micro sample of households according to this count and copies them to build the target population of households. Then he retrieves each person of the selected households by searching for the corresponding identifier, thus also creating the population of persons composed into households. 

This reweighing approach was applied in many different contexts \cite{bowman2004comparison}; its extensive analysis highlighted several difficulties relative to reweighing (other limitations due to the usage of samples will be discussed later in~\ref{indoc_stateart_pbsample}). 

% problems: zero cells
\textit{Zero cells} constitute a practical issue \cite{Guo2007,Ye2009,muller2010population,muller2011hierarchical}, as they technically forbid the convergence of IPF. Also, the semantics of these cells is debatable\cite{lovelace2015evaluating}: do they mean that these classes do not exist (by nature) in the real population, that they did not exist (by mistake) in the sample of this population? Practically, one can replace zero-cells with very low probabilities or adapt classes so that every cell contain a least a few records. \textit{Zero marginals} constitute another problem for the convergence of IPF which can be solved by adapting the classes to avoid them \cite{Ye2009}. Actual applications lead to large k-way tables which are computationally more difficult to tract \cite{Ma2015}, notably leading to sparse tables with many zeros for which specific data structures were proposed \cite{pritchard2012advances}. 

% problems: control both levels
A central, conceptual difficulty is to \textit{control both the distributions of households and persons}. In the Beckman's proposal, only the households' characteristics are controlled by marginals; the persons's characteristics are only indirectly controlled by the relationship household-person present in the original samples, and the underlying dependencies between households and persons characteristics. 
Many variations of the reweighing method were proposed to tackle this issue \cite{Arentze2007,Guo2007,Ye2009,bar2009estimating,auld2010efficient,muller2011hierarchical}.
Solutions include creating prototypes of households including persons' characteristics so fitting households and persons can be done in one pass \cite{Arentze2007}; 
selection of households only if adding the persons they are made of don't distord the target distribution \cite{Guo2007};
iterative updating of both household and persons k-way tables \cite{Ye2009,muller2011hierarchical};
reweigh both households and persons level using entropy maximization \cite{bar2009estimating};
bias the selection of households depending on the current distribution of households and persons' characteristics, the expected one and each household-with-person characteristics
\cite{auld2010efficient}. 
All of these methods follow the fit-and-generate approach from Beckman, yet biasing either fit or generation in order to match both household and persons marginals.
%We would like to emphasis that all these solutions end by twisting one or the other reference data: either the households distribution is enforced and the persons one is biased as in the Beckman's original approach; or both households and persons distributions are enforced, thus meaning the original relationship between households and persons was distorted; in fact, because sample and summary data are basically contradicting each other, \textit{the generation of synthetic population has to reconciliate both these inputs by accepting to bias those which are considered less important}.
We would like to underline how GoSP here fundamentally consists in \textit{transforming several pieces of data contradicting each other}, by \textit{biasing the less reliable input} (national micro samples) so that \textit{it becomes consistent with the most trusted one} (small area summary data).

% problems: integer
%Other authors identified that the integerization step might introduce biases \cite{pritchard2012advances,Lovelace2013,Ma2015}, with potential solutions such as the Quasirandom Integer Without-replacement Sampling method \cite{Smith2017}.
Most authors underline \cite{beckman1996creating,Guo2007,auld2010efficient,pritchard2012advances,Lovelace2013,Ma2015} that, because generation converts continuous probabilities to discrete counts of entities, \textit{rounding errors appear which require fixing by algorithmic workarounds}. 
Deterministic rounding might bias estimates, under or over-represent small probabilities \cite{pritchard2012advances}, and more generally bias the initial distribution \cite{Lovelace2013}; solutions include biasing the selection phases to correct rounding errors, stochastic rounding, or ad hoc rounding solutions to maintain totals. A few authors proposed original ideas to sample directly integers from distributions \cite{Lovelace2013,Smith2017}. 

\subsubsection{Combinatorial optimization of samples of households\label{indoc_stateart_co}}

%Instead of reweighting the kway table using IPF, 

To reach the same goal of generating a synthetic population of households made of persons, spatialized in small areas, based on a global sample of households and individuals and small area statistics, Williams proposed another formulation of the problem \cite{samuel_thiriot:bib_sma_simulation:williamson_1998_1}. 
The synthetic population should contain for each small area 0 or 1 of each record of the sample (meaning the same record can not be used twice for a zone). 
So the generation of a synthetic population might be seen as the search for the combination of 0 and 1 for each small area and each record which leads to the best fit of summary statistics. This best fit can be assessed as the minimization of the error, that is the difference between the expected summary statistics and the actual ones. 
The minimization of the error is an optimization problem which can be tackled with many optimization methods including genetic algorithms \cite{birkin2006synthetic}, simulated annealing or hill-climbing \cite{samuel_thiriot:bib_sma_simulation:williamson_1998_1}. 
Variations of this approach were proposed recently \cite{Ma2015}. 

This GoSP approach was assessed and compared by several authors \cite{voas2000evaluation,huang2001comparison,ryan2009population}. 
It probably leads to a better fit of summary statistics \cite{voas2000evaluation,huang2001comparison}. The choice of the variables to use for constraint was also discussed and shown to provide a better fit \cite{smith2009improving}. These benefits come at the cost of creating a combinatorial optimization problem which requires much time to be solved \cite{huang2001comparison}, and might even reveal intractable for a very large population. 
The Combinatorial Optimization remains so far less common than other "synthetic reconstruction" methods \cite{ryan2009population}.
It was only applied to households and persons, except applications to only one unique type of entity (e.g. for firms \cite{ryan2009population}).

\subsubsection{The limited scope of sample-based methods\label{indoc_stateart_pbsample}}

%Many limitations of sampleb were highlighted in the literature. 
% limit of data 
The main limitation of sample-based methods (based on reweighing or combinatorial optimization) is not their requirement of a sample, but rather the specific \textit{requirement that samples of households and persons should already contain the composition link}. 
% limit: application scope
This requirement \textit{limits the application scope} of these methods, as only few states provide such a sample (U.S.A, U.K, Switzerland were demonstrated in literature), but many other areas do not like Belgium or Canada \cite{samuel_thiriot:bib_sma_simulation:gargiulo_2010_1,pritchard2012advances,barthelemy2013synthetic}.
Note that \textit{this requirement also limits the type of entities which really can be generated using these methods}: such datasets often are available for households and persons because they are collected together at once during national census; but in the case of composition of other entity types like household and car, dwelling and household, such a common identifier is unlikely to exist. 
%As a consequence, generic generation methods (that is generation methods not limited to households and person) should accept as a parameter the constraints on which entities to associate. 
%
As a consequence, the claim of genericity of these methods appears us unlikely and was, by the way, not demonstrated to date. \textit{The existing sample-based reweighing methods were designed to tackle the specific case of the reweighing of samples of households already composed with persons},
%By design, these methods are designed to create populations made of $1:n$ relationships. 
and do not fulfill our more generic goal. 

%Some authors underlined that marginal data are not always available for every small area on all the variables \cite{Zhu2014}.

%
%% other data limitations
%Other difficulties with data availability were emphasised. Summary statistics might come from different institutes than the ones providing micro samples and might thus contradict each other \cite{barthelemy2013synthetic}. 
%Micro sample and summary statistics might no correspond on the attributes they address \cite{samuel_thiriot:bib_sma_simulation:gargiulo_2010_1} or their coding. 
%The micro sample also might not contain all the attributes required for the model \cite{birkin1988synthesis}.
%
%The very principle of using samples also might be discussed. Ye also advocates too much importance is given to a sample which is not necessarly representative of the local scale \cite{Ye2017}. 
%
%
%
%We also would like to highlight other reasons for not using reweighting methods. 
%First, the simulation experiments which requires a synthetic population as an input might intend to simulate a future population, that is a population for which no sample can exist as it does not exists today.

% TODO discuss the claims of genericity

\subsubsection{Sample-free methods\label{indoc_stateart_samplefree}}

As early as 1988, Birkin and Clarke had underlined how often the samples are not available or not suitable, and proposed a method which does not require a sample of persons to generate the synthetic population \cite{birkin1988synthesis}. They start with a sample of households which they disaggregate at the district scale. 
They then add incrementally attributes and persons' attributes to the household using conditional probabilities: country of birth given location, sex, marital status and age; then a spouse (or not) according to marital status; then sex of spouse according to sex of the head of family; then age of spouse given the age of the head; etc. 

% target pop data model 
Twenty years later, several authors  \cite{samuel_thiriot:bib_sma_simulation:gargiulo_2010_1,barthelemy2013synthetic,Huynh2016} underlined the data limitations of reweighing methods, and redeveloped independently alternative methods to generate households made of persons without a sample. 
They all adopt iterative solutions to build these populations. 
%Berthelemy et al. first fit joint a distribution for households and joint distributions for persons \cite{barthelemy2013synthetic}. 
They all first generate two "pools" of entities "households" and "persons" ready to be matched together, but propose different iterative algorithms to match them.

Gargiulo et al.~\cite{samuel_thiriot:bib_sma_simulation:gargiulo_2010_1} \textit{iterates every household and search for the relevant persons to compose inside it}. They first search for a relevant head according to probabilities of head's properties given households' properties. If the head is found, then other persons are searched for (if required) according to another distribution of probabilities to play the role of partner or children. When the right persons are not available in the pool of persons, then the current household is abandoned. At the end of the matching process, there might be persons not associated into households, or households for which relevant persons were not found. 
In their proposal, instead of giving up as soon as the expected person was not found for a household, Berthelemy and Toint \cite{barthelemy2013synthetic} rather search for this head in the households which were already built, and try to replace it with another relevant person. 

Huynh et al. \textit{start instead from persons and group them as households}~\cite{Huynh2016}: they select compliant persons to generate households made of married couples, then households made of a single person, then add students or children to households when required, etc. At the end of the process, the remaining persons which were not yet allocated a household are associated with households such as an error measure is minimized. The generated population ensures the expected count of entities of households and persons are enforced, as well as the combinations of attributes in these entities.
% Biases are present in the relationship between person and household, or between the persons. 

Earlier in 2008, Thiriot and Kant \cite{samuel_thiriot:bib_perso:thiriot_2008_3} had developed a sample-free method to generate entities structured as networks, with the links between entities being created conditional to the properties of the entities. This method was designed to create $n:n$ links between different or similar entity types, such as friendship networks (each individual might be linked with several other individuals) or company and firm (each firm has 0 to many employees). 
This method takes as input summary data provided in the form of Bayesian networks which describe the variables for entities A and B (for instance workers and firms), including conditional probabilities describing the count of links to create for each entity given its characteristics. The method also takes as input a Bayesian network describing matching probabilities, that is the probability to create a link given the properties of two entities A and B. 
The algorithm is also iterative: first the two pools of entities A and B are created; then each entity of A is iterated, the expected candidate's characteristics are randomly chosen given the pairing probabilities, and a corresponding entity B is searched for; if this entity is not found, then another possible candidate is searched with the same process until a valid candidate is identified. If no candidate can be found, then the creation of the link is abandoned.
This method was applied to the generation of a family structure with partners, friendship and work relationships \cite{samuel_thiriot:bib_perso:thiriot_2008_3}, and was applied to the creation of entities of workers composed into firms \cite{thiriot2011referral,lewkovicz2011detailed}. The actual semantics used to encode the probability to link two entities conditional to their characteristics and degrees was somehow unclear in practice.

% discuss these methods
These methods share several common points.
With the notable exception of the Thiriot et al. approach, they all were designed for the specific case of creating $1:n$ links for households made of persons.
%They were designed and applied to create households made of persons, but their principle might apply on other entities types without changing their methodology. 
Because they have to deal with the creation of links from scratch, these methods take as inputs constraints on \textit{how many links to create} (in the form of a household type and/or count of children given other types of attributes) and \textit{with who to create links} (in the form of probability distributions defining the characteristics of persons given households' ones). 
All these algorithms share a generate-match-fix approach: because the parameters for the generation of households and persons were not made consistent beforehand, there are inconsistencies between the count of households having various characteristics, the distributions of probabilities which constraint which persons to compose inside households given their characteristics, and the proportions of persons having various characteristics. \textit{These inconsistencies are solved during the iterative process}; depending on to the principle of the algorithms, the iterative solutions bias either the distributions of households, of persons or the matching probabilities.

\subsection{Approach \& Outline}

% et donc !
We base our study on the following analysis of this state of art. 
The sample-based methods first \textit{fit input data so it becomes consistent}, before generating out of this coherent solution; their approaches require the relationships to be known in the original samples and are therefore not generic. 
The sample-free methods are able to generate the links, but instead of fit they do solve iteratively the inconsistencies between datasets during generation, and have to detect and fix problems when they occur. 
We design a method which \textit{first solves the inconsistencies} between the pieces of input data, like the reweighing methods. Alike the sample-free methods, we will explicitly \textit{take in charge the creation of the links} between entities A and B based on summary data, in a fully generic setting. 

% outline
We first describe (section~\ref{indoc_inputs}) the input data required from the user and start introducing the core concepts of our approach. 
We then introduce in section~\ref{indoc_theoretical} (p.~\pageref{indoc_theoretical}) the theoretical framework to analyse the pairing problem, introduce the equations which lead to consistent solutions, and propose a solver to solve the original inconsistencies according to relaxation parameters. 
We then demonstrate the usage of this method (\ref{indoc_application} p~\pageref{indoc_application}) on a real-size case, and measure the accuracy of the solution when enforcing and relaxing different input datasets. 
As discussed later (\ref{indoc_discussion} p.~\pageref{indoc_discussion}), this solution appears relevant and generic, but would not be suitable for the generation of households and persons. 

\section{Inputs and core concepts\label{indoc_inputs}}

\begin{figure}[th!]
	\centering
	\tikzstyle{every picture}+=[remember picture,baseline]
	\tikzstyle{every node}+=[inner sep=1pt,anchor=base,
	minimum width=1.1cm,minimum height=1cm,align=right,text depth=0.5ex,outer sep=1pt,font=\footnotesize]
	\tikzstyle{every path}+=[thick, rounded corners, Black, line width=2pt]
	\tikzstyle{slot}=[circle,fill=black,scale=0.2,baseline=2em]
	\begin{tikzpicture}
	\matrix(a1)
	{
		\node[left,yshift=1.1em](titleEntitiesA) {\textbf{entities A}\\classes $i$\\frequencies $f_i$, size $n_A$}; &
		\node (home1){\pgfbox[center,bottom]{\pgfuseimage{home2}}}; & 
		\node (home2){\pgfbox[center,bottom]{\pgfuseimage{home3}}}; &
		& 
		\node (home3){\pgfbox[center,bottom]{\pgfuseimage{home2}}}; &
		& 
		\node (home4){\pgfbox[center,bottom]{\pgfuseimage{home1}}}; & 
		\node (home5){\pgfbox[center,bottom]{\pgfuseimage{home3}}}; 
		\\
		\node[left,yshift=0.7em](titleSlotsA) {\textbf{slots A}\\probabilities degrees\\$p(d_i=n)$}; &
		\node[slot](slotA1) {}; &
		\node[slot](slotA2) {}; & 
		\node[slot](slotA3) {}; &
		\node[slot](slotA4) {}; &
		\node[slot](slotA5) {}; & 
		\node[slot](slotA6) {}; & 
		\node[slot](slotA7) {}; \\[0.5em] 
		\\
		&
		\node[slot](slotC1) {}; &
		\node[slot](slotC2) {}; & 
		\node[slot](slotC3) {}; &
		\node[slot](slotC4) {}; &
		\node[slot](slotC5) {}; & 
		\node[slot](slotC6) {}; & 
		\node[slot](slotC7) {}; 
		\\
		\node[left](titleEdges) {\textbf{edges}\\pairing probabilities\\$p_{i,j}$}; & 
		&
		&
		&
		&
		&
		&
		\\
		&
		\node[slot](slotD1) {}; &
		\node[slot](slotD2) {}; & 
		\node[slot](slotD3) {}; &
		\node[slot](slotD4) {}; &
		\node[slot](slotD5) {}; & 
		\node[slot](slotD6) {}; & 
		\node[slot](slotD7) {}; \\[1em] 
		\\	
		\node[left,yshift=-0.8em,right delimiter={.}](titleSlotsB) {\textbf{slots B}\\probabilities degrees\\$p(d_j=n)$}; &	
		\node[slot](slotB1) {}; &
		\node[slot](slotB2) {}; & 
		\node[slot](slotB3) {}; &
		\node[slot](slotB4) {}; &
		\node[slot](slotB5) {}; &
		\node[slot](slotB6) {}; &
		\node[slot](slotB7) {};
		\\
		\tikzstyle{every node}+=[minimum height=1.4cm]
		\node[left,yshift=-1.65em](titleEntitiesB) {\textbf{entities B}\\classes $j$\\frequencies $f_j$, size $n_B$}; &
		\node (household1){\pgfbox[center,top]{\pgfuseimage{household2}}}; & 
		\node (household2){\pgfbox[center,top]{\pgfuseimage{household1}}}; &
		\node (household3){\pgfbox[center,top]{\pgfuseimage{household2}}}; &
		\node (household4){\pgfbox[center,top]{\pgfuseimage{household4}}}; & 
		\node (household5){\pgfbox[center,top]{\pgfuseimage{household3}}}; &
		\node (household6){\pgfbox[center,top]{\pgfuseimage{household4}}}; &
		\node (household7){\pgfbox[center,top]{\pgfuseimage{household2}}};
		\\
	};
	
	\draw ([yshift=-3pt]home1.center) -- (slotA1.center);
	\draw ([yshift=-3pt]home2.center) -- (slotA2.center);
	\draw ([yshift=-3pt]home2.center) -- (slotA3.center);
	\draw ([yshift=-3pt]home3.center) -- (slotA4.center);
	\draw ([yshift=-3pt]home3.center) -- (slotA5.center);
	\draw ([yshift=-3pt]home4.center) -- (slotA6.center);
	\draw ([yshift=-3pt]home5.center) -- (slotA7.center);
	
	\node[left delimiter=\lbrace,fit=(home1.north),yshift=0.5em](brace1){};
	\node[left delimiter=\lbrace,fit=(slotA1.north),yshift=0.9em](brace2){};
	\node[left delimiter=\lbrace,fit=(slotC1)(slotD1.north west),yshift=10pt](brace3){};
	\node[left delimiter=\lbrace,fit=(slotB1.north),yshift=-0.5em](brace4){};
	\node[left delimiter=\lbrace,fit=(household1.center),yshift=-0.9em](brace5){};

	%\draw [decorate,decoration={brace,amplitude=10pt},xshift=-4pt,yshift=0pt]
	%(0.5,0.5) -- (0.5,5.0) node [black,midway,xshift=-0.6cm] 
	%{\footnotesize $P_1$};
	
	\draw ([yshift=3pt]household1.center) -- (slotB1.center);
	\draw ([yshift=3pt]household2.center) -- (slotB2.center);
	\draw ([yshift=3pt]household3.center) -- (slotB3.center);
	\draw ([yshift=3pt]household4.center) -- (slotB4.center);
	\draw ([yshift=3pt]household5.center) -- (slotB5.center);
	\draw ([yshift=3pt]household6.center) -- (slotB6.center);
	\draw ([yshift=3pt]household7.center) -- (slotB7.center);
	
	\draw (slotC1.center) -- (slotD4.center);
	\draw (slotC2.center) -- (slotD1.center);
	\draw (slotC3.center) -- (slotD2.center);
	\draw (slotC4.center) -- (slotD3.center);
	\draw (slotC5.center) -- (slotD5.center);
	\draw (slotC6.center) -- (slotD7.center);
	\draw (slotC7.center) -- (slotD6.center);
	
	\end{tikzpicture}
	\caption{Illustration of the pairing problem with variable degrees of connectivity. From top to bottom: entities of the population A have several classes $i$, and are constrained by the frequencies $f_i$ passed as user parameters and the total count $n_A$. Each entity of A might have 0, 1 or more slots, that is candidates for edge connections, which are constrained by the probabilities of degrees $pd_i$ which define, for each entity type A, how many connections can be made with this entity. Edges are constrained by the pairing probabilities $p_{i,j}$. Each of the edges also should connect one available slot of population B constrained by $pd_j$. The population B also is constrained by the count $n_B$ and the frequencies $f_j$.}\label{fig_with_slots}
\end{figure}
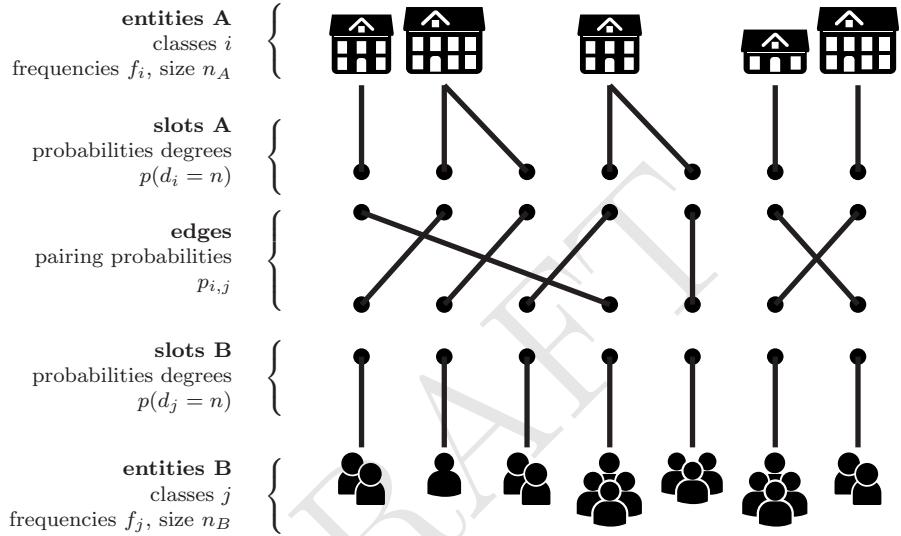

Our generation method should generate a population $\langle \hat{A},\hat{B},L \rangle$ with $\hat{A}$ and $\hat{B}$ entities representing different types of entities (such as dwellings and households in our example) each associated with different characteristics, and $L$ the links between entities $\hat{A}$ and $\hat{B}$. Links $K$ can encode $n:n$ relationships, so there might be 0, 1 or several links going out of the same $a \in \hat{A}$, and the same for $\hat{B}$.

The properties of this population are stylized in Figure~\vref{fig_with_slots} which illustrates this concept for dwellings and households. 
\begin{itemize}
	\item In this figure, the entities of A and B have different characteristics like surface and size which are defined as \textit{classes}. Information about these populations is provided in the form of weighted samples of A and B.
	\item Each entity might be connected to one or more links; this expected count of links for an entity is denoted count of slots. The user can parameter the expected count of links as probabilities conditional to entities' classes.
	\item Links between entities are created depending on the characteristics of the entities. They are constrained by a joint distribution of degrees named \textit{pairing probabilities}.
\end{itemize}

\subsection{Characteristics of entities: Variables, Modalities, Classes}

The user first defines the set of \textit{variables} which should be controlled in the resulting synthetic population, either because these variables influence the relative frequencies expected in the population (see below~\ref{indoc_inputs_frequencies}), how many entities can be connected to each entity (see below~\ref{indoc_inputs_degree}), or because these variables influence the pairing probabilities (see below~\ref{indoc_inputs_pairing}). In the example of pairing off dwellings with households, we consider that pairing depends on the surface of the dwellings (encoded over 3 modalities) and the size of the household (encoded over 4 modalities).
We denote $\text{Att}^A$ (resp. $\text{Att}^B$ ) the set of variables of interest for the generation process related to population $A$ (resp $B$). Each of these variables can take a discrete and finite set of modalities. 

%The definition of the variables is the responsibility of the user of the algorithm, and depends on data available, on data which might be collected, on the experimentations to be driven, etc. 

The \textit{classes} of entities to be studied for pairing, denoted $\text{Cla}^A_i$ for the population $A$ (resp. $\text{Cla}^B_j$ for $B$), define all the combinations of variable modalities which should be controlled in the synthetic population. 
%We denote $A_i \subset A$ the subset of $A$ whose attributes comply with $\text{Mod}_i(A)$; the combinations of the modalities should cover all the possibilities of samples of A and B, so $\bigcup_i A_i = A$ and $\bigcup_j B_j = B$.
%
In our example, we define:
$\text{Att}^A = \{\text{surface}\}$ 
with $\text{Cla}^A = \{\text{surface}=1,\text{surface}=2,\text{surface}=3\}$.
For the population of households B, we define $\text{Att}^B = \{\text{size}\}$ with 
$\text{Cla}^B = \{\text{size}=1,\text{size}=2,\text{size}=3,\text{size}=4\}$.
Note that several attributes might be used for each modality, as depicted in the application example (\vref{indoc_application}).
% We might have defined 
%$\text{Att}^X = \{\text{surface},\text{cost}\}$, with all the possible combinations of modalities: $\text{Mod}^X_i  = \{\text{surface}=1\&\text{cost}=1,\text{surface}=1\&\text{cost}=2,\text{surface}=2\&\text{cost}=1,\text{surface}=2\&\text{cost}=2,\text{surface}=3\&\text{cost}=1,\text{surface}=3\&\text{cost}=2\}$. We keep the examples simpler for shake of simplicity, but the algorithm works the same with more variables and modalities.
%
This formalism relies on the assumption only variables with finite sets of modalities, ordered or not, (categorical, logical or enumerated variables) can be used for pairing\footnote{Note that even numerical, even continuous, variables might be considered, as the values provided in the sample will necessarily include a finite set of values which might be processed by the algorithm. However the user has to provide other probabilities conditional to modalities which might be more difficult to provide for continuous variables.}.

%The generated set of entities $\hat{A}$ (respectively $\hat{B}$) are characterized by \textit{variables} which each have a discrete and finite set of possible \textit{modalities}. For instance a dwelling might have for variable "surface" the modalities "small", "medium" or "large"; a household might have for variable "size" and for modalities "1 person", "2 persons" and so on. The combinations of the possible modalities constitute the various \textit{classes} numbered $i$ (resp. $j$). In this first example with one variable for each entity type, the classes $i$ are thus "surface=small","surface=medium" and "surface=large". 

\subsection{Input samples $A$ and $B$\label{indoc_inputs_frequencies}}

\begin{table}[htp]
	\centering\footnotesize
	\begin{tabular}{cccc}
		\hline
		\multicolumn{4}{c}{\textbf{dwellings}} \\
		weight & surface & cost & ...\\ 
		\hline
		1 &   1 &   1 & ... \\ 
		1 &   2 &   7 & ... \\ 
		1 &   1 &   2 & ... \\ 
		1 &   3 &   4 & ... \\ 
		1 &   1 &   5 & ... \\ 
		1 &   1 &   8 & ... \\ 
		1 &   3 &   7 & ... \\ 
		1 &   2 &   5 & ... \\ 
		1 &   3 &   7 & ... \\ 
		1 &   2 &   5 & ... \\ 
		... &   ... &   ... & ... \\ 
		\hline
	\end{tabular}
	\quad \quad \quad \quad \quad
	\begin{tabular}{cccc}
		\hline
		\multicolumn{4}{c}{\textbf{households}} \\
		weight & size & income & ... \\ 
		\hline
		0.62 &   4 &   8  & ...\\ 
		0.05 &   1 &   7  & ...\\ 
		0.64 &   3 &   1  & ...\\ 
		0.58 &   4 &   2  & ...\\ 
		0.56 &   1 &   5  & ...\\ 
		0.54 &   1 &   1  & ...\\ 
		0.57 &   1 &   9  & ...\\ 
		0.21 &   4 &   1  & ...\\ 
		0.79 &   3 &   8  & ...\\ 
		0.81 &   4 &  10  & ...\\ 
		... &   ... &   ...  & ...\\ 
		\hline
	\end{tabular}
	\caption{Example of a weighted samples of dwellings and households. More variables would be present in actual samples. Note how the sample of dwelling actually represents entities without a weight.}\label{tab_example_sample}
\end{table}

We assume here that \textit{information on populations A and B is provided by the user as weighted samples}. Among the two possibilities identified in the state of the art (namely samples or summary data), weighted samples constitute the most generic data type: summary data might be used to generate samples without loss of information, whilst the reduction of sample data into summary data would loose information on the dependencies between records' variables. These micro samples of A and B can be totally independent and \textit{do not} require any common identifier nor specific relationship between each other.
These samples will be used as a source, and will be either reweighed, copied and probably resized during the pairing process.

The sample $A$ (respectively $B$) should contain all the variables $\text{Att}^A$ (resp. $\text{Att}^B$)
%and all the combinations of modalities $\text{Cla}^A_i$ (resp. $\text{Cla}^B_j$)
. In our example, as we want pairing to depend on the variable \textit{'surface'} of dwellings (A) and the variable \textit{'size'} of households (B), we obviously need dwellings to have a surface and households to have a size.
The fact these samples are weighted enables the use of lists of entities which are just a specific case of a weighted sample with all the weights being equivalent.
Table~\vref{tab_example_sample} provide examples for samples $A$ and $B$.

\begin{table}
	\centering\footnotesize
	\begin{tabular}{r|rrr}
		\hline
		& \multicolumn{3}{c}{\textbf{dwellings}} \\
		& surface=1 & surface=2 & surface=3 \\ 
		\hline
		$f_i$		& 0.33	& 0.33	& 0.33 \\
		\hline
	\end{tabular}
	\quad
	\begin{tabular}{r|rrrrr}
		\hline
		& \multicolumn{4}{c}{\textbf{households}} \\
		& size=1 & size=2 & size=3 & size=4 \\ 
		\hline
		$f_j$ & 0.50 & 0.30 & 0.15 & 0.05 \\ 
		\hline
	\end{tabular}
	\caption{Example of frequencies 
		%measured from sample for classes 
		based on variables "surface" for dwellings and "size" for households.}\label{tab_frequencies_example}
\end{table}

%We denote \textit{frequencies} $f_i$ and $f_j$ the relative frequencies of classes $\text{Cla}^A_i$ and $\text{Cla}^B_j$ expected in the synthetic sets $\hat{A}$ and $\hat{B}$. 
%The target synthetic populations will also be constrained by the expected counts of entities $n_A$ and $n_B$. 

Along with the weighted sample $A$ (resp $B$), the user also transmits the proportions
expected for each class $A_i$ (resp $B_j$), which often should be enforced during the generation process. 
We denote \textit{frequencies} the relative frequencies $f_i$ (resp. $f_j$) of the classes $\text{Cla}^A$ expected in the target population (resp. $\text{Cla}^B$ for population $B$). 
%These frequencies are computed as the sum of the weights of the entities $A_i \subset A$ of class $i$ (resp $B_j \subset B$ of class $j$), normalized by the total weight of the population.
%
%\begin{equation}
%\forall m_i \in Mod(A), f_i = \dfrac{\sum_{a \in A_i} \omega(a)}{\sum_{a' \in \mathcal{A}} \omega(a')} \text{ with } A_i \subset \mathcal{A} \text{ compliant with } m_i
%\end{equation}
%%
%\begin{equation}
%\forall m_j \in Mod(B), f_j = \dfrac{\sum_{b \in B_j} \omega(a)}{\sum_{b' \in \mathcal{B}} \omega(b')} \text{ with } B_j \subset \mathcal{B} \text{ compliant with } m_j
%\end{equation}
%
By definition and construction, these frequencies are summing up to $1$. 

\doubleequation{eq_fi_sum1}{eq_fi_sum2}{\sum_i f_i = 1}{\sum_j f_j = 1}

In our example, the table~\vref{tab_frequencies_example} depicts frequencies quantifying the relative proportions of small, medium and large surfaces of dwellings.

\subsection{Probabilistic distribution of degrees\label{indoc_inputs_degree}}

Each entity of $\hat{A}$ and $\hat{B}$ might be connected to 0, 1 or more entities of the other type. We here use the concept of \textit{degree of connectivity} (or more concisely "degree") to denote the "count of links" an entity has with other entities, as done in graph theory or social network analysis \cite{samuel_thiriot:bib_psycho:wasserman_1994_2}. 
An entity having degree 0 has no link; an entity having degree 1 has only one link, etc. In our example, a dwelling having degree 0 is a dwelling containing no household; a dwelling of degree two contains two households. In the same way, a household of degree 0 has no dwelling; a household of degree 1 has exactly one dwelling.

% TODO slots
It is as if each entity being generated in the synthetic population had a finite count of potential link connections, that we will denote here \textit{slots}. A slot can be used by one and only one link. Slots are constrained by the probabilistic distribution of degrees defined by the user. 
In the example of figure~\vref{fig_with_slots}, we depicted a few entities for which some dwellings have only one connection, and some others two, meaning they would be expected to connect with two households. On the side of population B, which is in this example made of households, we depicted that every household is expected to be connected to one and only one entity, meaning each of them has exactly one and only one slot.

\begin{table}[hbt]
	\centering\footnotesize
	\begin{tabular}{r|ccccc}
		\hline
		\textbf{degree}	& \multicolumn{5}{c}{\textbf{A}} \\
		$n$ & $\text{Cla}^A_1$ & \textemdash & $\text{Cla}^A_i$ & \textemdash & $\text{Cla}^A_m$ \\ 
		\hline
		0 & $p(d_1=0)$ & \textemdash & $p(d_i=0)$ & \textemdash & $p(d_m=0)$ \\ 
		\textbar & \textbar &  & \textbar &  & \textbar \\ 
		$n$ & $p(d_1=n)$ & \textemdash & $p(d_i=n)$ & \textemdash & $p(d_m=n)$ \\ 
		\textbar & \textbar &  & \textbar &  & \textbar \\ 
		\hline
		\textit{total}	& 1.00	& \textemdash & 1.00	& \textemdash & 1.00 \\
		\hline
	\end{tabular}
	\caption{General structure of the distribution of degrees $pd_i$ for entities of population A. For each class $\text{Cla}^A_i$, for each of the $n$ potential degrees, the table contains the conditional probabilities of an entity having characteristics $i$ to have $n$ connections.\label{tab_probabilistic_pdi}}
\end{table}

In general, the degree of an entity depends on its characteristics; for instance bigger dwellings are more likely to contain several households.
In a probabilistic setting, we propose to encode this dependency as a \textit{distribution of probability of an entity having each possible degree conditional to its characteristics} (as done before by \cite{samuel_thiriot:bib_perso:thiriot_2008_3} or \cite{samuel_thiriot:bib_sma_simulation:gargiulo_2010_1,barthelemy2013synthetic,Huynh2016}). 
We denote this \textit{probabilistic distribution of degree} $pd(n,i)$, with $n \in \mathbb{N}$ and $i$ the class of the population. We shorten this notation as $pd_i$ and $pd_j$ the probability distributions of degrees for population A and B. 
In practice, the user provides this distribution in the form of a table as depicted in table~\vref{tab_probabilistic_pdi}. Being conditional probability distributions, $pd_i$ and $pd_j$ should sum up to 1 vertically:
\doubleequation{eq_pdi_sum1}{eq_pdi_sum2}{\forall i, \sum_{n \in \mathbb{N}} pd(n,i) = 1}{\forall j, \sum_{n \in \mathbb{N}} pd(n,j) = 1}

We depict in tables~\vref{tab_example_pdi} and \vref{tab_example_pdj} examples of probability distribution of degrees for dwellings and households, where larger dwellings contain more households, and households are contained by exactly one dwelling.
Note that entities of the population B also might be connected to several entities of A. This might be the case in this example, as a given household might hold several dwellings (principal and secondary residences).

\begin{table}[pth]
	\centering\footnotesize
	\begin{tabular}{r|rrr}
		\hline
		\textbf{degree}	& \multicolumn{3}{c}{\textbf{dwellings}} \\
		%$i$ & 1 & 2 & 3 \\
		$n$ & surface=1 & surface=2 & surface=3 \\ 
		\hline
		0 & 0.20 & 0.15 & 0.05 \\ 
		1 & 0.80 & 0.80 & 0.80 \\ 
		2 & 0.00 & 0.05 & 0.10 \\ 
		3 & 0.00 & 0.00 & 0.05 \\ 
		4 & 0.00 & 0.00 & 0.00 \\ 
		\hline
		total	& 1.00	& 1.00	& 1.00 \\
		\hline
		average $\tilde{d}_i$ & 0.80 & 0.90 & 1.15 \\
		\hline
	\end{tabular}
	\caption{Example of distributions of degrees for dwellings. It is read as: 20\% of the small dwellings (encoded here with surface=1) are empty (degree $n=0$) and the remaining 80\% only contain one link $n=1$ (that is, only one household lives in it)}\label{tab_example_pdi}
\end{table}

\begin{table}[pth]
	\centering\footnotesize
	\begin{tabular}{r|rrrrr}
		\hline
		\textbf{degree}	& \multicolumn{4}{c}{\textbf{households}} \\
		$n$ & size=1 & size=2 & size=3 & size=4 \\ 
		\hline
		0 & 0.0 & 0.0 & 0.0 & 0.0 \\ 
		1 & 1.0 & 1.0 & 1.0 & 1.0 \\ 
		\hline
		\textit{total} & 1.0 & 1.0 & 1.0 & 1.0 \\ 
		\hline
		\textit{average} $\tilde{d}_j$ & 1.0 & 1.0 & 1.0 & 1.0 \\
		\hline
	\end{tabular}
	\caption{Example of distributions of degrees for households. We suppose here that every household in the resulting sample should be in one and only one dwelling.}\label{tab_example_pdj}
\end{table}

In the table for probability distribution of degrees, \textit{a zero is considered structural, meaning this value is not possible and never should be generated}. The pairing algorithm will fail rather than adding even a low probability during the generation. As a consequence, a user considering a link being unlikely but still possible should provide a very low probability rather than null for the corresponding cell. 

% average degree
The distribution of probabilities $pd_i$ and $pd_j$ provided by the user implicitly defines the \textit{average degree} for each class $i$ and $j$ denoted $\tilde{d}_i$ and $\tilde{d}_j$ for populations A and B. The average degree is a positive real which describes, for the entities of given characteristics, how many links would be created for them on average. We introduce the notion of average degree because it is easier to deal with than the distributions of probabilities; as a consequence, the average degree will be used as a proxy in later computations. The average degree is computed as the sum of the degree times the probability of this degree. 
\doubleequation{eq:di_pdi}{eq:dj_pdj}{\tilde{d}_i = \sum_{n \in \mathbb{N}} n.p(d_i=n)}{\tilde{d}_j = \sum_{n \in \mathbb{N}} n.p(d_j=n)}

For instance in table~\vref{tab_example_pdi}, for dwellings of class "surface=2", 15\% have degree $0$ (and will this lead to $0*0.15=0$ links), 80\% have degree $1$ (will give born to $1*0.8=0.8$ links) and 5\% have degree 2 ($2*0.05=0.1$ links created). So the total average degree for this class is $\tilde{d}_2=0+0.8+0.1=0.9$.  It means that for a hundred dwellings having surface 2 to be created, we should generate on average 90 links for them: 80 links connecting entities having only one link, and 10 links connecting 5 entities having two links each. 
%The basic intuition is: the count of entities of class $A_i$, multiplied by the average degree $\tilde{d}_i$ of this class, corresponds to the total count of slots available for connection. 

% TODO min and max di dj

%The table~\vref{tab_example_pdi_di} illustrates this computation. On this example, for dwellings having surface 2, 15\% would create no link ($n=0$) and would thus lead to no link (degree $0.0$). 80\% other percent would create one link and would thus add on average $0.8$ links per entity. The remaining 5\% of this column do lead to 2 links, and would thus lead to an average of $0.05*2=0.1$ links. The cumulated links created by all the dwellings having surface 2 would thus sum up to $0.0+0.8+0.1=0.9$, which stands as the average degree for class \textit{"surface=2"}.

\subsection{Pairing probabilities: constraints of pairing\label{indoc_inputs_pairing}}

At the central row of the figure~\vref{fig_with_slots}, we depicted the edges which should enforce the \textit{pairing probabilities} defined by the user. Each link connects a slot of an entity A and a slot of an entity B. 
%
%This figure highlights intuitively the need for consistency between the various values we mentioned: for a link to be created with an entity A of a given class, there should be an entity of this class existing in A having a slot available (which depends on frequencies $f_i$ and distribution of degrees $pd_i$). In the same way, the link also requires a slot available for an entity of B of the valid class. 
%We will detail later these constraints in a more formal way.
%
The links between entities from A and B depend on the characteristics of the two linked entities; for instance larger dwellings tend to be occupied by bigger households, and luxurious dwellings are more hosting wealthier households. As done before in iterative algorithms for matching populations (see~\vref{indoc_stateart_samplefree}), we require as an input a distribution of probabilities to encode these dependencies.
The \textit{pairing probabilities} denoted $p_{i,j}$ define, for a link to be created in the synthetic population, the probability for this link to pair an entity from population A of class $\text{Cla}^A_i$ with an entity of population B of class $\text{Cla}^B_j$. It takes the form of a two-dimensional table having the classes of population A as columns and the classes of population B as rows. This table contains a joint probability distribution which enforces by definition $\sum_i \sum_j p_{i,j} = 1$.
Table~\vref{tab_example_pij_pi_pj} provides an example of pairing probabilities for pairing dwellings and households based on the surface of the dwellings and the sizes of the households.

As for the probability distribution of degrees, a zero in the pairing probabilities table means this value is not possible at all and should never be generated. Therefore, a user considering a probability to be unlikely but still possible should use a low probability instead of 0.

%
%\begin{table}
%\centering
%\footnotesize
%\begin{tabular}{r|rrr|r}
%\hline
%\textbf{household}	& \multicolumn{3}{c}{\textbf{dwellings}} & \textit{total}\\
%			& surface=1 & surface=2 & surface=3 &  \\ 
%\hline
%size=1 		& 0.20 		& 0.04 		& 0.01 & 0.25 \\ 
%size=2 		& 0.10 		& 0.12 		& 0.03 & 0.25 \\ 
%size=3 		& 0.05 		& 0.10 		& 0.10 & 0.25 \\ 
%size=4 		& 0.03 		& 0.05 		& 0.17 & 0.25 \\ 
%\hline
%\textit{total} & 0.38 & 0.31 & 0.31 & 1.0 \\
%
%\end{tabular}
%\caption{Example of pairing probabilities to match two populations of dwellings and households based on the surface of the dwelling and the size of the household. These probabilities describe how, when a household is in a dwelling, there tends to be smaller households in smaller dwellings. Note the entire table probabilities sum up to 1.}\label{tab_example_pij}
%\end{table}

\begin{table}
	\centering\footnotesize
	\tikzstyle{every picture}+=[remember picture,baseline]
	\tikzstyle{every node}+=[inner sep=0pt,anchor=base,
	minimum width=0.7cm,align=right,text depth=0.5ex,outer sep=1pt]
	\tikzstyle{every path}+=[thick, rounded corners]
	\begin{tabular}{r|ccc|c}
		\hline
		\textbf{household}	& \multicolumn{3}{c|}{\textbf{dwellings}} & \\
		& surface=1 & surface=2 & surface=3 & \textit{totals} \\ 
		\hline
		size=1 		& \tikz\node (c11){0.20}; & 0.04 		& 0.01 	& 0.25 \\ 
		size=2 		& \tikz\node (c21){0.10}; 		& 0.12 		& \tikz\node (c23){0.03}; 	& \tikz\node (c24){0.25}; \\ 
		size=3 		& \tikz\node (c31){0.05};		& 0.10 		& 0.10 	& 0.25 \\ 
		size=4 		& \tikz\node (c41){0.03}; & 0.05 		& 0.17 	& 0.25 \\ 
		\hline
		\textit{totals}		& \tikz\node (c51){0.38};		& 0.31		& 0.31	& 1 \\ 
	\end{tabular}
	\caption{Example of the computation of probabilities $p_i$ and $p_j$ from $p_{i,j}$: totals are just the sums of rows and columns. The probabilities $p_{i,j}$ are read as: a link has 20\% of chances to link a dwelling having surface 1 and a dwelling having size 1. Marginal probabilities $p_i$ are read as: 38\% of the links originating from dwellings should come from dwellings having surface 1, 31\% other percents should originate from dwellings with surface 2.}\label{tab_example_pij_pi_pj}
	%	\begin{tikzpicture}[overlay]
	%	\tikzcol{c11}{c41}{red}
	%	\tikzaround{c51}{red}
	%	\path[-latex] (c31.north west) edge [bend right=50,red](c51.west);
	%	
	%	\tikzline{c21}{c23}{blue}
	%	\tikzaround{c24}{blue}
	%	\path[-latex] (c23.north west) edge [bend left=50,blue](c24.north west);
	%	\end{tikzpicture}
\end{table}

When we study the pairing probabilities $p_{i,j}$, we can sum the rows and columns to obtain the marginals of this table. These sums constitute a constraint on the proportion of the slots which have to exist for each class of populations A and B. For instance in table~\vref{tab_example_pij_pi_pj}, if we want to respect the probabilities contained in the table, then a proportion of exactly 38\% of the dwellings should have surface=1. If it is not the case, then these probabilities can not be satisfied. We denote $p_i$ and $p_j$ the proportions of slots from entities having for classes $i$ and $j$. Variables $p_i$ and $p_j$ are governed by equations:
\doubleequation{eq:pi_pij}{eq:pj_pij}{\hat{p}_i = \sum_j \hat{p}_{i,j}}{\hat{p}_j = \sum_i \hat{p}_{i,j}}

\section{Theoretical framework\label{indoc_theoretical}}

%Our generation method is based on the assumption that the main difficulty, and task, of a pairing algorithm is to ensure these different values are consistent.

\subsection{Probabilist perspective of the pairing problem}

At this stage of the formalization of user inputs, it appears that all the inputs we defined have to be consistent with each other for the generation of a synthetic population to be possible. In order to generate the expected links for classes $Cla^A_i$, the proportions of slots $p_i$ should match the pairing probabilities; yet these proportions of slots for each class depend on the frequencies of classes $f_i$ and how many slots are created for each class (average degree $\tilde{d}_i$.

%The relative frequencies of classes $f_i$ describe the proportion of entities A which should exist for each class. We also derived the average degrees $\tilde{d}_i$ from the user defined distribution of degrees $pd_i$ for each of these classes. The average degree is linked with the relative frequencies. For instance if the average degree of a class is 0, then the proportion of slots for this class will also be null (as no slot is created for this class). If the average degree of a class is 2, then the corresponding proportion of slots will be bigger than other classes having degree 1. We saw before that the proportion of slots $p_i$ is related to the content of table $p_{i,j}$, which is itself linked with the proportion of slots $p_j$, and the relative frequencies $f_j$ through the average degree $\tilde{d}_j$. 

\begin{table}
	\raggedleft
	\tikzstyle{every picture}+=[remember picture,baseline]
	\tikzstyle{every node}+=[inner sep=0pt,anchor=base,
	minimum width=0.7cm,align=right,text depth=0.5ex,outer sep=1pt]
	\tikzstyle{every path}+=[thick, rounded corners]
	\vspace{3cm}
	\begin{tabular}{cccc|ccccc|c}
		&					&				&	\tikz\node (pijfi){$f_i$};			& $f_1$			& \textemdash	& $f_i$			& \textemdash	& $f_n$			& 1 \\
		&					&				&	\tikz\node (pijdi){$\tilde{d}_i$};	& $\tilde{d}_1$	& \textemdash	& $\tilde{d}_i$	& \textemdash	& $\tilde{d}_n$ & \\
		& 					& 				&	\tikz\node (pijpi){$p_i$};		& $p_1$ 		& \textemdash	& $p_i$			& \textemdash	& $p_n$			& 1 \\
		\tikz\node (pijfj){$f_j$};			& \tikz\node (pijdj){$\tilde{d}_j$}; & \tikz\node (pijpj){$p_j$};		&	indices			& $1$ 			& \textemdash 	& $i$ 			& \textemdash 	& $n$ 			& \\
		\hline
		$f_1$			& $\tilde{d}_1$		& $p_1$			&	$1$				& $p_{1,1}$		& \textemdash	& $p_{i,1}$		&  \textemdash 	& $p_{n,1}$ 	& \\
		\textbar		& \textbar			& \textbar		&	\textbar		& \textbar		&				& \textbar		&				& \textbar		& \\
		$f_j$			& $\tilde{d}_j$		& $p_j$			&	$j$				& $p_{1,j}$		& \textemdash	& $p_{i,j}$		& \textemdash	& $p_{n,j}$ 	&\\
		\textbar		& \textbar			& \textbar		&	\textbar		& \textbar		&				& \textbar		&				& \textbar		&\\
		$f_m$			& $\tilde{d}_m$		& $p_m$			&	$m$				& $p_{1,m}$		& \textemdash	& $p_{i,m}$		& \textemdash	& $p_{n,m}$ 	&\\
		\hline
		1				&					& 1				&					&				&				&				&				&				& 1 \\
	\end{tabular}
	\begin{tikzpicture}[overlay]
	\tikzstyle{every node}+=[font=\footnotesize]
	
	\node [above left = 1.0cm of pijfi,left,text=blue] (pipj) {relative frequencies of entities};
	\path[-latex] (pipj.east) edge [bend left=50,blue](pijfi);
	\path[-latex] (pipj.south west) edge [bend right=50,blue](pijfj);
	
	\node [above left = 0.8cm of pijdi,left,text=red] (didj) {average degree of entities};
	\path[-latex] (didj.east) edge [bend left=30,red](pijdi.west);
	\path[-latex] (didj.south west) edge [bend right=30,red](pijdj.north);
	
	\node [above left = 0.7cm of pijpi,left,text=blue](pipj) {relative frequencies\\of links};
	\path[-latex] (pipj.south east) edge [bend right=30,blue](pijpi.west);
	\path[-latex] (pipj.south east) edge [bend right=30,blue](pijpj.north);
	\end{tikzpicture}
	\caption{Probabilistic vision of the pairing problem: pairing probabilities $p_{i,j}$, corresponding probabilities for an entity in $A$ or $B$ being linked to have each combination of characteristics $p_i$, $p_j$, average degree $\delta_i$, $\delta_j$, frequencies of the different classes in A and B $f_i$ and $f_j$. The variables $pd_i$ and $pd_j$ are not presented here, but are part of the statistical view of the pairing problem, and are here present through the proxy values of average degrees $\tilde{d}_i$ and $\tilde{d}_j$}\label{tab_probabilistic}
\end{table}

\newcommand{\probabilistperspective}{$\langle f_i,\allowbreak pd_i,\allowbreak \tilde{d}_i,\allowbreak p_{i,j},\allowbreak \tilde{d}_j,\allowbreak pd_j,\allowbreak f_j \rangle$}

We represent these dependencies in figure~\vref{tab_probabilistic}, which depicts altogether the probabilistic variables related to populations A and B, disposed around the pairing probabilities $p_{i,j}$. 
This table depicts the \textit{probabilistic perspective of the pairing problem}, which is made of variables \probabilistperspective. 
This table contains the essence of the pairing problem, and stands as the intuition we rely on to elaborate our method. Figure~\vref{tab_probabilistic_example} represents the same table filled with the values we already introduced for our dwellings/household example.

Note that this probabilistic perspective contains the variables underlying the schema introduced in figure~\vref{fig_with_slots}: the proportions of each class of A are represented at the top of the table, the distribution of degree is encoded as average degrees, the pairing probabilities describe the proportions of links linking each combination of classes $i$ and $j$, etc.

%
%According to the form we defined for the user inputs, we listed several variables:
%\begin{itemize}
%\item we measured from sample A and sample B the frequencies of each set of characteristics $f_i$ and $f_j$, which represent the proportions of each class present in the original samples A and B (and thus assumed to be expected in the generated population)
%\item we measured from input distribution of degrees $pd_i$ and $pd_j$ the corresponding expected average degrees $\tilde{d}_i$ and $\tilde{d}_j$ which describe for each class of A and B how many links should be created on average for the entities of these classes. 
%\item we measured from the pairing probabilities $p_{i,j}$ the proportions of links $p_i$ and $p_j$ originating from entities of each class for A and B.
%\end{itemize}

We already introduced most of the relationship between the variables of this table. 
The novel relationship introduced in this probabilistic vision is the link between proportions of slots and the frequencies and average degrees.  
\textit{The proportions $p_i$ and $p_j$ of slots originating from each class, which are required by the pairing probabilities, should match the proportions of slots created from the population itself.} For each class $i$ of $A$, the relative frequencies $f_i$ define the proportion of each class in the target population. In the example of Table~\vref{tab_probabilistic_example}, 33\% of dwellings have surface 1. According to the distribution of degree $pd_i$, an average of 80\% of them will require a link. For the 33\% of dwellings having surface 3, they will require more links. In fact, the proportions $p_i$ correspond to the relative frequency $f_i$ multiplied by the average degree $\tilde{d}_i$ (normalized to reach a probability). The relationship between them should thus be: 
\doubleequation{eq:di_fi_pi}{eq:dj_fj_pj}{p_i = \dfrac{f_i . \tilde{d}_i}{\sum_{i'}f_{i'}\tilde{d}_{i'}}}{p_j = \dfrac{f_j . \tilde{d}_j}{\sum_{j'}f_{j'}\tilde{d}_{j'}}}

\begin{table}
\centering\footnotesize
\begin{tabular}{rrrr|rrr|c}
\multicolumn{3}{c}{\textbf{households}}	&					& \multicolumn{3}{c|}{\textbf{dwellings}} & \\ \hline
				&					&				&					& surface=1		& surface = 2 	& surface = 3	& \textit{totals} \\
				&					&				&	$f_i$			& 0.33			& 0.33			& 0.33			& 1 \\
				&					&				&	$\tilde{d}_i$	& 0.80			& 0.90			& 1.15 			& - \\
				& 					& 				&	$p_i$			& 0.38 			& 0.31			& 0.31			& 1 \\
$f_j$			& $\tilde{d}_j$ 	& $p_j$			& $p_{i,j}$			& 	 			& 				&				& 	\\
\hline
0.50			& 1.00				& 0.25			&					& 0.20			& 0.04			& 0.01			& \\
0.30			& 1.00				& 0.25			&					& 0.10			& 0.12			& 0.03 			& \\
0.15			& 1.00				& 0.25			&					& 0.05			& 0.10			& 0.10			& \\
0.05			& 1.00				& 0.25			&					& 0.03			& 0.05			& 0.17 			& \\
\hline
1				& -				& 1				&					&				&				&				& 1 \\
\end{tabular}
\caption{Probabilistic representation of the example before resolution. These values are not consistent and do not constitute a possible solution for generation.}\label{tab_probabilistic_example}
\end{table}

%In general, the entire system should be consistent: the relative frequencies $f_i$, related through the degree $\tilde{d}_i$ to $p_i$, should correspond the sum of columns of $p_{i,j}$. But also the sums of lines of $p_{i,j}$, which correspond $p_j$, should trough the average degrees $\tilde{d}_j$ be compliant with the relative frequencies $f_j$. 

The probabilistic perspective of the pairing system is entirely tied together by the equations we highlighted before. 
This explains why the previous iterative methods (see~\vref{indoc_stateart_samplefree}) always had to deal with matching issues: it is unlikely that the initial user parameters are naturally consistent together. A generation algorithm which relies on this system without solving it is bound to introduce biases in one or the other values. In our approach, we propose to solve this system analytically prior to the generation step, so the biases will be explicit and mastered instead of appearing implicitly because of the algorithmic process. 

\textit{A probabilistic view of a given pairing problem $\langle f_i,\allowbreak pd_i,\allowbreak \tilde{d}^i,\allowbreak p_{i,j},\allowbreak \tilde{d}^j,\allowbreak pd_j,\allowbreak f_j \rangle$ is said consistent} if all the equations \ref{eq:di_fi_pi}, \ref{eq:dj_fj_pj} (linking frequencies, degrees and pairing probabilities), \ref{eq:pi_pij} and \ref{eq:pj_pij} (linking pairing probabilities with slot probabilities) are satisfied; else the problem is said to be inconsistent. 
%
%One goal of the solving stage is to define approximate values $<\hat{f}_i,\allowbreak \hat{pd}_i,\allowbreak \hat{\tilde{d}}^i,\allowbreak \hat{p}_{i,j},\allowbreak \hat{\tilde{d}}^j,\allowbreak \hat{pd}_j,\allowbreak \hat{f}_j>$ such as these relationships are satisfied. 

\subsection{Discrete perspective of the pairing problem}

The statistical view of the pairing problem brings together the probabilities provided by the user as parameters. Yet an actual generation process should lead to a discrete version of this system: a discrete count of entities of each class $i$ will be generated; $n_A$ and $n_B$ entities A and B in the synthetic population ( $n_A$ and $n_B$ are parameters provided by the user); a finite count of $n_L$ links will be created to link these entities. This discrete aspect is build explicitly in IPF-based solutions in a so-called integerization stage (see~\ref{indoc_stateart_reweighting}). In iterative-based solutions  (see~\ref{indoc_stateart_samplefree}), the discrete aspect is only reached when the entities are generated and linked. In our method, we prefer the explicit solving of the discrete counterpart of the probabilistic perspective of the pairing problem. The explicit resolution will enable to explicitly deal with rounding issues, and to ensure the rounding are consistent between the counts of entities in each of the classes $i$ and $j$, the distribution of degrees, the counts of slots and the count of links between each of the classes $i$ and $j$.

\begin{table}[th]
	\centering
	\begin{tabular}{cccc|ccccc|c}
		
		%frequencies		
		&					&				&	$c_i$			& $c_1$			&  \textemdash	& $c_i$			& \textemdash	& $c_n$ 		& $n_A$ \\
		&					&				&	$\tilde{d}_i$	& $\tilde{d}_1$	& \textemdash	& $\tilde{d}_i$	& \textemdash	& $\tilde{d}_n$ & \\
		& 					& 				&	$n_i$			& $n_1$ 		& \textemdash	& $n_i$			& \textemdash	& $n_n$			& $n_L$ \\
		$c_j$			& $\tilde{d}_j$ 	& $n_j$			&	indices			& $1$ 			& \textemdash 	& $i$ 			& \textemdash 	& $n$ 			&  \\
		\hline
		$c_1$			& $\tilde{d}_1$		& $n_1$			&	$1$				& $n_{1,1}$		& \textemdash	& $n_{i,1}$		&  \textemdash & $n_{n,1}$ & \\
		\textbar		& \textbar			& \textbar		&	\textbar		& \textbar		&				& \textbar		&				& \textbar	& \\
		$c_j$			& $\tilde{d}_j$		& $n_j$			&	$j$				& $n_{1,j}$		& \textemdash	& $n_{i,j}$		& \textemdash	& $n_{n,j}$ & \\
		\textbar		& \textbar			& \textbar		&	\textbar		& \textbar		&				& \textbar		&				& \textbar	& \\
		$c_m$			& $\tilde{d}_m$		& $n_m$			&	$m$				& $n_{1,m}$		& \textemdash	& $n_{i,m}$		& \textemdash	& $n_{n,m}$ & \\
		\hline
		$n_B$			&					& $n_L$			&					&				&				&				&				&			& $n_L$ \\
	\end{tabular}
	\caption{Discrete vision of the pairing process, structured as the probabilistic vision with the discrete counterparts. Note the discrete representation of the pairing problem also includes tables $nd_i$ and $nd_j$ which are  represented here by the proxy variables of average degrees $\tilde{d}_i$ and $\tilde{d}_j$.}\label{tab_contigencies}
\end{table}

\newcommand{\discreteperspective}{$\langle c_i,\allowbreak nd_i,\allowbreak n_i,\allowbreak n_{i,j},\allowbreak n_j,\allowbreak nd_{j}, \allowbreak c_j \rangle$}

We name \textit{discrete perspective on the pairing problem} the discrete variables \discreteperspective. 
We represent in table~\vref{tab_contigencies} the discrete perspective of the pairing problem, in a table similar to the probabilistic perspective.
Each variable of the probabilistic perspective of the pairing problem has a discrete counterpart in the discrete perspective; we will list them below, as well as the relationships between the probabilistic and discrete variables and the relationships between the discrete variables.
%frequencies $f_i$ and $f_j$ of classes correspond to finite cardinalities of classes $A_i$ and $B_i$ denoted $c_i$ and $c_j$. Proportions of slots $p_i$ and $p_j$ have corresponding numbers of slots $n_i$ and $n_j$. The pairing probabilities $p_{i,j}$ also should have integer cardinalities $n_{i,j}$. The distribution of degrees $pd_i$ and $pd_j$ also will have discrete counterparts $nd_i$ and $nd_j$ which represent how many links of each class should lead to each count of slots $n$. 
%
\textit{A discrete perspective $\langle \hat{n}_A,c_i,\allowbreak nd_i,\allowbreak n_i,\allowbreak n_{i,j},\allowbreak n_j,\allowbreak nd_{j}, \allowbreak c_j, \allowbreak \hat{n}_B \rangle$ of a pairing problem is said to be consistent} 
if and ony if all the equations \ref{eq:ci_fi_nA}-\ref{eq:cj_ndi_pdi} are satisfied.

%iif equations \ref{eq:ci_fi_nA}, \ref{eq:cj_fj_nB}, \ref{eq:ci_fi} and \ref{eq:cj_fj} (related to cardinalities, degrees and slots), \ref{eq:ni_nij} and \ref{eq:nj_nij} (link number of slots with the number of links $n_{i,j}$), \ref{eq:ni_nij_pij} and \ref{eq:nj_nij_pij} (link the probabilist distribution of degrees and the count of entities given every ), \ref{eq:ci_nA} and \ref{eq:cj_nB} (related to counts of entities of each class), \ref{eq:ni_di_ci} and \ref{eq:nj_dj_cj} (proportion of entities and slots of each class given average degree) and \ref{eq:ci_ndi_pdi} and \ref{eq:cj_ndi_pdi} (related to degrees) \textit{are satisfied}. 

%The discretisation process which translates the probabilistic view into a discrete view is related to user parameters representing the expected count of entities $n_A$ and $n_B$, and is governed by several formula we list in the following paragraphs.

%The actual generation of the actual populations A and B, and of the links between entities of A and B, requires a discrete vision of the probabilistic problem. We do not need relative frequencies for generation (such as: TODO\% of the dwellings are small, TODO\% are medium, etc.), but contingencies (positive integers) for these values (such as: TODO small dwellings, TODO medium dwellings, etc.).

%The discrete representation of a problem is defined as the vector $< c_i, nd_i, n_i,\allowbreak n_{i,j},\allowbreak n_j,\allowbreak nd_{j}, \allowbreak c_j>$. Its resolution also implies the average degrees $\tilde{d}_i$ and $\tilde{d}_j$ 

%TODO ndi !!!

The discrete counterparts of the relative frequencies of classes $f_i$ (resp. $f_j$) are the \textit{cardinalities of each class} $c_i$ (resp. $c_j$).
They represent how many entities of each class should be generated in the synthetic population. 
Cardinalities $c_i$ are obtained from the relative frequencies $f_i$ multiplied by the total of entities to create of type A $n_A$ (and rounded). This relationship is governed by the equations: 
\doubleequation{eq:ci_fi_nA}{eq:cj_fj_nB}{\hat{c}_i = \text{round}\left(\hat{n}_A.\hat{f}_i\right)}{\hat{c}_j = \text{round}\left(\hat{n}_B.\hat{f}_j\right)}

\doubleequation{eq:ci_nA}{eq:cj_nB}{\hat{n}_A = \sum_i \hat{c}_i}{\hat{n}_B = \sum_j \hat{c}_j}

Note that if the counts of entities $c_i$ (resp. $c_j$) are known, we can infer directly the relative frequencies $f_i$ (resp. $f_j$)  and the total count of entities $n_A$ (resp. $n_B$). 
% Reciprocally, if we know frequencies $f_i$ and $n_A$ (resp $f_j$), then we can compute $c_i$ (resp. $c_j$).
\doubleequation{eq:ci_fi}{eq:cj_fj}{\hat{f}_i = \dfrac{\hat{c}_i}{\sum_{i'} \hat{c}_{i'}} = \dfrac{\hat{c}_i}{n_A}}{\hat{f}_j = \dfrac{\hat{c}_j}{\sum_{j'} \hat{c}_{j'}} = \dfrac{\hat{c}_j}{n_B}}

The \textit{numbers of slots} $n_i$ (respectively $n_j$) are the absolute frequencies of links originating from A (respectively reaching entities of B) for each class. If the counts of entities $c_i$ and  the average degrees $\tilde{d}_i$ are known, then the number of links $n_i$ for each class can be computed using: 
\doubleequation{eq:ni_di_ci}{eq:nj_dj_cj}{\hat{n}_i = \text{round}\left(\hat{c}_i.\hat{\tilde{d}}_i\right)}{\hat{n}_j = \text{round}\left(\hat{c}_j.\hat{\tilde{d}}_j\right)}

The \textit{numbers of links connecting each class of A and B} $n_{i,j}$ constitute the discrete counterpart of the relative frequencies of links $p_{i,j}$. As for the probabilistic vision, the counts of links originating from each source $n_i$ and each destination correspond the column (resp. line) totals of $n_{i,j}$: 
\doubleequation{eq:ni_nij}{eq:nj_nij}{\hat{n}_i = \sum_j \hat{n}_{i,j}}{\hat{n}_j = \sum_i \hat{n}_{i,j}}

If the relative frequencies of links $p_{i,j}$ and the counts of links $n_i$ are known, then a switch from the probabilistic and the discrete vision can be done using the following equations: % (TODO check):
\doubleequation{eq:ni_nij_pij}{eq:nj_nij_pij}{\hat{n}_{i,j} = \text{round}\left(\hat{n}_i.\hat{p}_{i,j}\right)}{\hat{n}_{i,j} = \text{round}\left(\hat{n}_j.\hat{p}_{i,j}\right)}

% TODO hats everywhere !!!

The last elements to discretized are the distributions of probabilities $pd_i$ and $pd_j$. They contain the number of slots $n$ to create for each class $i$ (and $j$). 
\doubleequation{eq:ci_ndi_pdi}{eq:cj_ndi_pdi}{\hat{nd}_i = \hat{c}_i.\hat{pd}_i}{\hat{nd}_j = \hat{c}_j.\hat{pd}_j}

\subsection{Relaxation of constraints}

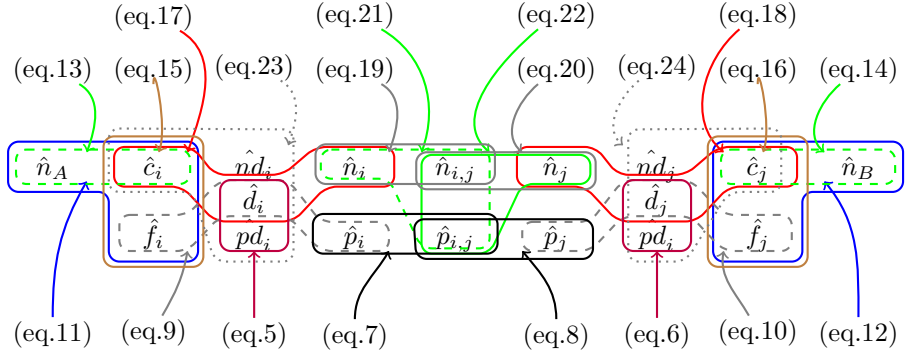
\begin{figure}[ht!]
	% Some options common to all the nodes and paths
	\tikzstyle{every picture}+=[remember picture,baseline]
	\tikzstyle{every node}+=[inner sep=0pt,anchor=base,
	minimum width=0.8cm,align=center,text depth=.25ex,outer sep=1.5pt]
	\tikzstyle{every path}+=[thick, rounded corners]
	\vspace{2cm}
	\begin{tabular}{cccccccccc}
		\tn{$\hat{n}_A$} 	& \tn{$\hat{c}_i$}	& \tn{$\hat{nd}_i$}	& \tn{$\hat{n}_i$}		& \tn{$\hat{n}_{i,j}$} 	& \tn{$\hat{n}_j$}	& \tn{$\hat{nd}_j$} & \tn{$\hat{c}_j$}	& \tn{$\hat{n}_B$} \\
		&					& \tn{$\hat{d}_i$}	&						&						&					& \tn{$\hat{d}_j$}	&					& \\
		& \tn{$\hat{f}_i$}	& \tn{$\hat{pd}_i$}	& \tn{$\hat{p}_i$}		& \tn{$\hat{p}_{i,j}$}	& \tn{$\hat{p}_j$} 	& \tn{$\hat{pd}_i$}	& \tn{$\hat{f}_j$}	& \\
	\end{tabular}
	\vspace{1cm}
	
	\begin{tikzpicture}[overlay]
	% nA * fi = ci 
	\draw [blue]([xshift=-4pt,yshift=4pt]1.north west) -- ([xshift=4pt,yshift=4pt]2.north east) -- ([xshift=4pt,yshift=-4pt]12.south east) -- ([xshift=-4pt,yshift=-4pt]12.south west) -- ([xshift=-4pt,yshift=-4pt]2.south west) -- ([xshift=-4pt,yshift=-4pt]1.south west) -- cycle;
	\draw [blue]([xshift=-4pt,yshift=4pt]8.north west) -- ([xshift=4pt,yshift=4pt]9.north east) -- ([xshift=4pt,yshift=-4pt]9.south east) -- ([xshift=4pt,yshift=-4pt]8.south east) -- ([xshift=4pt,yshift=-4pt]18.south east) -- ([xshift=-4pt,yshift=-4pt]18.south west) -- cycle;
	
	\node [left=2cm,below=2cm,minimum width=0pt] at (1) (eq_ci_fi_nA) {(eq.\ref{eq:ci_fi_nA})};
	\draw [blue,->,out=90,in=240] (eq_ci_fi_nA) to (1.south east);
	\node [left=2cm,below=2cm,minimum width=0pt] at (9) (eq_cj_fj_nB) {(eq.\ref{eq:cj_fj_nB})};
	\draw [blue,->,out=90,in=310] (eq_cj_fj_nB) to (9.south west);
	
	% ci * di = ni
	\draw [red]([xshift=-2pt,yshift=2pt]2.north west) -- ([xshift=2pt,yshift=2pt]2.north east) -- ([xshift=-2pt,yshift=2pt]10.north west) -- ([xshift=-2pt,yshift=2pt]10.north east) -- ([xshift=-2pt,yshift=2pt]4.north west)-- ([xshift=2pt,yshift=2pt]4.north east) -- ([xshift=2pt,yshift=-2pt]4.south east) -- ([xshift=2pt,yshift=-2pt]4.south west) -- ([xshift=2pt,yshift=-2pt]10.south east)  -- ([xshift=-2pt,yshift=-2pt]10.south west)  -- ([xshift=-2pt,yshift=-2pt]2.south east) -- ([xshift=-2pt,yshift=-2pt]2.south west) -- cycle;
	
	\draw [red]([xshift=-2pt,yshift=2pt]6.north west) -- ([xshift=2pt,yshift=2pt]6.north east) -- ([xshift=-2pt,yshift=2pt]11.north west) -- ([xshift=-2pt,yshift=2pt]11.north east) -- ([xshift=-2pt,yshift=2pt]8.north west)-- ([xshift=2pt,yshift=2pt]8.north east) -- ([xshift=2pt,yshift=-2pt]8.south east) -- ([xshift=2pt,yshift=-2pt]8.south west) -- ([xshift=2pt,yshift=-2pt]11.south east)  -- ([xshift=-2pt,yshift=-2pt]11.south west)  -- ([xshift=-2pt,yshift=-2pt]6.south east) -- ([xshift=-2pt,yshift=-2pt]6.south west) -- cycle;
	
	\node [right=1cm,above=1.81cm,minimum width=0pt] at (2) (eq_ni_di_ci) {(eq.\ref{eq:ni_di_ci})};
	\draw [red,->,out=330,in=70] (eq_ni_di_ci) to (2.north east);
	\node [right=1cm,above=1.83cm,minimum width=0pt] at (8) (eq_nj_dj_cj) {(eq.\ref{eq:nj_dj_cj})};
	\draw [red,->,out=220,in=140] (eq_nj_dj_cj) to (8.north west);
	
	% ni * pij = nij
	\draw [green,dashed]([yshift=1pt]4.north west) -- ([yshift=1pt]5.north east) -- ([yshift=1pt]15.south east) -- ([yshift=1pt]15.south west) -- ([yshift=1pt]4.south east) -- ([yshift=1pt]4.south west) -- cycle;
	
	\draw [green]([yshift=-1pt]5.north west) -- ([yshift=-1pt]6.north east) -- ([yshift=-1pt]6.south east) -- ([yshift=-1pt]6.south west) -- ([yshift=-1pt]15.south east) -- ([yshift=-1pt]15.south west) -- cycle;
	
	\node [right=1cm,above=1.83cm,minimum width=0pt] at (4) (eq_ni_nij_pij) {(eq.\ref{eq:ni_nij_pij})};
	\draw [green,->,out=330,in=70] (eq_ni_nij_pij) to (5.north west);
	\node [right=1cm,above=1.83cm,minimum width=0pt] at (6) (nj_nij_pij) {(eq.\ref{eq:nj_nij_pij})};
	\draw [green,->,out=220,in=140] (nj_nij_pij) to (5.north east);
	
	% fi, di, pi
	\draw [gray,dashed](12.north west) -- (12.north east) -- (10.north west) -- (10.north east) -- (14.north west) -- (14.north east) -- (14.south east) -- (14.south west) -- (10.south east) -- (10.south west) -- (12.south east) -- (12.south west) -- cycle;
	\draw [gray,dashed](16.north west) -- (16.north east) -- (11.north west) -- (11.north east) -- (18.north west) -- (18.north east) -- (18.south east) -- (18.south west) -- (11.south east) -- (11.south west) -- (16.south east) -- (16.south west) -- cycle;
	\\
	
	% ni, di, ci
	\node [right=1cm,below=1.1cm,minimum width=0pt] at (12) (eq_ni_fi_pi) {(eq.\ref{eq:di_fi_pi})};
	\draw [gray,->,out=60,in=260] (eq_ni_fi_pi) to (12.south east);
	\node [right=1cm,below=1.1cm,minimum width=0pt] at (18) (eq_nj_fj_pj) {(eq.\ref{eq:dj_fj_pj})};
	\draw [gray,->,out=120,in=260] (eq_nj_fj_pj) to (18.south west);

	% pi, pij
	\draw [black]([xshift=-3pt,yshift=3pt]14.north west) -- ([xshift=3pt,yshift=3pt]15.north east) -- ([xshift=3pt,yshift=-1pt]15.south east) -- ([xshift=-3pt,yshift=-1pt]14.south west) -- cycle;
	\draw [black]([xshift=-3pt,yshift=1pt]15.north west) -- ([xshift=1pt,yshift=1pt]16.north east) -- ([xshift=1pt,yshift=-3pt]16.south east) -- ([xshift=-2pt,yshift=-3pt]15.south west) -- cycle;
	
	\node [right=1cm,below=1.1cm,minimum width=0pt] at (14) (eq_pi_pij) {(eq.\ref{eq:pi_pij})};
	\draw [black,->,out=90,in=240] (eq_pi_pij) to (14.south east);
	\node [right=1cm,below=1.1cm,minimum width=0pt] at (16) (eq_pj_pij) {(eq.\ref{eq:pj_pij})};
	\draw [black,->,out=90,in=310] (eq_pj_pij) to (16.south west);
	
	% di, pdi
	\draw [purple](10.north west) -- (10.north east) -- (13.south east) -- (13.south west) -- cycle;
	\draw [purple](11.north west) -- (11.north east) -- (17.south east) -- (17.south west) -- cycle;
	
	\node [left=1cm,below=1.13cm,minimum width=0pt] at (13) (eq_di_pdi) {(eq.\ref{eq:di_pdi})};
	\draw [purple,->,out=90,in=270] (eq_di_pdi) to (13);
	\node [left=1cm,below=1.13cm,minimum width=0pt] at (17) (eq_dj_pdj) {(eq.\ref{eq:dj_pdj})};
	\draw [purple,->,out=90,in=270] (eq_dj_pdj) to (17);
	
	% na, ci
	\draw [green,dashed]([xshift=-1pt,yshift=1pt]1.north west) -- ([xshift=1pt,yshift=1pt]2.north east) -- ([xshift=1pt,yshift=-1pt]2.south east) -- ([xshift=-1pt,yshift=-1pt]1.south west) -- cycle;
	\draw [green,dashed]([xshift=-1pt,yshift=1pt]8.north west) -- ([xshift=1pt,yshift=1pt]9.north east) -- ([xshift=1pt,yshift=-1pt]9.south east) -- ([xshift=-1pt,yshift=-1pt]8.south west) -- cycle;
	
	\node [right=1cm,above=1.1cm,minimum width=0pt] at (1) (eq_na_ci) {(eq.\ref{eq:ci_nA})};
	\draw [green,->,out=330,in=70] (eq_na_ci) to (1.north east);
	\node [right=1cm,above=1.1cm,minimum width=0pt] at (9) (eq_nb_cj) {(eq.\ref{eq:cj_nB})};
	\draw [green,->,out=220,in=140] (eq_nb_cj) to (9.north west);
	
	% ni nij
	\draw [gray]([xshift=-2pt,yshift=3pt]4.north west) -- ([xshift=2pt,yshift=3pt]5.north east) -- ([xshift=2pt,yshift=-1pt]5.south east) -- ([xshift=-2pt,yshift=-1pt]4.south west) -- cycle;
	\draw [gray]([xshift=-2pt,yshift=0pt]5.north west) -- ([xshift=2pt,yshift=0pt]6.north east) -- ([xshift=2pt,yshift=-3pt]6.south east) -- ([xshift=-2pt,yshift=-3pt]5.south west) -- cycle;
	
	\node [right=1cm,above=1.1cm,minimum width=0pt] at (4) (eq_ni_nij) {(eq.\ref{eq:ni_nij})};
	\draw [gray,->,out=330,in=70] (eq_ni_nij) to (4.north east);
	\node [right=1cm,above=1.1cm,minimum width=0pt] at (6) (eq_nj_nij) {(eq.\ref{eq:nj_nij})};
	\draw [gray,->,out=220,in=140] (eq_nj_nij) to (6.north west);
	
	% ci fi
	\draw [brown]([xshift=-6pt,yshift=6pt]2.north west) -- ([xshift=6pt,yshift=6pt]2.north east) -- ([xshift=6pt,yshift=-6pt]12.south east) -- ([xshift=-6pt,yshift=-6pt]12.south west) -- cycle;
	
	\draw [brown]([xshift=-6pt,yshift=6pt]8.north west) -- ([xshift=6pt,yshift=6pt]8.north east) -- ([xshift=6pt,yshift=-6pt]18.south east) -- ([xshift=-6pt,yshift=-6pt]18.south west) -- cycle;
	
	\node [right=1cm,above=1.1cm,minimum width=0pt] at (2) (eq_ci_fi) {(eq.\ref{eq:ci_fi})};
	\draw [brown,->,out=90,in=70] (eq_ci_fi.south) to (2);
	\node [right=1cm,above=1.1cm,minimum width=0pt] at (8) (eq_cj_fj) {(eq.\ref{eq:cj_fj})};
	\draw [brown,->,out=90,in=70] (eq_cj_fj.south) to (8);
	
	% ci ndi pdi 
	\draw [gray,dotted]([yshift=9pt,xshift=-4pt]2.north west) -- ([yshift=6pt,xshift=2pt]3.north east) -- ([yshift=-4pt,xshift=2pt]13.south east) -- ([yshift=-4pt,xshift=-2pt]13.south west) -- ([yshift=-4pt,xshift=2pt]2.south east) -- ([yshift=-4pt,xshift=-4pt]2.south west) -- cycle;
	
	\draw [gray,dotted]([yshift=9pt,xshift=-4pt]8.north east) -- ([yshift=6pt,xshift=2pt]7.north west) -- ([yshift=-4pt,xshift=2pt]17.south west) -- ([yshift=-4pt,xshift=-2pt]17.south east) -- ([yshift=-4pt,xshift=2pt]8.south west) -- ([yshift=-4pt,xshift=-4pt]8.south east) -- cycle;
	
	\node [right=1cm,above=1.1cm,minimum width=0pt] at (3) (eq_ci_ndi_pdi) {(eq.\ref{eq:ci_ndi_pdi})};
	\draw [gray,dotted,->,out=330,in=70] (eq_ci_ndi_pdi) to (3.north east);
	
	\node [right=1cm,above=1.1cm,minimum width=0pt] at (7) (eq_cj_ndi_pdi) {(eq.\ref{eq:cj_ndi_pdi})};
	\draw [gray,dotted,->,out=220,in=140] (eq_cj_ndi_pdi) to (7.north west);
	
	% TODO remaining ?
	\end{tikzpicture}

	\caption{Variables to solve and their relationships. The figure is symmetrical, with on the left variables related to population A, on the right the variables related to population B, and in the center the variables constraining pairing. The first row lists the variables from the probabilistic view of the pairing problem; the last row contains the variables of the discrete problem; the middle line contains the proxy variables of average degree common to both understandings.}\label{tab_variables_relationships}
\end{figure}

We introduced \arabic{equation}~equations which define the probabilistic and discrete views of the pairing problem, and we defined the relationships between the probabilistic view, the discrete view, the data inputs and the user parameters. 
We depict all the variables and all the relationships defined by the equations in figure~\vref{tab_variables_relationships}. 
This synthetic view highlights how all the variables required for the resolution of the problem are covered by equations, and that these equations are all connected together; so having values for on variable should enable us, by applying equations, to obtain the results for all the other variables. 

\newcommand{\inputdata}{$\langle n_A,\allowbreak  f_i,\allowbreak  pd_i,\allowbreak  p_{i,j},\allowbreak  pd_j,\allowbreak  f_j,\allowbreak n_B \rangle$}

This view also underlines how most of the variables are covered by more than one equation.  
Given the user provides as inputs the target sizes $n_A$ and $n_B$, the distributions of degrees $pd_i$ and $pd_j$, the pairing probabilities $p_{i,j}$, and the relative frequencies $f_i$ and $f_j$, it means \textit{the system is over-constrained by nature}. Yet the relationships we identified between the variables reflect relationships which have to exist for a pairing solution to exist; it also makes sense to let the user provide the size of the population, the distributions of degrees, the frequencies of the classes in the target population, and the preferences for pairing. We claim the problem is not over-constrained because of the way we defined it - we just analyzed what the problem is - but is over-constrained by nature. 
Each input data constitutes a constraint of the problem. 
For the problem to be solved, we need to accept to relax some input data . 
In order to reach a consistent solution (that is a solution which respects all the equations which reflect the equalities necessary for generation to be possible), we need to \textit{relax some constraints provided by the user} in the form of data \inputdata. Relaxing constraints might also be understood as deciding explicitly were the errors should preferably occur. The user might prefer to relax given constraints in one case and others in another case.

\begin{table}
	\footnotesize
	\begin{tabular*}{\textwidth}{p{3.5cm}|ccccccc}
		\textbf{input data} & 
		$n_A$ & $f_i$ & $p(d_i=n)$& $p_{i,j}$ &  $p(d_j=n)$ & $f_j$ & $n_B$  \\
		\hline	
		\textbf{relaxation parameter} & 
		$\nu_A$ & 	$\phi_A$	& $\delta_A$ 	& $\gamma$ 		& $\delta_B$ 	& $\phi_B$ 	& $\nu_B$  \\
		\hline \hline
		no error anywhere; probably impossible & 
		0		&	0			& 0				& 0				& 0				& 0			& 0 \\
		\hline
		create exactly as many entities as specified, and distort if necessary the other elements &
		0		&	1			& 1				& 1				& 1				& 1			& 0 	\\
		\hline
		consider A as a list of entities, but relax pairing probabilities (proposed by experts) and frequencies for B (not statistically representative for small area) & 
		0		&	0			& 0				& 1				& 1				& 1			& 1 	 \\
		\hline
		complete flexibility: with equal repartition of biases &
		1		&	1			& 1				& 1				& 1 			& 1			& 1		\\
		complete flexibility, but errors on A are 100 times less important than on B &
		100		&	100			& 100			& 1				& 1 			& 1			& 100		\\
%		\hline
%		report all errors on $f_j$ and $n_B$ & 
%		0		&	0			& 0				& 0				& 0				& 1			& 1 	 \\
%		report all errors on $f_i$ and $n_A$ & 
%		1		&	1			& 0				& 0				& 0				& 0			& 0 	 \\
%		report all errors on $f_j$ and $n_B$  &
%		0		&	1			& 0				& 0				& 0				& 1			& 1 	\\
%		
		\hline
	\end{tabular*}
	\caption{Examples of parameter setting for the relaxation (line 2) of the various constraints (line 1). A relaxation of 0 means the resulting value should be equal. For instance, if $\nu_A=0$, then we should have $n_A=\hat{n}_A$.}\label{tab_relaxation_params}
\end{table}

\newcommand{\relaxationparameters}{$\langle \nu_A,\allowbreak \phi_A,\allowbreak  \delta_A,\allowbreak \gamma,\allowbreak \delta_B,\allowbreak \phi_B,\allowbreak \nu_B \rangle$}

We define relaxation parameters \relaxationparameters ~such as each relaxation parameter is a positive real taking value 0 if the approximated variable should equal or as close as possible to the input data, and have higher values if the importance of respecting this constraint is lower. For instance $\phi_A=0$ means that $\hat{f}_i = f_i$, whilst $\phi_A>0$ means that the input data should be enforced as much as possible but might be quiet different $\hat{f}_i \stackrel{?}{\simeq} f_i$. The relative values between two relaxation parameters describe the relative importance of errors: a relaxation parameter having value $2$ means the error on the corresponding variable is $2$ times less important than the one having a relaxation parameter of $1$. 

The user might sometimes accept to relax the relative frequencies $f_j$, for instance because these frequencies are indicative but are related to another scale (for instance national census) which is not relevant for the scale of interest. In this case, we might forget about the original frequencies $f_j$ (thus accept $\hat{f}_j \neq f_j$) and compute them from the probabilities $\hat{p}_j = p_j$ and the target degree $\hat{\tilde{d}}_j = \tilde{d}_j$. If the user prefers to relax the degrees instead, we might only consider the original $\hat{f}_i = f_i$ and $\hat{p}_j = p_j$ and infer the degrees from it. The table~\vref{tab_relaxation_params} illustrates the meaning of a few combinations of relaxation parameters.

\newcommand{\probabilistperspectivehat}{$\langle \hat{f_i},\allowbreak \hat{pd}_i,\allowbreak \hat{\tilde{d}}_i,\allowbreak \hat{p}_{i,j},\allowbreak \hat{\tilde{d}}^j,\allowbreak \hat{pd}_j,\allowbreak \hat{f}_j \rangle$}

\newcommand{\discreteperspectivehat}{$\langle \hat{n}_A,\allowbreak \hat{c}_i,\allowbreak \hat{nd}_i,\allowbreak \hat{n}_i,\allowbreak \hat{n}_{i,j},\allowbreak \hat{n}_j,\allowbreak \hat{nd}_j, \allowbreak \hat{c}_j,\allowbreak \hat{n}_A \rangle $}

The goal of a solver of the pairing problem is, by taking input data \inputdata ~, relaxation parameters \relaxationparameters, and using the equations 1- \arabic{equation}, to find approximate solutions for both the probabilistic \probabilistperspectivehat
~and the discrete perspectives \discreteperspectivehat ~minimizing the error $E$.

\subsection{Measure the quality of solutions}

\newcommand{\inputdatasolved}{$\langle \hat{n}_A,\allowbreak  \hat{f}_i,\allowbreak  \hat{pd}_i,\allowbreak  \hat{p}_{i,j},\allowbreak  \hat{pd}_j,\allowbreak  \hat{f}_j,\allowbreak \hat{n}_B \rangle $}

In our method, we measure the accuracy of a solution as the difference between the initial user data \inputdata ~and the solved solution \inputdatasolved. 
%Note that this is different from other measures in the state of the art. In the reweighting of micro samples approach, the measure of quality is 
We need this measure both to \textit{assess the quality of the solution after solving}, but also to decide which solution to prefer when several ones are available \textit{during the solving process}.

% chi2
In a review of the best measures to use for GoSP, Voas \cite{Voas2001} emphasis the Chi squared based values for goodness of fit. This family of measures computes both a $\chi^2$ (the higher the better the correlation) and a p-value which conveys the probability for such a correlation to appear by luck (the smaller the better goodness of fit). While these solutions provide an interesting semantic to assess goodness of fit in GoSP, their actual computation is less easy than it appears. The computation using the Pearson approach is not suitable for small values (contingencies $< 5$, low probabilities) and can not be computed in case of zero cells. The Freeman-Tukey approach \cite{read1993freeman} can be applied on low probabilities or null cells, but is not tractable on large tables nor on null marginals. Yet our solution precisely relies on the usage of large tables (to enable the usage of many criteria), potentially small probabilities, possible null values for structural zeros, etc. 
%
%  AAPD: not for us (no zero cells)
Average Absolute Percentage Deviation (AAPD) was sometimes used to assess the difference between the estimated (i.e. after fit) and generated distributions \cite{guo2007population} \cite{barthelemy2013synthetic}. 
This does not makes sense in our case, as the difference between the generated and solved values always is 0. Moreover, this measure involved a division by the initial required probability and thus renders impossible the measure of a method able to deal with zero cells. 

% => RMSE and NRMSE
A standard solution to measure goodness of fit in the Rooted Mean Squared Error (RMSE).
RMSE is an established measure of a model fitting in GoSP \cite{muller2011hierarchical,pritchard2012advances,Farooq2013,Zhu2014}.
%,  hydrology \cite{Besbeas2014} or TODO.
This method can deal with zero cells in expected and/or generated values; it penalizes the large differences, so the measured error will be bigger if a few cells are very different (outliers) than if many cells have little differences (this seems us more suitable to generation). 
RMSE is a value in $\mathbb{R}+$, the smaller the better. 

RMSE give results on the same scale as the measured numbers; before comparing RMSE, they first have to be normalized as a Normalized Root Mean Square Error (NRMSE). All the measures on probability tables or frequencies are by definition defined on a scale 0:1, and will not be scaled. We normalize the error rate on $n_A$ and $n_B$ by the expected size, meaning an error of 0.5 on $n_A$ means we generate half too much or not enough individuals compared to expectations. The errors are quantified as: 

\begin{equation}
\text{NRMSE}(\hat{p}_{i,j}) = \text{MSE}(\hat{p}_{i,j}) =  \sqrt{\dfrac{\sum_{i,j} (\hat{p}_{i,j} - p_{i,j})^2}{max(i) max(j)}}
\end{equation}

\begin{equation}
\text{NRMSE}(\hat{pd}_i) = \text{MSE}(\hat{pd}_i) =  \sqrt{\dfrac{\sum_{i,n} (\hat{pd}(n|i) - pd(n|j))^2}{max(i) max(n)}}
\end{equation}

\begin{equation}
\text{NRMSE}(\hat{f}_i) = \text{MSE}(\hat{fi}_i) =  \sqrt{\dfrac{\sum_i (\hat{f}_i - f_i)^2}{max(i)}}
\end{equation}

\begin{equation}
\text{NRMSE}(\hat{n}_A) = \dfrac{|\hat{n}_A - n_A|}{n_A}
\end{equation}

Given the relaxation parameters introduced before, we compute the weighted error of a solution $S$ as:
\begin{equation}\label{eq_error_solution}
\begin{aligned}
E(S) = & v(n_A,\nu_A) + v(f_i,\phi_i) + v(pd_i,\delta_i) + \\ 
& v(p_{i,j},\gamma) + \\
& v(pd_j,\delta_j) + v(f_j,\phi_j) + v(n_B,\nu_B) \\
\end{aligned}
\end{equation}
where 
\begin{equation*}
v(e,w) := \begin{cases} 0 & w = 0 \\ \text{NRMSE}(e)/w & w > 0 \end{cases} 
\end{equation*}
Meaning that for each variable having a non null relaxation parameter, the error is divided by the relaxation value (the higher the relaxation, the lower the importance of the corresponding error). Parameters for which weight is 0 are already enforced to the smallest possible error because of the resolution process, so their errors are not considered in the final result.

\subsection{Resolution of the system by inference}

Solving the system of equations might be done in many ways. 
A simple and intuitive one is to start from variables having relaxation parameter being 0 (no freedom), infer other variables using the equations, explore missing values by trying to consider them as having no freedom, and ensure consistency of the explored solutions during the whole process. We explain how the solving works manually, as it reflects exactly how we implemented the automatic solver algorithm.

First, if the user defined relaxation parameters to 0, it means s/he requires no error on the corresponding parameters. For instance if the user sets $\phi_i=0$, $\delta_i=0$ and $\gamma=0$, we can state $\hat{f}_i=f_i$, $\hat{pd}_i=pd_i$ and $\hat{p}_{i,j}=p_{i,j}$. Depending on the initial parameters, these assumptions might already create inconsistencies, that is some equations would not be verified. 

If the solution is consistent so far, we can try to \textit{infer} novel values using the equations we listed before. Given the probabilistic distribution of degrees $\hat{pd}_i$ we can compute the average degree $\tilde{d}_i$ using equation~\ref{eq:di_pdi}. Given pairing probabilities $\hat{p}_{i,j}$, we can compute the proportions of slots $\hat{p}_i$ and $\hat{p}_j$ using equations \ref{eq:pi_pij} and \ref{eq:pj_pij}. We have to ensure the system is still consistent, as our inference led to the availability of frequencies of classes $\hat{f}_i$, average degree $\hat{\tilde{d}}_i$ and the proportions of slots $\hat{p}_i$, which should be enforcing equation \ref{eq:di_fi_pi}. More generally, every time we infer a novel value, we should ensure this novel value is not invalidating any equation. If an equation is not satisfied at this stage, the resolution is said failed because the system is too constrained: the user asked for the satisfaction of too many constraints. Else the solving continues. 

At this stage, we have information for  $\hat{f}_i$, $\hat{\tilde{d}}_i$, $\hat{pd}_i$, $\hat{p}_i$, $\hat{p}_{i,j}$ and $\hat{p}_j$. No further equations apply directly, as they all would require more variables to be known for being applied. Yet the fact the user did not explicitly states that s/he wants to enforce $\hat{\tilde{d}}_i=\tilde{d}_i$, or $\hat{f}_i=f_i$, does not means those would be bad solutions; the user just leaves freedom to the solver, and lets it find a solution minimizing errors. So the solver has to explore the remaining hypothesis on the free variables: 
we might state $\hat{pd}_j=pd_j$ and then compute $\tilde{d}_j$ and $\tilde{f}_j$. 
Or, we might state $\hat{f}_j=f_j$, then compute the average degree $\tilde{d}_j$ required given the frequencies and slots proportions. 
Stating both $\hat{pd}_j=pd_j$ and $\hat{f}_j=f_j$ might fail if the system was not consistent before solving, so it would not lead a possible solution.
Or, we might state $\hat{n}_B=n_B$ and $\hat{f}_j=f_j$, then compute $\hat{c}_j$ (eq.~\ref{eq:cj_fj_nB}), the proportions of slots $\hat{n}_j$ (eq.~\ref{eq:nj_dj_cj}) and the rest of the system. 

More generally, we can \textit{explore all the possible combinations of hypothesis}, with an hypothesis being defined as stating the approximate solution of a variable $x$ being assumed to be the expected one: $\hat{x}=x$, and $x$ being one of $n_A$,$f_i$,$pd_i$,$p_{i,j}$,$pd_j$,$f_j$ or $n_B$. All these combinations of solutions are automatically generated and explored, as the maximum number of $k$-combinations remains small for a computer (in the worst case, $\sum_{0 \leq k < 7}\binom{7}{k}=2^7=128$). For each combination of hypothesis, the consistency is checked, inference is driven, and consistency is checked again. Consistent solutions are kept aside for later comparison.

If no valid solution is found, the solving process fails. If only one solution was found, it is returned. If several solutions are found, we only retain the solution having the minimal error\footnote{If several solutions have the same error, then one of them is chosen randomly (this constitutes the only stochastic case of the resolution process which is else deterministic). In practice, experience shows that exactly similar errors correspond to different hypothesis leading to the same conclusions, so this resolution of apparently multiple solutions often falls back the selection of the unique solution.}.

\begin{table}
	\centering\footnotesize
	\begin{tabular}{rrrr|rrr|c}
		
		%frequencies		
		\multicolumn{3}{c}{households}	&					& \multicolumn{3}{c|}{dwellings} & \\
		&					&				&					& surface=1		& surface = 2 	& surface = 3	& \textit{totals} \\
		&					&				&	$\hat{f}_i$			& 0.330			& 0.351			& 0.319			& 1 \\
		&					&				&	$\hat{\tilde{d}}_i$	& 0.9339394			& 0.6958405			& 0.7760502			& - \\
		& 					& 				&	$\hat{p}_i$			& 0.38525 			& 0.30530			& 0.30945			& 1 \\
		$\hat{f}_j$			& $\hat{\tilde{d}}_j$ 	& $\hat{p}_j$			& $\hat{p}_{i,j}$			& 	 			& 				&				& 	\\
		\hline
		0.27465			& 1.00				& 0.27465		&					& 0.219725		& 0.041200		& 0.013725			& \\
		0.22320			& 1.00				& 0.22320		&					& 0.089275		& 0.111600		& 0.022325		& \\
		0.26035			& 1.00				& 0.26035		&					& 0.052075		& 0.104125		& 0.104150		& \\
		0.24180			& 1.00				& 0.24180		&					& 0.024175		& 0.048375		& 0.169250		& \\
		\hline
		1				& -				& 1				&					&				&				&				& 1 \\
	\end{tabular}
	\caption{Probabilistic representation of the example after resolution (note the variable now have an hat). These values are consistent.}\label{tab_probabilistic_example_solved}
\end{table}

\begin{table}
	\centering\footnotesize
	\begin{tabular}{rrrr|rrr|c}
		%frequencies		
		\multicolumn{3}{c}{\textbf{households}}	&					& \multicolumn{3}{c|}{\textbf{dwellings}} & \\
		&					&				&					& surface=1		& surface = 2 	& surface = 3	& \textit{totals} \\
		&					&				&	$\hat{c}_i$			& 16,500			& 17,550			& 15,950			& $50,000=\hat{n}_A$ \\
		&					&				&	$\hat{\tilde{d}}_i$	& 0.9339394			& 0.6958405			& 0.7760502 			& - \\
		& 					& 				&	$\hat{n}_i$			& 15,410			& 12,212			& 12,378			& 40,000 \\
		$\hat{c}_j$			& $\hat{\tilde{d}}_j$ 	& $\hat{n}_j$			& $\hat{n}_{i,j}$			& 	 			& 				&				& 	\\
		\hline
		$10,986$		& 1.00				& $10,986$			&					& 8,789			& 1,648			& 549			& \\
		$8,928$			& 1.00				& $8,928$			&					& 3,571			& 4,464			& 893 			& \\
		$10,414$				& 1.00				& $10,414$			&					& 2,083			& 4,165			& 4,166			& \\
		$9,672$			& 1.00				& $9,672$			&					& 967			& 1,935			& 6,770 			& \\
		\hline
		$\hat{n}_B=40,000$ & -				& $40,000$				&					&				&				&				& 40,000 \\
	\end{tabular}
	\caption{Discrete representation of the example after resolution depicted in Table~\vref{tab_probabilistic_example_solved}.}\label{tab_contingencies_example}
\end{table}

Applying this process to the pairing example of dwellings and households depicted in Table~\vref{tab_probabilistic_example}, we obtain as a solution the probabilistic perspective~\vref{tab_probabilistic_example_solved} and the discrete perspective \vref{tab_contingencies_example}.

%TODO masks
%Prior to the resolution of the system, we first have to deal with \textit{structural zeros}, that is zero values in frequencies, average degrees and pairing probabilities which are strictly equal to zero and encompass the semantics of \textit{a case which should not exist} (for frequencies, average degree and pairing information} or \textit{a case which can not exist} (in case a class is not represented in the input samples). 
%These structural zeros are processed as follows: a zero in the frequencies of A (resp. B) leads to a zero in the corresponding pairing probabilities column (resp. line), because a class which should not be represented can not be the source nor destination of a link. 
%A column (resp. line) of the pairing probabilities summing to zero means no slot should be allocated, so if the average degree is not 
% TODO is that thanks to formula ?

Note that the discrete view is not computed after the probabilistic view, as an integerization post-processing step as done in literature (\ref{indoc_stateart_reweighting}). Here the discrete perspective of the problem contributes to solve the same time of the probabilistic perspective. The fact the discrete version of the problem is solved analytically also means the rounded values are ensured to be consistent among values.

\subsection{Automatic resolution of the system}

The manual resolution of the \arabic{equation}~equations would be too tedious and error prone to be applied manually in practice. 
As a consequence, we formalized the aforementioned process as an algorithm~\vref{algo_resolve} and implemented the process as a simple solver. We developed it as a package of the R statistical software \cite{RCoreTeam2018} \opensource{and released it as an opensource software}. 

\begin{algorithm}[thp]
	\caption{Resolve}
	\label{algo_resolve}
	\begin{algorithmic} % enter the algorithmic environment
		%    \REQUIRE $n \geq 0 \vee x \neq 0$
		%    \ENSURE $y = x^n$
		%    \STATE $y \Leftarrow 1$
		
		\Function{resolve}{
			% input data
			$f_i$, $f_j$, $pd_i$, $pd_j$, $p_{i,j}$,
			$n^A \in \mathbb{N}$, $n^B \in \mathbb{N}$,
			% relaxation
			$\phi^A \in \mathbb{R}^+$, $\phi^B \in \mathbb{R}^+$, $\delta^A \in \mathbb{R}^+$, $\delta^B \in \mathbb{R}^+$, $\gamma \in \mathbb{R}^+$
		}
		
		\State $c \leftarrow$ \Call{set\_initial\_values}{$f_i$, $f_j$, $pd_i$, $pd_j$, $p_{i,j}$, $n^A$, $n^B$,
			$\phi^A$, $\phi^B$, $\delta^A$, $\delta^B$, $\gamma$}
		
		\If{\textbf{not} \Call{consistent}{c}} 
		\State \Call{fail}{"case over-constrained: try relaxing parameters"}
		\EndIf 		
		\State  $c \leftarrow$ \Call{inference}{c} \Comment{Apply the equations to solve other variables}
		\If{\textbf{not} \Call{consistent}{c}} 
		\State \Call{fail}{"case over-constrained: try relaxing parameters"}
		\EndIf 	
		
		\If{\Call{is\_complete}{c}}
		\State\Return sol
		\Else										\Comment{Case not constrained enough: Formulate hypothesis}
		\State $S \leftarrow \emptyset$ 
		\State $H \leftarrow$ \Call{generate\_hypothesis}{c}
		\For{$h \in H$} 					\Comment{Test every hypothesis}
		\State $h \leftarrow$ \Call{inference}{$h$}
		\If{\Call{consistent}{$h$} \textbf{and} \Call{is\_complete}{$h$}}
		\State $S \leftarrow S \cup h$
		\EndIf
		\EndFor  
		\If{$|S| = 0$} 							
		\State \Call{fail}{"case over-constrained: no valid set of hypothesis"}
		%			\ElsIf{$|S|=1$}
		%				\State\Return H
		\Else
		\State\Return \Call{best\_solution}{S}
		\EndIf
		\EndIf
		\EndFunction
	\end{algorithmic}
\end{algorithm}

The actual implementation of this solver involves many technical details. 
As an example, the equations~\ref{eq:pi_pij} and~\ref{eq:pj_pij} describe the relationship between the pairing probabilities and the proportions of slots for A and B, which also are the marginals of the pairing probabilities. Such an equation might in practice lead to distinct resolution options:
\begin{itemize}
	\item if only $\hat{p}_i$ is known, then initial pairing probabilities can be reweighed so that they comply with these marginal $\hat{p}_{i,j}=p_{i,j}/\hat{p}_i$. 
	\item if only $\hat{p}_j$ is known, then pairing probabilities can be adapted in the same way. 
	\item if both $\hat{p}_i$ and $\hat{p}_j$ are known, then the reweighing of the pairing probabilities requires the usage of Iterative Proportional Fitting to adapt the pairing probabilities so that they match the totals.
\end{itemize}

Many other technical or methodological details have to be solved, such as 
the implementation of each equation in all the possible directions, 
rounding of matrices so to preserve vertical, horizontal or total sums, 
heuristic solutions to "reweigh" the probabilistic distributions of degrees in order to increase or decrease the average degrees, etc. 
These technical solutions are not presented in detail, as this solver constitutes only an example of how to deal with the theoretical problem introduced under the names of probabilistic and discrete perspectives of the pairing problem. \opensource{The solutions used for this paper can be directly analyzed for reproduction in the source code of the solver released in open source.}

Not all the equations can be translated to operational computation in all the directions.
For instance computing the frequencies $f_i$ based on the degrees $\delta{d}_i$ and proportions of slots $p_i$ using equation~\ref{eq:di_fi_pi}) is not feasible if the average degree is zero (division by zero), and not usable if the expected degree is very low (as it comes to divide by nearly zero and leads to very high figures). The fact the solver explores various hypothesis and then "paths" to solve the problem enables the resolution of complex cases by first assigning a value to $n_A$, then $n_i$, then $f_i$, thus enabling the computation of variables using workarounds in difficult cases.

\subsection{Direct generation from the discrete perspective}
%
%An actual population is defined as $<\hat{A}, \hat{B}, L>$. 
%
%
%\begin{algorithm}
%	\caption{Generation of a population for a solved pairing case}
%	\label{algo_generation}
%	\begin{algorithmic} % enter the algorithmic environment
%		%    \REQUIRE $n \geq 0 \vee x \neq 0$
%		%    \ENSURE $y = x^n$
%		%    \STATE $y \Leftarrow 1$
%		
%		\Function{match\_populations}{
%			$\mathcal{A}$, $\mathcal{B}$,
%			$p(Att(B)|Att(A))$,
%			$d^{in}(a)$,
%			$d^{out}(b)$,
%			$p(\mathcal{B})$}
%		
%		% TODO init actual degree 0
%		\State $\hat{A} \Leftarrow \bigcup_i sample\_n(A,c_i,w)$ with A without degrees \Comment{Add entities from samples}
%		\State $\hat{B} \Leftarrow \bigcup_i sample\_n(B,c_j,w)$
%		
%		\For{$i$, $n$} \Comment{Set target degree}
%		\State set target degree $sample\_n(A_i, nd_i(n)) \Leftarrow n$
%		\EndFor
%		
%		\For{$i$, $j$} \Comment{Create links}
%		\State $sample\_n(A_i, n_{i,j})$ with $A_i$ the subset of entities which have target degree lower than actual degree
%		\State $sample\_n(B_i, n_{i,j})$ with $A_i$ the subset of entities which have target degree lower than actual degree
%		\State create the link between a and b
%		\EndFor
%		
%		\Return  $<\hat{A}, \hat{B}, L>$
%		
%		% TODO repair 
%		
%		\EndFunction
%		
%	\end{algorithmic}
%\end{algorithm}
%
%Based on the discretised case described before, the algorithm~\ref{algo_generation} generate a synthetic population. 

The generation process is based on the solved discrete perspective on the pairing problem \discreteperspectivehat. The process is \textit{direct} (as denoted in the name of the method), because the consistency of the data \textit{and} the integerization were already solved beforehand. 

The steps of the generation process follows the intuition depicted on figure\ref{fig_with_slots}:
\begin{enumerate}
	\item \textbf{Generate entities A and B}: exactly $c_i$ and $c_j$ entities of each class $i$ and $j$ for populations A and B are copied out of the samples. 
	We take at once this count of entities out of the micro sample according to their weights.
	This process is stochastic, and is similar to the usage of a roulette biased by weights: each entity of the micro sample has a probability to be selected proportional to its weight divided by the sum of the weights of the other candidates. The same record of the micro sample might be reused several times; this will be the case for sure when the sample is up-sized because we generate more numerous entities than in the original sample. 
	Because we do not take these entities one after each other, we do not have to formulate a specific method in order to guarantee the statistical distribution of the properties of the entities; this problem was solved already. This sampling with replacement of $c_i$ entities out of a weighted sample is in practice delegated to the \textsc{sample\_n} method of the dyplr R package \cite{Wickham2018}. 
	\item \textbf{Generate slots A and B}: among the $c_i$ entities of each class $i$ of A, we know that exactly $pd_i(n=0)$ should have 0 slot, $nd_i(n=1)$ should have 1 slot, and so on for all the possible $n$ in the table $pd_i$. As a consequence, for each $i$ and $n$, we select $nd_i(n)$ random entities of class $i$ which had no target degree defined, and define their target degree to $nd_i(n)$. 
	\item \textbf{Generate links between slots of A and B}: the system was solved so that the count of slots A and B of each class exactly match the count of links to create from and to these slots; so the generation algorithm has no problem to deal with. For each class of $i$ and $j$, we select $nd_{i,j}$ random entities of class $i$ from A which do not yet have their degree equal to their target degree, and we select $nd_{i,j}$ random entities of class $j$ from B which do not yet have enough links. We add these links to the pool of links, and increase the degree of the corresponding entities A and B.	
\end{enumerate}

We only start the generation process after solving the problem as described in the previous step. So it means no error can be measured on this process, as it only directly matches the constraints defined by the user and solved to make them consistent. 
The only element to check is the distribution of the variables which are not controlled by the algorithm (not involved in classes $i$ and $j$). 

Note that the generation process, unlike the solving process, is stochastic, so two runs will lead to the selection of different records of the micro sample and different distributions of other characteristics' frequencies. However, every generation will enforce exactly the same proportions of classes $i$ and $j$. 

% complexity
The complexity of this process directly depends on the size of the population, the count of slots and the count of links. There is no additional cost due to iterating several times to find a relevant candidate as in iterative methods identified in the state of the art (cf~\ref{indoc_stateart_samplefree}).

\section{Experimental application\label{indoc_application}}

\subsection{Description of the case}

% data
As an illustration of our method, we generate synthetic populations of dwellings and households in the city of Lille in France. The micro samples for dwellings and households were collected during the 2014 census information by the French national institute for statistics (INSEE). These data sets are independent, in the sense \textit{they do not share any common identifier matching the dwelling and households prior to the generation process} (instead of the micro samples of type PUMS used in reweighing methods~\ref{indoc_stateart_pbsample}). We describe in annex \ref{annex_samples} p.~\pageref{annex_samples} the  preprocessing applied on these data sets. 

% micro samples
In the micro sample of dwellings (see excerpt in Annex Table~\ref{tab_sample_lille_dwellings} p.~\pageref{tab_sample_lille_dwellings}), dwellings are notably characterized by several categorical variables which include the surface SURF, the occupancy status CATL, and a weight IPONDL for each record. 
The micro sample of households (excerpt in Annex table~\ref{tab_sample_lille_households} p.~\pageref{tab_sample_lille_households}) contains one line per household's head, and describe the size of the household INPER, the age AGEREV or the employment status EMPL. 
We expect the target population to be made of dwellings (A) and households (B) holding 
the same characteristics as in the initial samples. 
We intend to generate a population representative of Lille in 2014, which is estimated by the institute of statistics to $n^A \sim 130000$ dwellings, and $n^B \sim 120000$ households. 

\newcommand{\countA}{51480}
\newcommand{\countB}{46138}

\begin{table}[pth]
	\centering
	% latex table generated in R 3.4.4 by xtable 1.8-3 package
% Tue Oct  2 13:23:08 2018
\begingroup\footnotesize
\begin{tabular}{rrrrrr}
  \hline
 & CATL=1 & CATL=2 & CATL=3 & CATL=4 & CATL=Z \\ 
  \hline
0 &   0 &   1 &   1 &   1 &   1 \\ 
  1 & 0.95 &   0 &   0 &   0 &   0 \\ 
  2 & 0.05 &   0 &   0 &   0 &   0 \\ 
   \hline
\end{tabular}
\endgroup

	\caption{Distribution of degrees for population A of dwellings.
		The count of households to add into each dwelling depends on variable CATL which encodes "household category", with
		1=main residence, 2=occasional residence, 3=secondary residence, 4=vacant residence, Z=not an ordinary buildings.}\label{lille_pdi}
\end{table}

\begin{table}[pth]
	\centering
	% latex table generated in R 3.4.4 by xtable 1.8-3 package
% Tue Oct  2 13:22:21 2018
\begingroup\footnotesize
\begin{tabular}{rrrrrrrl}
  \hline
 & INPER=1 & INPER=2 & INPER=3 & INPER=4 & INPER=5 & INPER=6 & ... \\ 
  \hline
0 &   0 &   0 &   0 &   0 &   0 &   0 & ... \\ 
  1 &   1 &   1 &   1 &   1 &   1 &   1 & ... \\ 
   \hline
\end{tabular}
\endgroup

	\caption{Distribution of degrees for population B of households.
		Households are always attached to exactly one unique dwelling, so this table does not really represent a dependency to this variable TACT (activity type).}\label{lille_pdj}
\end{table}

% degrees.
Depending on their occupancy status and surface, the dwellings might contain 0, 1 or 2 households, as encoded in the distribution of degrees table~\ref{lille_pdi}. "Occasional", "vacant" or "secondary residences" will contain no household. Following summary statistics from the statistics institute, 95\% of the "main residence" dwellings contain one dwelling, and only 5\% of them contain two of them. 
The dwellings are expected to always be connected to 1 and only 1 dwelling (we do not represent secondary residence nor homeless households in this study), as encoded in the corresponding table~\ref{lille_pdj}.
In this example, we expect both dwellings and households to enforce the frequencies found in these micro samples, as the samples delivered by INSEE are weighted at the small area scale (IRIS) \cite{insee_ponderations_2018}. The expected frequencies are depicted in figures~\vref{fig_lille_all_free_dwellings} and~\vref{fig_lille_all_free_households}.

% include the last version of the file generated by LaTeX
\begin{table}[htp]
	\begin{adjustbox}{width=1\textwidth}
	% latex table generated in R 3.4.4 by xtable 1.8-3 package
% Tue Oct  2 08:31:05 2018
\begingroup\footnotesize
\begin{tabular}{rrrrrrrr}
  \hline
 & SURF=1 & SURF=2 & SURF=3 & SURF=4 & SURF=5 & SURF=6 & SURF=7 \\ 
  \hline
INPER=1 & 0.15 & 0.10 & 0.14 & 0.07 & 0.04 & 0.01 & 0.01 \\ 
  INPER=2 & 0.01 & 0.02 & 0.07 & 0.07 & 0.04 & 0.02 & 0.02 \\ 
  INPER=3 & 0.00 & 0.00 & 0.01 & 0.03 & 0.03 & 0.01 & 0.01 \\ 
  INPER=4 & 0.00 & 0.00 & 0.01 & 0.01 & 0.02 & 0.01 & 0.01 \\ 
  INPER=5 & 0.00 & 0.00 & 0.00 & 0.01 & 0.01 & 0.01 & 0.01 \\ 
  INPER=6 & 0.00 & 0.00 & 0.00 & 0.00 & 0.00 & 0.00 & 0.00 \\ 
  INPER=7 & 0.00 & 0.00 & 0.00 & 0.00 & 0.00 & 0.00 & 0.00 \\ 
  INPER=8 & 0.00 & 0.00 & 0.00 & 0.00 & 0.00 & 0.00 & 0.00 \\ 
  INPER=9 & 0.00 & 0.00 & 0.00 & 0.00 & 0.00 & 0.00 & 0.00 \\ 
  INPER=10 & 0.00 & 0.00 & 0.00 & 0.00 & 0.00 & 0.00 & 0.00 \\ 
  INPER=12 & 0.00 & 0.00 & 0.00 & 0.00 & 0.00 & 0.00 & 0.00 \\ 
  INPER=14 & 0.00 & 0.00 & 0.00 & 0.00 & 0.00 & 0.00 & 0.00 \\ 
  INPER=Y & 0.01 & 0.00 & 0.00 & 0.00 & 0.00 & 0.00 & 0.00 \\ 
   \hline
\end{tabular}
\endgroup

	\end{adjustbox}
	\caption{Pairing probabilities for the Lille case, which define the probability for one generated link to associate a dwelling of a given surface SURF with a household of a given size INPER.,
	SURF is encoded as 1: Less than 30 m2; 2: from 30 to 40 m2; 3: from 40 to 60 m2; 4: from 60 to 80 m2; 5: from 80 to 100 m2; 6: from 100 to 120 m2; 7:120 m2 or more; Z:Out of standard categories.
	INPER is encoded as a count of individuals in the household, or Y for "out of standard housing".}\label{tab_lille_pij}
\end{table}
The pairing probabilities presented in table~\vref{tab_lille_pij} define the joint probability for linking dwellings and households given the surface SURF of the dwelling and the size of the household INPER. This simple correlation was extracted from INSEE data.
The classes $i$ for dwellings are made of the combinations of values for modalities SURF (surface) and CATL (occupancy), which are respectively necessary to compute the degree of dwellings and pairing probabilities. The tables for dwellings are thus expanded to represent these combinations. 
The classes $j$ are limited to the various counts of persons INPER.

This pairing problem can be seen as a table depicting the probabilistic perspective, depicted in annex table~\ref{tab_lille_allfree_probaview_init} (page~\pageref{tab_lille_allfree_probaview_init}). 

\subsection{Solution for a fully relaxed case}

\begin{figure}[thp]
	\includegraphics[width=\linewidth]{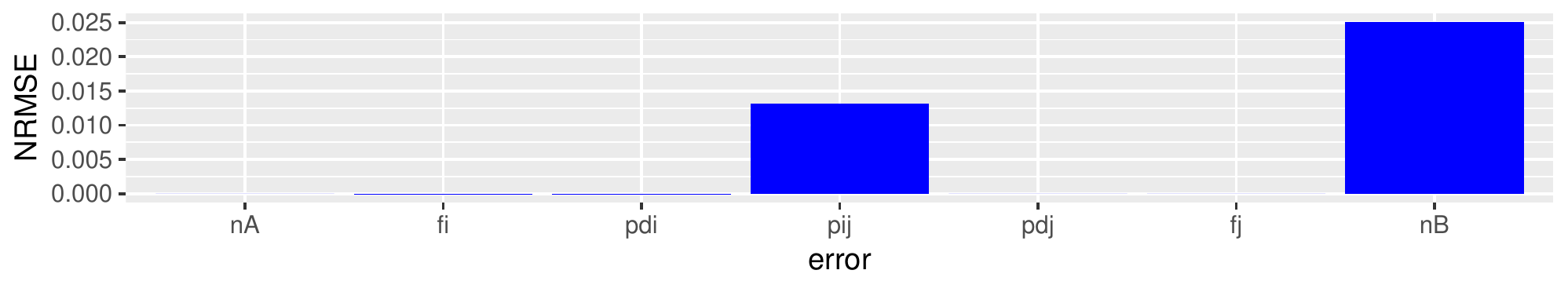}%
	\caption{Error rates of the case solved with all parameters relaxed.\label{fig_lille_all_free_errors}}
\end{figure}

We first run the solving of this pairing problem with all the relaxation parameters relaxed: $\nu^A=\allowbreak\phi^A=\allowbreak\delta^A=\allowbreak\gamma=\allowbreak\delta^B=\allowbreak\phi^B=\allowbreak\nu^B=1$. The solver explores all the possible 128 combinations of hypothesis. 8 valid solutions are found, with the one minimizing the weighted error being based on hypothesis: $\hat{n}_A=n_A,\allowbreak \hat{f}_i=f_i,\allowbreak \hat{pd}_i=pd_i,\allowbreak \hat{pd}_j=pd_j$ and $\hat{f}_j=f_j$. This solution accepts the required count of dwellings $\hat{n}_A=n_A$, preserves the frequencies for dwellings and households $f_i$ and $f_j$, the distribution of degrees $pd_i$ and $pd_j$, but does not preserve the pairing probabilities $p_{i,j}$ nor the count of households $\hat{n}_B$. The algorithm generates a population of exactly $\hat{n}_A=130000$ dwellings but $\hat{n}_B=123016$ households (slightly more than expected). The repartition of error rates (Fig.~\vref{fig_lille_all_free_errors}) shows that the solving process reported biases on pairing probabilities and count of entities $n_B$. 
Tables~\ref{tab_lille_freeall_perspective_proba_solved} and~\ref{tab_lille_freeall_perspective_discrete} in annex, in pages \pageref{tab_lille_freeall_perspective_proba_solved} and~\pageref{tab_lille_freeall_perspective_discrete} depict excerpts of the solved probabilistic perspective and the discrete perspective. 
We depict in table~\vref{tab_lille_all_free_synpop} an excerpt of the generated population.

\begin{figure}[htp]
	\includegraphics[width=\linewidth]{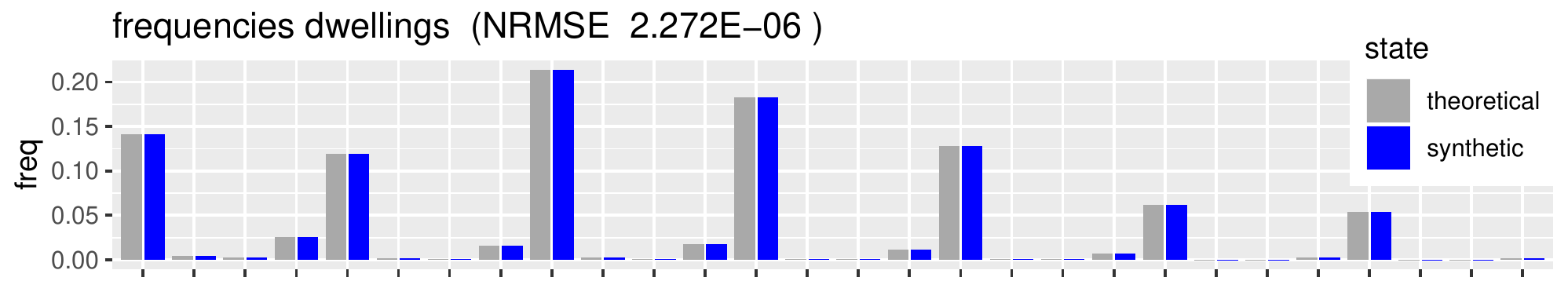}\\
	%\newline
	\includegraphics[width=\linewidth]{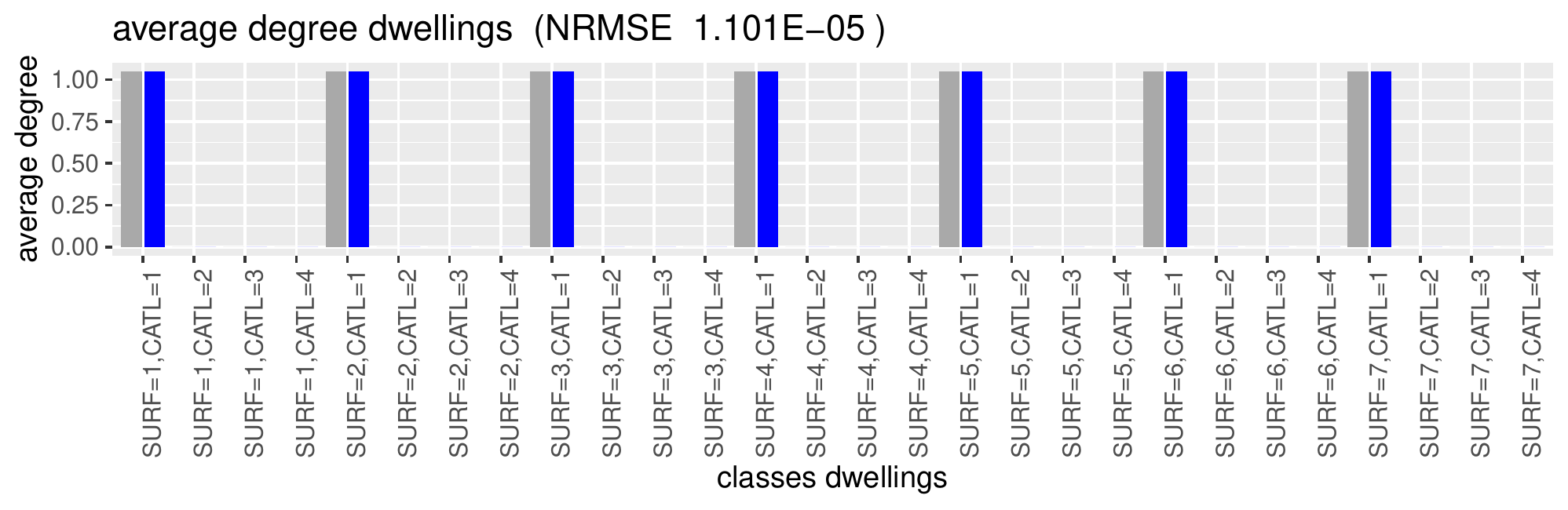}\\
	%\vspace{-2em}
	\caption{Comparison between expected and solved variables for dwellings in the fully relaxed experiment: \textit{(top)} frequencies for classes, \textit{(bottom)} average degrees.}\label{fig_lille_all_free_dwellings}
\end{figure}

The frequencies and average degrees of dwellings (Fig~\vref{fig_lille_all_free_dwellings}) are preserved with a very high precision. 
Even the numerous classes of dwellings for which the degree is 0 are represented as expected, demonstrating the capability of the solver based on our theoretical framework to deal with the zero cells. 
The measured NRMSE($f_i$) and NRMSE($d_i$) are very low, and correspond to the necessary rounding of probabilities introduced during solving.

\begin{figure}[htp]
	\hfil\includegraphics[width=0.99\linewidth]{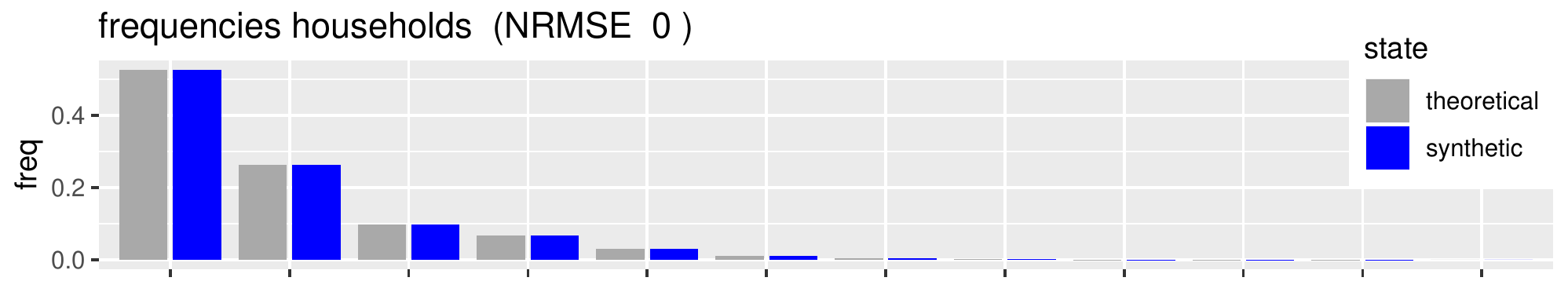}
	\includegraphics[width=\linewidth]{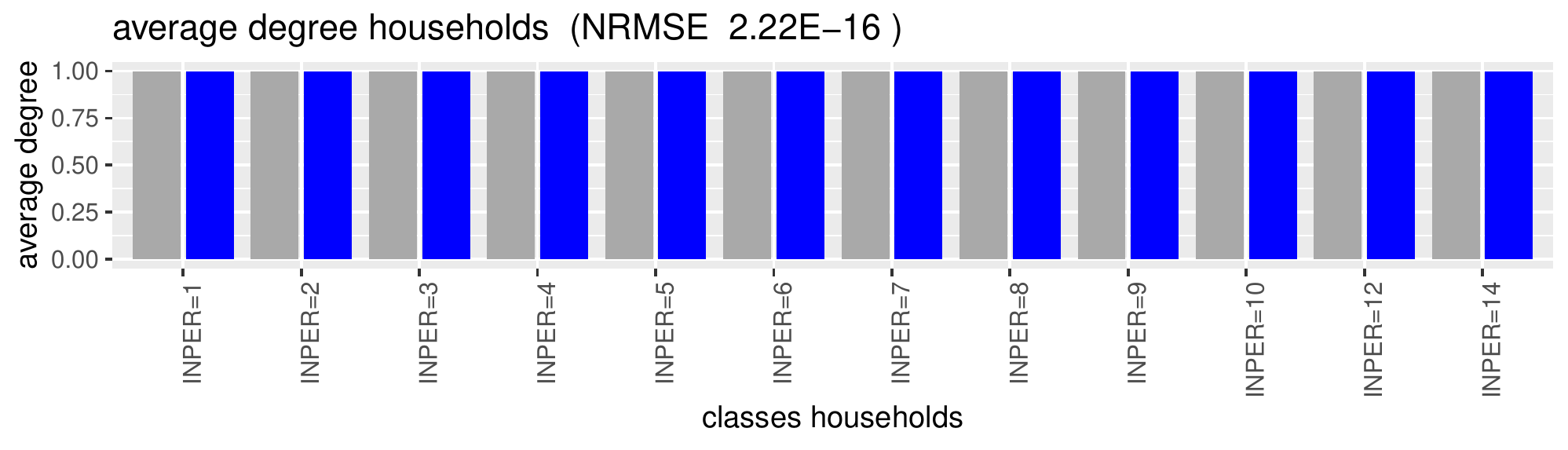}
	\caption{Comparison between expected and solved variables for households in the fully relaxed experiment: \textit{(top)} frequencies for classes, \textit{(bottom)} average degrees.}\label{fig_lille_all_free_households}
\end{figure}

On the side of households depicted in Table~\vref{fig_lille_all_free_households}, the frequencies and average degrees are enforced exactly; here even rounding did not lead to any error, as the probabilities in the distribution of degree were binary (1 or 0) rather than continuous. Note that even the frequencies which were null or nearly null were processed without specific workaround.

\begin{figure}[htp]
	\includegraphics[width=\linewidth]{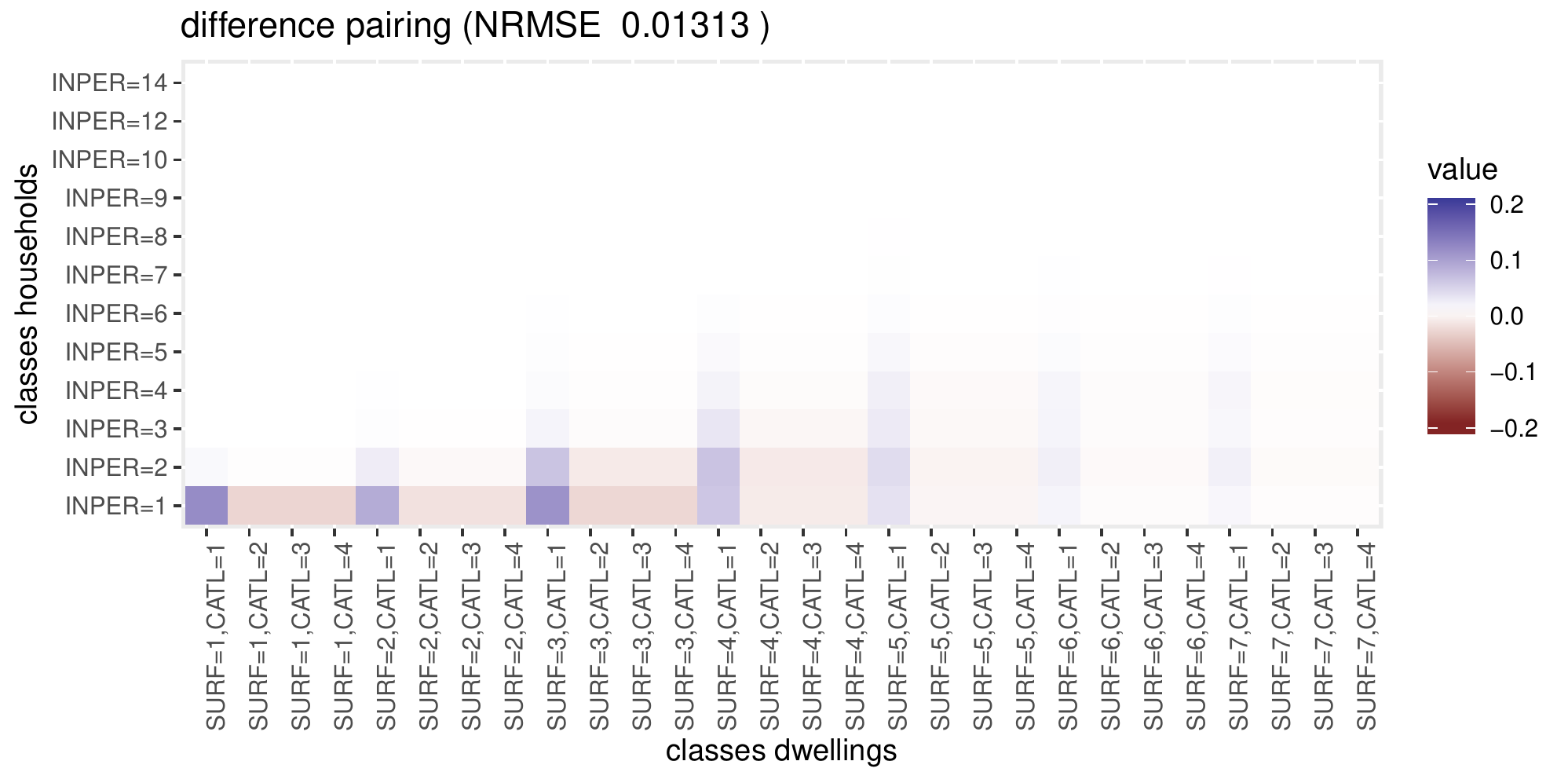}
	\caption{Pairing probabilities: heat-map of the differences (red) for less values than expected, blue for more.}\label{fig_lille_all_free_pairing}
\end{figure}
The pairing probabilities (Fig.~\vref{fig_lille_all_free_pairing}) show where the probabilities were mainly modified: there are slightly more links created between dwellings having CATL=1 (main residences) and small households (INPER=1 or 2). In order to keep the frequencies of households classes similar, this additional proportion of links was balanced during computation by a small diminution of the other links created for each line in order to enforce the marginals (and thus the frequencies for households); those last have no impact, as they only modify the proportions of slots for classes which have degree 0; as empty slots are not visible in the synthetic population, this will be neutral for the model. This correction of these probabilities was done because we provided contradictory information as an input: the expected degrees for dwellings depend on the category CATL which determines whether they are empty or not, but the pairing probabilities we provided only do depend on the surface SURF of the dwelling; we should have provided pairing probabilities with a null probability of a link connecting any dwelling having CATL different from 1. This constitutes an example of a correction of a bias which is desirable, and does not really introduce any bias in the synthetic population.

\begin{figure}[tp]
	\includegraphics[width=\linewidth]{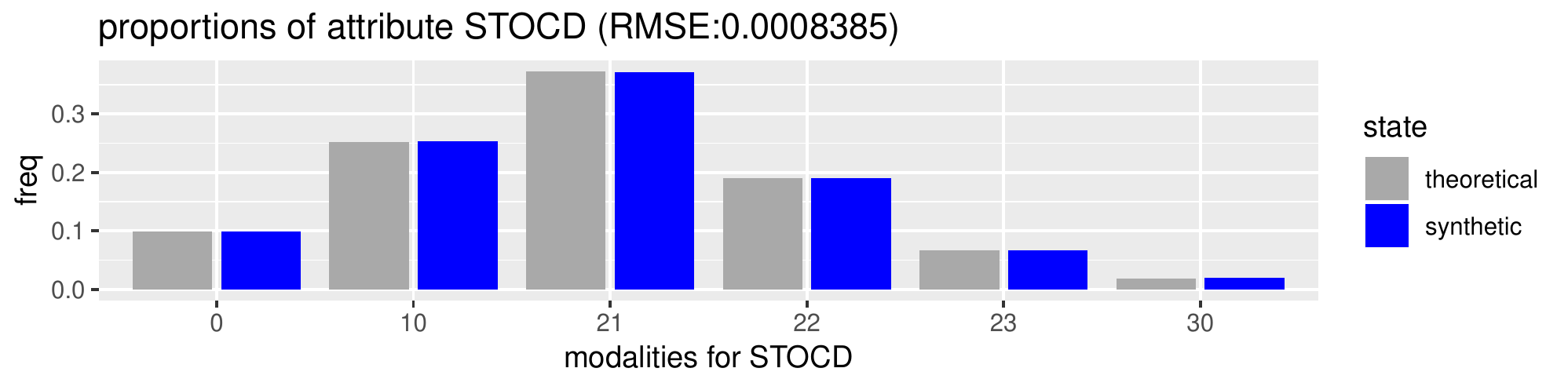}\newline
	\includegraphics[width=\linewidth]{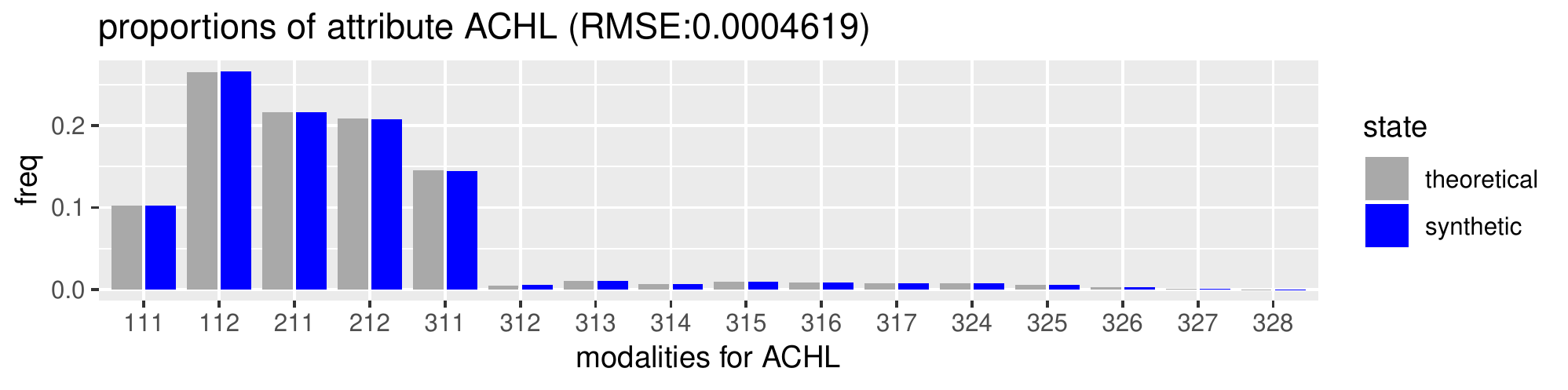}\newline
	\includegraphics[width=\linewidth]{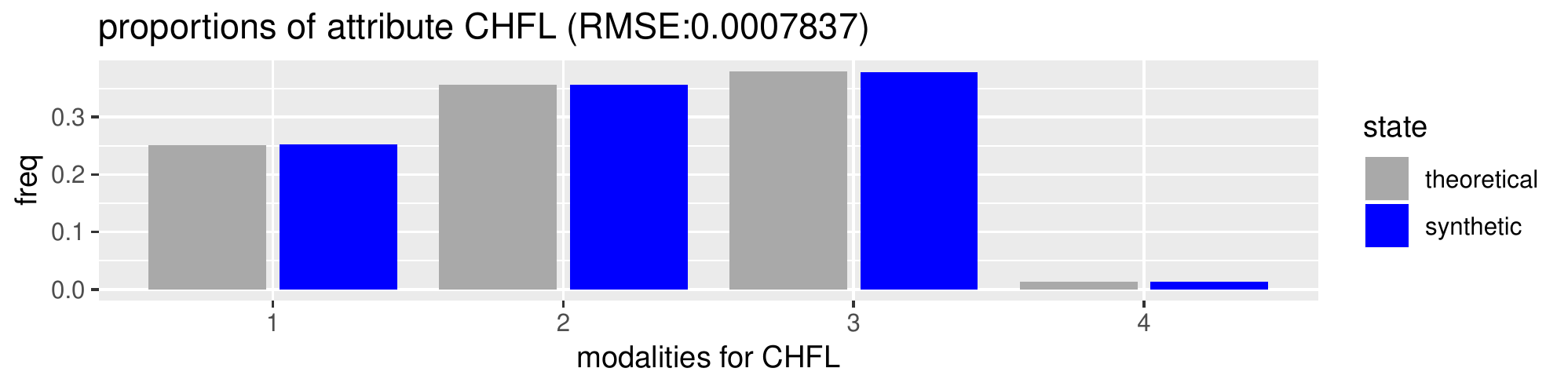}
	\caption{Comparison between the initial and synthetic distributions of free variables for dwellings in the fully relaxed case. From top to bottom: detailed occupancy status (STOCD), date of building construction (ACHL), type of heater (CHFL).
	}\label{fig_lille_all_free_vars_dwelling}
\end{figure}

\begin{figure}[tp]
	\includegraphics[width=\linewidth]{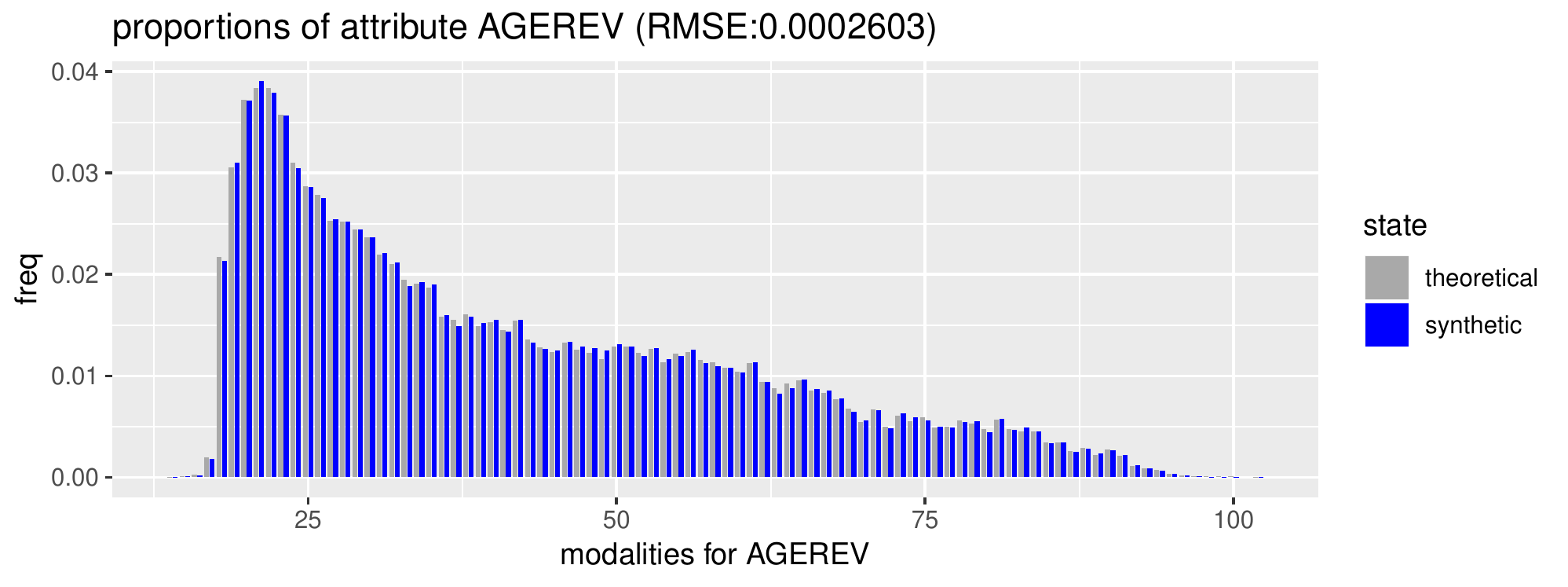}\newline
	\includegraphics[width=\linewidth]{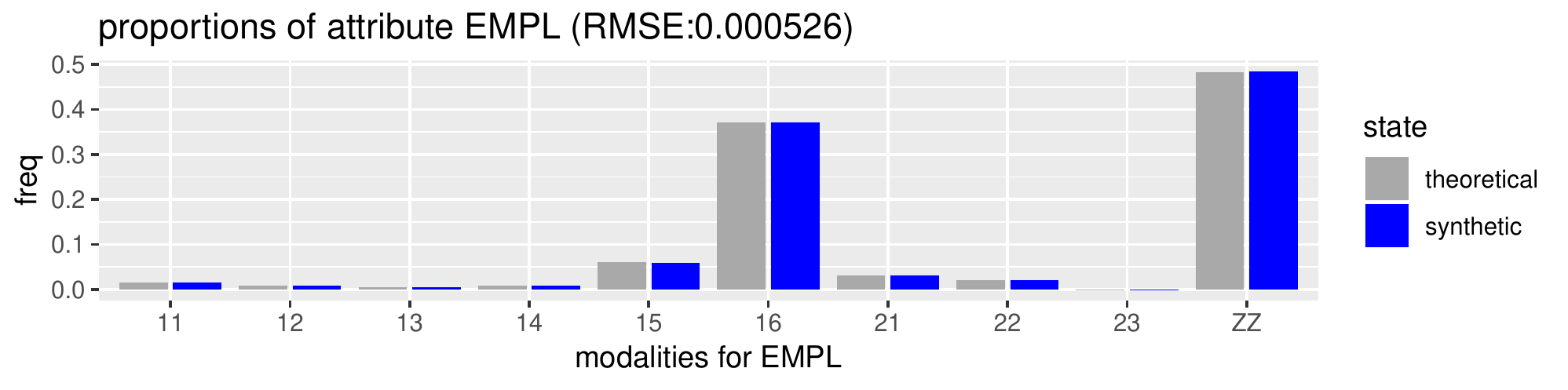}\newline
	\includegraphics[width=\linewidth]{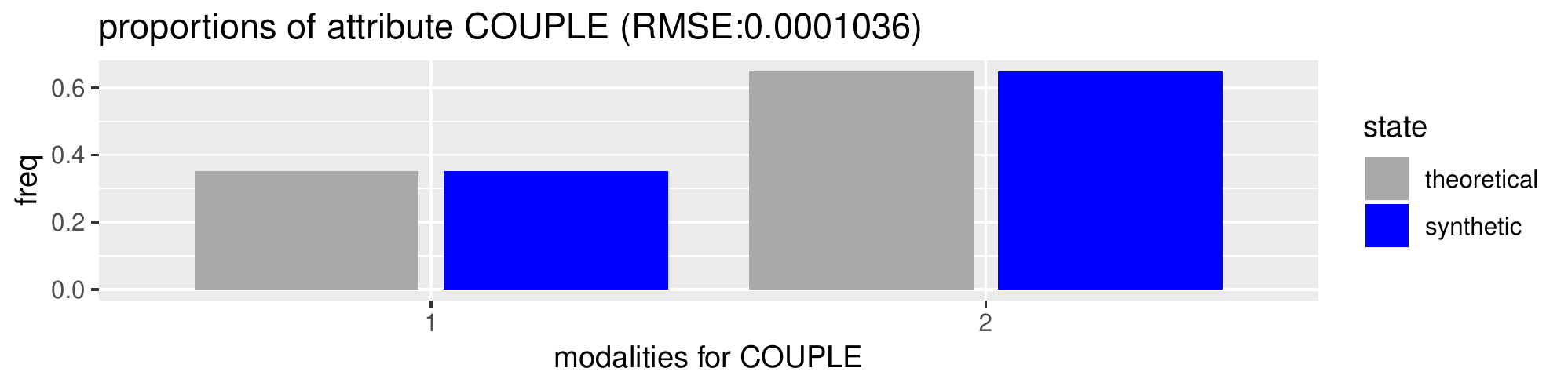}\newline
	\includegraphics[width=\linewidth]{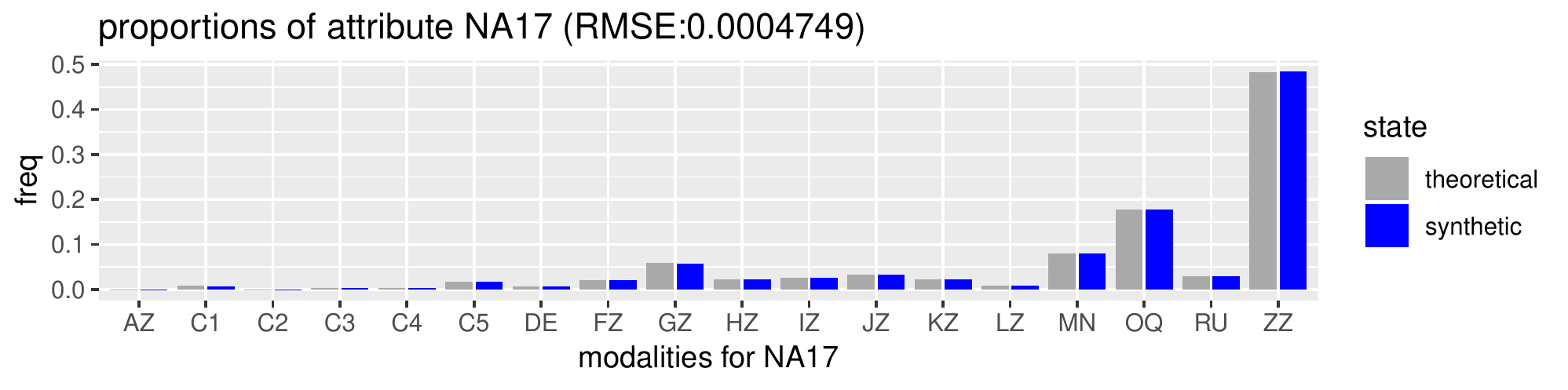}
	\caption{Comparison between the initial and synthetic distributions of free variables for households in the fully relaxed case. From top to bottom: 
		Detailed age of the head of the household (AGEREV),
		employment status (EMPL),
		marital status (COUPLE),
		type of job  (NA17)
	}\label{fig_lille_all_free_vars_households}
\end{figure}

If the solving process preserves as expected the constrained distributions of classes for the classes controlled during solving, there are other variables for dwellings and households which are not controlled. We depict in Fig.~\vref{fig_lille_all_free_vars_dwelling} and~\vref{fig_lille_all_free_vars_households}  the difference between initial distributions in the sample and the generated ones for dwellings. The error quantification are low enough for any usage. The absence of difference for the detailed variable AGEREV is of interest, as it has very similar distributions and very good aggregate statistics despite its many classes. The initial distribution of these weights is maintained because the generation phases selects randomly the entities to copy proportionally to their weights, and also because the frequencies of the controlled variables are enforced. If the proportion of empty dwellings was to be modified by the pairing algorithm, this distribution would naturally be biased in the same way according to the statistical dependencies present in the micro sample. 
%thus enforcing the statistical dependancies within variables.

\begin{sidewaystable}[pth]
	\center
	% latex table generated in R 3.4.4 by xtable 1.8-3 package
% Tue Oct  2 11:54:27 2018
\begingroup\tiny
\begin{tabular}{llllll}
  \hline
id.A & ACHL & CATL & CHFL & NBPI & SURF \\ 
  \hline
102565 & 112 & 1 & 4 & 5 & 5 \\ 
  63140 & 211 & 1 & 1 & 3 & 3 \\ 
  84201 & 112 & 1 & 1 & 3 & 4 \\ 
  110738 & 111 & 1 & 2 & 5 & 5 \\ 
  108720 & 211 & 1 & 1 & 8 & 5 \\ 
  10711 & 212 & 1 & 3 & 1 & 1 \\ 
  31670 & 211 & 1 & 2 & 3 & 2 \\ 
  74560 & 212 & 1 & 2 & 5 & 4 \\ 
  58291 & 325 & 1 & 3 & 1 & 3 \\ 
  39660 & 212 & 4 & 3 & 2 & 2 \\ 
  103030 & 212 & 1 & 2 & 5 & 5 \\ 
  78253 & 212 & 1 & 1 & 3 & 4 \\ 
  87708 & 212 & 1 & 2 & 1 & 4 \\ 
  52353 & 212 & 1 & 1 & 2 & 3 \\ 
  64265 & 211 & 1 & 3 & 1 & 3 \\ 
  85993 & 112 & 1 & 2 & 3 & 4 \\ 
  27427 & 111 & 1 & 3 & 2 & 2 \\ 
  88197 & 211 & 1 & 2 & 4 & 4 \\ 
  66125 & 211 & 1 & 1 & 1 & 3 \\ 
  114772 & 212 & 1 & 1 & 5 & 6 \\ 
  89387 & 212 & 1 & 3 & 3 & 4 \\ 
  91534 & 112 & 1 & 3 & 4 & 4 \\ 
  98405 & 211 & 1 & 1 & 5 & 5 \\ 
  79872 & 211 & 1 & 1 & 3 & 4 \\ 
  26858 & 112 & 1 & 1 & 2 & 2 \\ 
  106756 & 212 & 1 & 1 & 5 & 5 \\ 
  33034 & 111 & 1 & 2 & 2 & 2 \\ 
  9885 & 311 & 1 & 3 & 1 & 1 \\ 
  58510 & 212 & 1 & 3 & 2 & 3 \\ 
  9148 & 311 & 1 & 3 & 1 & 1 \\ 
  28266 & 212 & 1 & 3 & 2 & 2 \\ 
  13022 & 324 & 1 & 3 & 2 & 1 \\ 
  3756 & 311 & 1 & 3 & 1 & 1 \\ 
  33036 & 212 & 1 & 3 & 2 & 2 \\ 
  36126 & 112 & 1 & 3 & 2 & 2 \\ 
  128526 & 111 & 1 & 2 & 6 & 7 \\ 
  127527 & 112 & 1 & 3 & 7 & 7 \\ 
  10483 & 311 & 1 & 3 & 1 & 1 \\ 
  56975 & 211 & 1 & 1 & 2 & 3 \\ 
  124214 & 112 & 1 & 2 & 3 & 7 \\ 
  113171 & 211 & 1 & 2 & 4 & 5 \\ 
  1440 & 315 & 1 & 1 & 1 & 1 \\ 
  51318 & 212 & 1 & 2 & 1 & 3 \\ 
  39821 & 112 & 4 & 2 & 1 & 2 \\ 
  56995 & 211 & 1 & 3 & 2 & 3 \\ 
  69621 & 211 & 4 & 2 & 3 & 3 \\ 
  109885 & 211 & 1 & 1 & 3 & 5 \\ 
  103719 & 112 & 1 & 2 & 5 & 5 \\ 
  6418 & 325 & 1 & 1 & 1 & 1 \\ 
  39588 & 212 & 4 & 1 & 2 & 2 \\ 
  ... & ... & ... & ... & ... & ... \\ 
   \hline
\end{tabular}
\endgroup
	\quad\quad\quad
	% latex table generated in R 3.4.4 by xtable 1.8-3 package
% Tue Oct  2 11:54:27 2018
\begingroup\tiny
\begin{tabular}{ll}
  \hline
id.A & id.B \\ 
  \hline
102565 & 234203 \\ 
  63140 & 226070 \\ 
  84201 & 197463 \\ 
  110738 & 209788 \\ 
  108720 & 236952 \\ 
  10711 & 173346 \\ 
  31670 & 157397 \\ 
  74560 & 148814 \\ 
  58291 & 146090 \\ 
  39660 &  \\ 
  103030 & 234345 \\ 
  78253 & 130227 \\ 
  87708 & 148775 \\ 
  52353 & 246276 \\ 
  64265 & 248992 \\ 
  85993 & 136302 \\ 
  27427 & 175290 \\ 
  88197 & 132970 \\ 
  66125 & 161881 \\ 
  114772 & 236386 \\ 
  89387 & 211231 \\ 
  91534 & 176206 \\ 
  98405 & 226251 \\ 
  79872 & 149952 \\ 
  26858 & 146315 \\ 
  106756 & 240091 \\ 
  33034 & 175591 \\ 
  9885 & 156578 \\ 
  58510 & 197153 \\ 
  9148 & 150927 \\ 
  28266 & 161260 \\ 
  13022 & 143012 \\ 
  3756 & 163184 \\ 
  33036 & 179380 \\ 
  36126 & 185435 \\ 
  128526 & 249872 \\ 
  127527 & 208320 \\ 
  10483 & 150333 \\ 
  56975 & 194862 \\ 
  124214 & 250556 \\ 
  113171 & 205322 \\ 
  1440 & 174686 \\ 
  51318 & 204638 \\ 
  39821 &  \\ 
  56995 & 184359 \\ 
  69621 &  \\ 
  109885 & 234174 \\ 
  103719 & 199647 \\ 
  6418 & 193356 \\ 
  39588 &  \\ 
  ... & ... \\ 
   \hline
\end{tabular}
\endgroup
	\quad\quad\quad
	% latex table generated in R 3.4.4 by xtable 1.8-3 package
% Tue Oct  2 11:54:27 2018
\begingroup\tiny
\begin{tabular}{llllll}
  \hline
id.B & AGEREV & COUPLE & EMPL & INPER & NA17 \\ 
  \hline
234203 & 58 & 1 & 16 & 3 & OQ \\ 
  226070 & 41 & 2 & 16 & 2 & GZ \\ 
  197463 & 37 & 1 & 12 & 2 & MN \\ 
  209788 & 64 & 1 & ZZ & 2 & ZZ \\ 
  236952 & 38 & 1 & 16 & 3 & OQ \\ 
  173346 & 37 & 2 & ZZ & 1 & ZZ \\ 
  157397 & 41 & 2 & 16 & 1 & OQ \\ 
  148814 & 50 & 2 & ZZ & 1 & ZZ \\ 
  146090 & 21 & 2 & ZZ & 1 & ZZ \\ 
   &  &  &  &  &  \\ 
  234345 & 31 & 1 & 15 & 3 & OQ \\ 
  130227 & 69 & 2 & ZZ & 1 & ZZ \\ 
  148775 & 29 & 2 & 16 & 1 & MN \\ 
  246276 & 27 & 1 & ZZ & 4 & ZZ \\ 
  248992 & 33 & 2 & 16 & 5 & OQ \\ 
  136302 & 54 & 2 & 16 & 1 & HZ \\ 
  175290 & 20 & 2 & ZZ & 1 & ZZ \\ 
  132970 & 28 & 2 & 16 & 1 & GZ \\ 
  161881 & 44 & 2 & 12 & 1 & MN \\ 
  236386 & 32 & 2 & ZZ & 3 & ZZ \\ 
  211231 & 37 & 1 & 16 & 2 & C5 \\ 
  176206 & 58 & 2 & 16 & 1 & GZ \\ 
  226251 & 64 & 1 & ZZ & 2 & ZZ \\ 
  149952 & 74 & 2 & ZZ & 1 & ZZ \\ 
  146315 & 69 & 2 & ZZ & 1 & ZZ \\ 
  240091 & 54 & 2 & 21 & 4 & OQ \\ 
  175591 & 55 & 2 & ZZ & 1 & ZZ \\ 
  156578 & 61 & 1 & 16 & 1 & KZ \\ 
  197153 & 33 & 1 & 16 & 2 & MN \\ 
  150927 & 65 & 2 & ZZ & 1 & ZZ \\ 
  161260 & 23 & 2 & 16 & 1 & JZ \\ 
  143012 & 20 & 2 & ZZ & 1 & ZZ \\ 
  163184 & 33 & 1 & 16 & 1 & JZ \\ 
  179380 & 21 & 2 & ZZ & 1 & ZZ \\ 
  185435 & 39 & 2 & ZZ & 1 & ZZ \\ 
  249872 & 34 & 1 & ZZ & 5 & ZZ \\ 
  208320 & 36 & 2 & ZZ & 2 & ZZ \\ 
  150333 & 23 & 2 & ZZ & 1 & ZZ \\ 
  194862 & 38 & 1 & ZZ & 2 & ZZ \\ 
  250556 & 47 & 1 & 16 & 5 & OQ \\ 
  205322 & 32 & 2 & 16 & 2 & LZ \\ 
  174686 & 21 & 2 & 16 & 1 & GZ \\ 
  204638 & 33 & 2 & 16 & 2 & C5 \\ 
   &  &  &  &  &  \\ 
  184359 & 61 & 2 & ZZ & 1 & ZZ \\ 
   &  &  &  &  &  \\ 
  234174 & 32 & 1 & ZZ & 3 & ZZ \\ 
  199647 & 68 & 2 & ZZ & 2 & ZZ \\ 
  193356 & 70 & 2 & ZZ & 1 & ZZ \\ 
   &  &  &  &  &  \\ 
  ... & ... & ... & ... & ... & ... \\ 
   \hline
\end{tabular}
\endgroup
	\caption{Excerpt of the synthetic population generated for Lille. \textit{(left)} Population A (dwellings). \textit{(center)} Links between A and B (here: composition links). \textit{(right)} Population B (households). 
		Empty lines correspond to empty dwellings which have no corresponding household.}
	\label{tab_lille_all_free_synpop}
\end{sidewaystable}

%
%\begin{figure}[tp]
%%	\includegraphics[width=\linewidth]{130000_120000_1111111_plot_error_rates.pdf}\newline
%%	\includegraphics[width=\linewidth]{130000_120000_1111111_plot_frequencies_A.pdf}\newline
%%	\includegraphics[width=\linewidth]{130000_120000_1111111_plot_degrees_A.pdf}\newline
%	\includegraphics[width=\linewidth]{130000_120000_1111111_plot_frequencies_B.pdf}\newline
%	\includegraphics[width=\linewidth]{130000_120000_1111111_plot_degrees_B.pdf}\newline
%	\includegraphics[width=\linewidth]{130000_120000_1111111_plot_pij.pdf}
%	\caption{For the resolution of the case.
%		\protect\input{130000_120000_1111111_conditions.tex}. 
%		From top to bottom: 
%		comparison of error rates,
%		comparisons between expected and obtained frequencies of classes of dwellings, 
%		of households, 
%		heatmap for pairing.
%	}\label{fig_lille_all_free}
%\end{figure}
%

We applied this solution at the scale of Lille with only one constraint for summary statistics $f_i$ and $f_j$ for the entire city, because our initial sample is weighted so to be relevant at the local scale. The same approach might be used on each distinct statistical small area with different values if the initial sample is not statistically representative, as was done in the reweighing solutions~\ref{indoc_stateart_reweighting}.

\subsection{Impact of relaxation parameters}

We saw the result of resolution with all the relaxation parameters being relaxed, so the solver was free to explore all the possible solutions and retain the one minimizing the weighted error. 
We now test what happens if we constrain the case on pairing probabilities, so relaxation parameters are $\gamma=0$ and  $\nu^A=\allowbreak\phi^A=\allowbreak\delta^A=\allowbreak\phi^B=\allowbreak\delta^B=\allowbreak\nu^B=1$. 

\begin{figure}[hpt]
	\includegraphics[width=\linewidth]{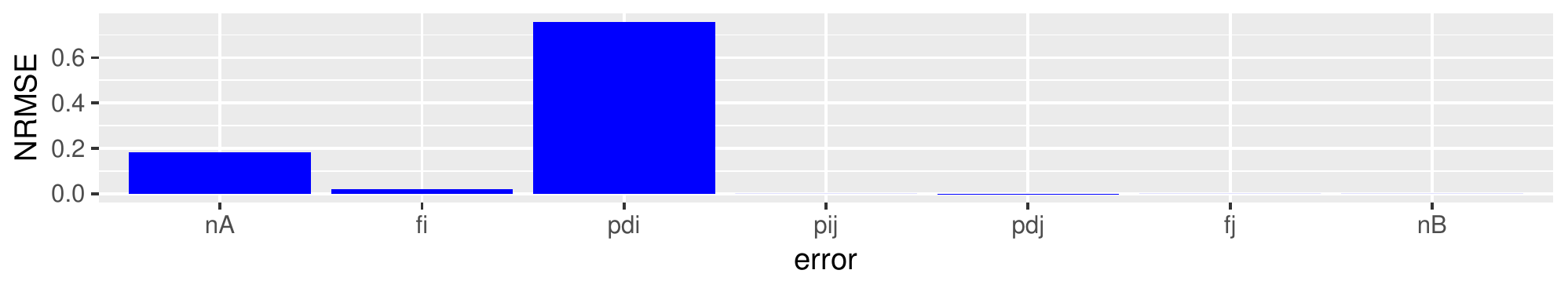} %
	\caption{Error rates of the case solved with the constraint $p_{i,j}=\hat{p}_{i,j}$.}\label{fig_lille_pairing_errors}
\end{figure}

% TODO \input{130000_120000_1110011_conditions.tex}.

\begin{figure}[htp]
	\flushright
	\includegraphics[width=\linewidth]{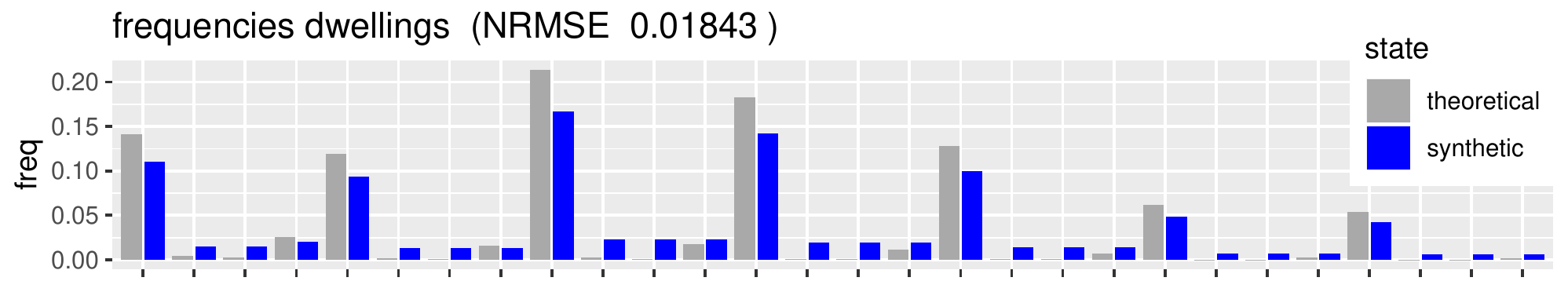}
	\hfil\includegraphics[width=0.99\linewidth]{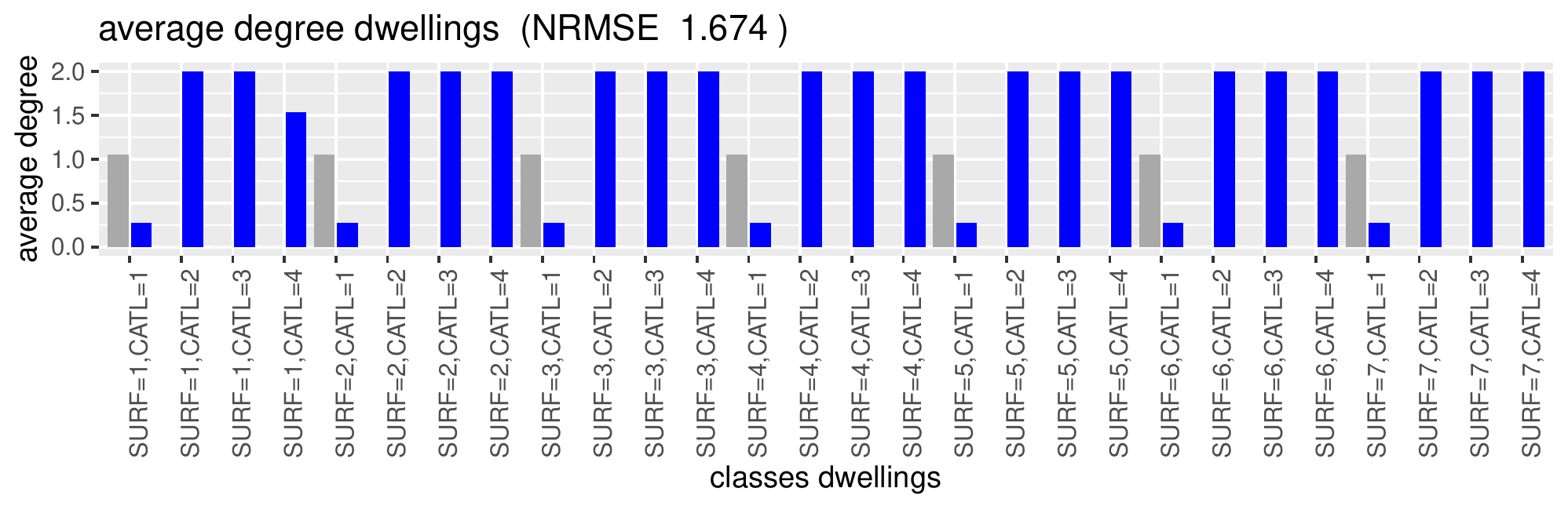}
	\includegraphics[width=\linewidth]{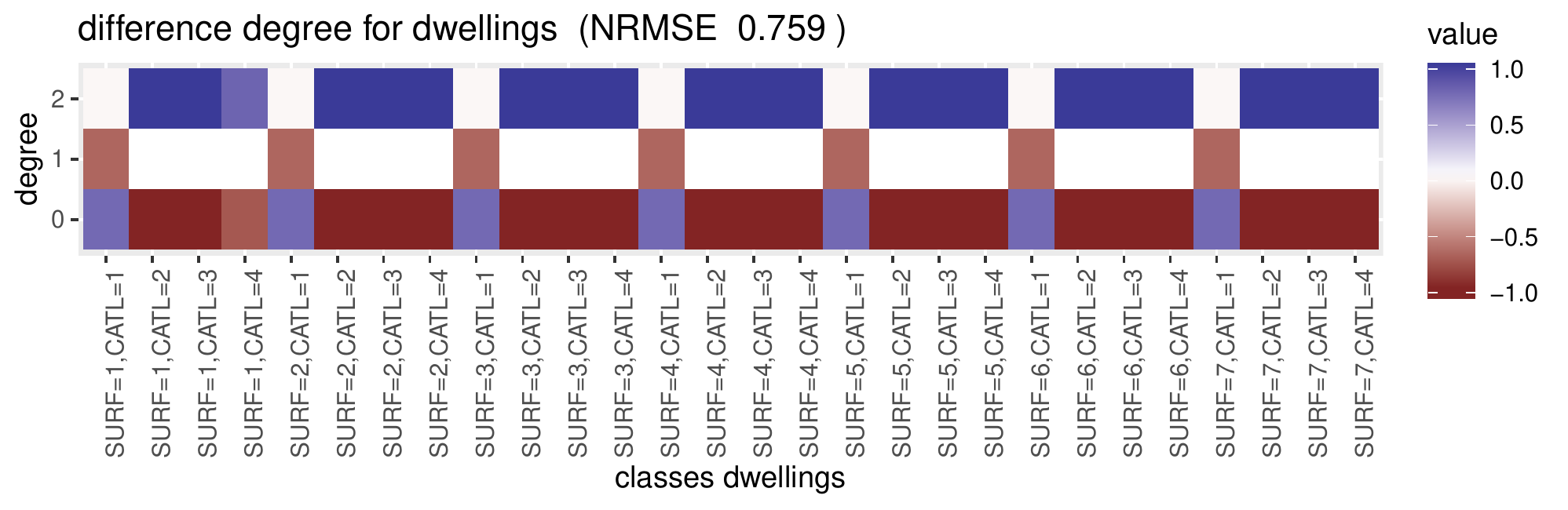}
	\caption{Comparison between expected and solved variables for dwellings when the pairing probabilities are not relaxed: \textit{(top)} frequencies for classes, \textit{(middle)} average degrees, \textit{(bottom)} detailed distribution of degrees}\label{fig_lille_pairing_degree_dwelling}
\end{figure}

This time, after the analysis of 16 valid solutions, the solver ends with a best solution based on hypothesis $\hat{f}_i=f_i$,$\hat{p}_{i,j}=p_{i,j}$, $\hat{f}_j=f_j$ and $\hat{n}^B=n^B$. The synthetic population contains $\hat{n}_A=153850$ (more than expected) and $\hat{n}_B=120000$. We depict in figure~\vref{fig_lille_pairing_errors} the errors obtained at the end of the process. 
The constraint on the pairing probabilities is enforced, with only a very low error rate due to rounding. But the errors obtained this time are high where the relaxation parameters allowed it. We plot in figure~\vref{fig_lille_pairing_degree_dwelling} the detail of the average degree and distribution of degrees. The frequencies of classes were modified a lot: all the classes leading to degree 0 (those with CATL!=1) are slightly over represented in relative frequencies, and their theoretical degree was shifted from 0 to 2. In other terms, because the pairing probabilities were requiring proportions of links even when no or few entities and slots were supposed to be created for them, the algorithm distorted these probabilities in order to create the necessary slots and links.
This huge distortion of the input parameters is probably not desirable in practice, as we try to enforce the pairing probabilities which are not consistent for some classes. 
In order to use a population, we would fix the pairing probabilities. 
In the scope of this paper however, this experiment demonstrated how the solver and theoretical frameworks provide the user with the freedom do define where to introduce biases, and enables to quantify the quality of the result.

%
%\begin{figure}[htp]
%	\includegraphics[width=\linewidth]{130000_120000_1110111_plot_pij.pdf}
%	\caption{Pairing probabilities: heatmap of the differences (red) for less values than expected, blue for more (TODO fix rank)}\label{fig_lille_pairing_probas}
%\end{figure}

% load the sideways table from its file
\renewcommand*\rot{\rotatebox{80}}
% latex table generated in R 3.4.4 by xtable 1.8-3 package
% Tue Oct  2 11:59:39 2018
\begin{sidewaystable}[ht]
\centering
\begingroup\tiny
\begin{tabular}{|cc|ccccccc|cc|ccccccc|cc|}
  \hline
$n^A$ & $n^B$ & $\nu^A$ & $\phi^A$ & $\delta^A$ & $\gamma$ & $\delta^B$ & $\phi^B$ & $\nu^B$ & $\hat{n}_A$ & $\hat{n}_B$ & \rot{nrmse.nA} & \rot{nrmse.fi} & \rot{nrmse.pdi} & \rot{nrmse.pij} & \rot{nrmse.pdj} & \rot{nrmse.fj} & \rot{nrmse.nB} & \rot{solving time} & \rot{generation time} \\ 
  \hline
130000 & 120000 & 1 & 1 & 1 & 1 & 1 & 1 & 1 & 130000 & 123016 & 0.00 & 0.00 & 0.00 & 0.01 & 0.00 & 0.00 & 0.03 & 4.14 & 14 \\ 
  130000 & 120000 & 1 & 1 & 1 & 1 & 1 & 1 & 0 & 114286 & 120000 & 0.12 & 0.00 & 0.00 & 0.01 & 0.00 & 0.00 & 0.00 & 4.17 & 10 \\ 
  130000 & 120000 & 1 & 1 & 1 & 1 & 1 & 0 & 1 & 130000 & 123016 & 0.00 & 0.00 & 0.00 & 0.01 & 0.00 & 0.00 & 0.03 & 3.19 & 14 \\ 
  130000 & 120000 & 1 & 1 & 1 & 1 & 0 & 1 & 1 & 130000 & 123016 & 0.00 & 0.00 & 0.00 & 0.01 & 0.00 & 0.00 & 0.03 & 3.47 & 14 \\ 
  130000 & 120000 & 1 & 1 & 1 & 0 & 1 & 1 & 1 & 153841 & 120000 & 0.18 & 0.02 & 0.76 & 0.00 & 0.00 & 0.00 & 0.00 & 4.97 & 39 \\ 
  130000 & 120000 & 1 & 1 & 0 & 1 & 1 & 1 & 1 & 130000 & 123016 & 0.00 & 0.00 & 0.00 & 0.01 & 0.00 & 0.00 & 0.03 & 2.21 & 14 \\ 
  130000 & 120000 & 1 & 0 & 1 & 1 & 1 & 1 & 1 & 130000 & 123016 & 0.00 & 0.00 & 0.00 & 0.01 & 0.00 & 0.00 & 0.03 & 3.33 & 15 \\ 
  130000 & 120000 & 0 & 1 & 1 & 1 & 1 & 1 & 1 & 130000 & 123016 & 0.00 & 0.00 & 0.00 & 0.01 & 0.00 & 0.00 & 0.03 & 4.20 & 15 \\ 
  130000 & 120000 & 0 & 1 & 0 & 1 & 1 & 1 & 1 & 130000 & 123016 & 0.00 & 0.00 & 0.00 & 0.01 & 0.00 & 0.00 & 0.03 & 2.63 & 15 \\ 
  130000 & 120000 & 1 & 1 & 0 & 1 & 1 & 1 & 0 & 114286 & 120000 & 0.12 & 0.00 & 0.00 & 0.01 & 0.00 & 0.00 & 0.00 & 2.44 & 11 \\ 
  130000 & 120000 & 0 & 0 & 1 & 1 & 1 & 1 & 1 & 130000 & 123016 & 0.00 & 0.00 & 0.00 & 0.01 & 0.00 & 0.00 & 0.03 & 3.86 & 15 \\ 
  130000 & 120000 & 0 & 0 & 1 & 1 & 1 & 1 & 1 & 130000 & 123016 & 0.00 & 0.00 & 0.00 & 0.01 & 0.00 & 0.00 & 0.03 & 3.85 & 14 \\ 
  130000 & 120000 & 1 & 0 & 0 & 1 & 1 & 1 & 1 & 130000 & 123016 & 0.00 & 0.00 & 0.00 & 0.01 & 0.00 & 0.00 & 0.03 & 2.92 & 15 \\ 
  130000 & 120000 & 1 & 1 & 0 & 0 & 1 & 1 & 1 & 130000 & 123016 & 0.00 & 0.00 & 0.00 & 0.01 & 0.00 & 0.00 & 0.03 & 2.86 & 16 \\ 
  130000 & 120000 & 1 & 1 & 1 & 0 & 0 & 1 & 1 & 153865 & 120000 & 0.18 & 0.02 & 0.76 & 0.00 & 0.00 & 0.00 & 0.00 & 4.53 & 42 \\ 
  130000 & 120000 & 1 & 1 & 1 & 1 & 0 & 0 & 1 & 130000 & 123016 & 0.00 & 0.00 & 0.00 & 0.01 & 0.00 & 0.00 & 0.03 & 4.69 & 15 \\ 
  130000 & 120000 & 0 & 0 & 0 & 1 & 1 & 1 & 1 & 130000 & 123016 & 0.00 & 0.00 & 0.00 & 0.01 & 0.00 & 0.00 & 0.03 & 2.67 & 16 \\ 
  130000 & 120000 & 1 & 0 & 0 & 0 & 1 & 1 & 1 & 130000 & 123016 & 0.00 & 0.00 & 0.00 & 0.01 & 0.00 & 0.00 & 0.03 & 3.50 & 16 \\ 
  130000 & 120000 & 1 & 1 & 0 & 0 & 0 & 1 & 1 & 130000 & 123016 & 0.00 & 0.00 & 0.00 & 0.01 & 0.00 & 0.00 & 0.03 & 2.22 & 15 \\ 
  130000 & 120000 & 1 & 1 & 0 & 0 & 0 & 1 & 1 & 130000 & 123016 & 0.00 & 0.00 & 0.00 & 0.01 & 0.00 & 0.00 & 0.03 & 2.25 & 15 \\ 
  130000 & 120000 & 1 & 1 & 1 & 0 & 0 & 0 & 1 & 153865 & 120000 & 0.18 & 0.02 & 0.76 & 0.00 & 0.00 & 0.00 & 0.00 & 5.96 & 40 \\ 
  130000 & 120000 & 1 & 1 & 1 & 1 & 0 & 0 & 0 & 114286 & 120000 & 0.12 & 0.00 & 0.00 & 0.01 & 0.00 & 0.00 & 0.00 & 4.92 & 12 \\ 
  130000 & 120000 & 1 & 0 & 1 & 0 & 0 & 0 & 0 & - & - & - & - & - & - & - & - & - & 0.03 & - \\ 
  130000 & 120000 & 1 & 0 & 0 & 0 & 0 & 0 & 0 & - & - & - & - & - & - & - & - & - & 0.04 & - \\ 
  130000 & 120000 & 1 & 1 & 1 & 0 & 0 & 0 & 0 & 153865 & 120000 & 0.18 & 0.02 & 0.76 & 0.00 & 0.00 & 0.00 & 0.00 & 4.15 & 41 \\ 
  130000 & 120000 & 0 & 1 & 1 & 0 & 0 & 1 & 1 & 130000 & 56633 & 0.00 & 0.02 & 0.76 & 0.01 & 0.00 & 0.00 & 0.53 & 5.32 & 25 \\ 
  130000 & 120000 & 0 & 1 & 1 & 1 & 1 & 1 & 1 & 130000 & 123016 & 0.00 & 0.00 & 0.00 & 0.01 & 0.00 & 0.00 & 0.03 & 4.32 & 16 \\ 
  130000 & 120000 & 0 & 0 & 1 & 1 & 1 & 1 & 1 & 130000 & 123016 & 0.00 & 0.00 & 0.00 & 0.01 & 0.00 & 0.00 & 0.03 & 3.97 & 16 \\ 
  130000 & 120000 & 0 & 0 & 0 & 1 & 1 & 1 & 1 & 130000 & 123016 & 0.00 & 0.00 & 0.00 & 0.01 & 0.00 & 0.00 & 0.03 & 2.76 & 15 \\ 
  130000 & 120000 & 1 & 0 & 0 & 1 & 0 & 0 & 1 & 130000 & 123016 & 0.00 & 0.00 & 0.00 & 0.01 & 0.00 & 0.00 & 0.03 & 4.29 & 16 \\ 
  130000 & 120000 & 1 & 1 & 1 & 0 & 1 & 1 & 1 & 153841 & 120000 & 0.18 & 0.02 & 0.76 & 0.00 & 0.00 & 0.00 & 0.00 & 5.18 & 40 \\ 
   \hline
\end{tabular}
\endgroup
\caption{Exploration of the space of solvers' parameters applied on the Lille case:  target population sizes $n_A$ and $n_B$, relaxation parameters,  resulting synthetic population sizes $\hat{n}_A$ and $\hat{n}_B$,  Normalized Rooted Mean Squarred Errors, solving and generation time in seconds.} 
\label{tab_relax_errors}
\end{sidewaystable}

We test other combinations of relaxation parameters to illustrate the potential results and demonstrate the capability of our method to assign the biases in different places of the problem, thus enabling a user to enforce at least part of his constraints. 
The results are depicted in table~\vref{tab_relax_errors}.
When a relaxation parameter set to 0 is leading to fail, then all the combinations involving the same parameter also fails.
Note that the computation is sometimes not possible because the case if considered over-constrained, and is then refused. Note the generation time is also depicted in this table, and remains below one minute. 

% TODO creuser: pourquoi ça marche pas avec nB ? hein ? 
%
%During this exploration of the space, we also observe that (i) variation between several generations of populations are neglictible in the error rates, (ii) performance is about linear with time TODO.

\section{Discussion\label{indoc_discussion}}

\subsection{Summary}

We tackle the problem of \textit{generating a synthetic population made of entities of type A and B in which entities A and B can be connected together with 0, 1 or more links according to their characteristics, and A and B enforce given distributions of frequencies}. 
%from micro samples of A and B and summary statistics, 

% framed inputs 
We proposed the \textit{semantics and formats for the input data} (\ref{indoc_inputs}) made of weighted samples of A and B, definition of classes $i$ and $j$ made of combinations of modalities of variables for A and B, expected frequencies for these classes $f_i$ and $f_j$, distribution of degrees encoded as conditional probabilities for an entity A or B to have degree $n$ given it belongs to class $i$ or $j$, and pairing constraints in the form of a joint probability table between classes $i$ and $j$. 
% analytic framework
We proposed a \textit{theoretical framework} (\ref{indoc_theoretical}) to analyze and solve the pairing problem, which is made of variables describing the \textit{probabilistic perspective of the pairing problem} and the variables describing the \textit{discrete counterpart required for generation}. We defined how each variable is produced from input data and/or its relationship with other variables in the form of \ref{eq:cj_ndi_pdi} equations. 
% intuition: solve then generate 
Our approach is based on the \textit{principle that the pairing problem is an over-constrained problem which requires an approximate solution relaxed according to relaxation parameters, which can be found by an analytic process}. 
% solver
We proposed an example of a solver which was demonstrated able to solve the toy and real-size pairing problems. We illustrated how this solver proposes solutions enforcing various combinations of relaxation parameters on the same user data. 
% generation 
We explained and demonstrated (\ref{indoc_application}) how the generation of a synthetic population based on this theoretical framework becomes a \textit{direct process}, mainly because our approach involves the removal of any inconsistency between data in the probabilistic \textit{and} discrete perspectives at the solving stage.

% applied
We illustrated this approach with two examples. A toy example of dwellings used for the illustration of the pairing problems, in which the initial samples were actually generated randomly. 
We then applied our methodology to a real-size case for the reconstruction of dwellings and households in the city of Lille in France. We demonstrated that the controlled variables are enforced when the relaxation parameters require and, and that the variables not controlled by the algorithm are also enforced accordingly.

% open source package
\opensource{In order to facilitate replication and reusal of this method, we publish along this paper the open-source, documented R package which was used for the experiments. Instead of some past methods which delivered hard-coded software \cite{Guo2007,muller2010population} or none, we propose it as a fully generic software. It is shared in \url{https://github.com/samthiriot/gosp.dpp}.}

We denoted this method Direct Probabilistic Pairing. "Pairing" stands as the focus of the algorithm, which is not to be based or not on samples, but focuses on the creation of links between entities. "Direct" refers to the direct generation of the population after solving, which avoid the iterative linking solutions proposed in past sample-free methods. "Probabilistic" stands because it accepts several inputs formulated as probabilities and analyses the pairing problem in a probabilistic framework. 

%
%This paper is quiet ambitious, as it proposes a novel point of view on a problem, 
%a novel way to deal with it, had to introduce a formalism and many equations, 
%and also describe a solver to deal with these equations, and an application on real data. 
%Many details are still hidden, as rounding routines or TODO. 
%Tackling only a subset of the problem would not be scienfici: the novel approach is worthless without an experimentation, the experimentation requires the solver, etc. 
%
%Unfortunately, such a presentation can hardly be made more progressively. 
%As a consequence, this type of paper gives grab to critisicm on all these many points, including the approach, the theory, the formalism, the solver, etc. 
%

\subsection{Position in the state of the art}

Positioning our proposal in the state of the art in the generation of synthetic populations stands as an open discussion, as we do not tackle the very same problem as these studies. Our proposal notably differs in its \textit{goal}, \textit{input data}, the \textit{method}, and \textit{application scope}.

Regarding the \textit{goal}, we focus on the creation of any population made of entities A and B linked together with $n:n$ relationships (that is, each a and b can have 0 to n links with the other type). 
The SR methods so far focus on the only generation of $1:n$ links with $n > 0$ (such as between households and persons). 
The notable exception is the method from Thiriot et al. \cite{samuel_thiriot:bib_perso:thiriot_2008_3} which generates $n:n$ links. 
Note that our method enables the creation of no link for some classes whilst preserving the initial distribution of entities.

About the \textit{input data}, sample-based SR methods do require micro samples of A and B which should share a common identifier; we underlined (\ref{indoc_stateart_pbsample}) how this constraints is likely to be satisfied in very specific datasets only. 
Sample-free methods require as inputs summary data for A and B and joint probabilities for pairing. 
We here ask for a mix between both, as we do require weighted micro samples for A and B, and summary data for pairing probabilities and distributions of degrees. 
Note that the micro samples we require, as they do not have a strong constraint of link between them, might be generated from summary data (as was done for our initial example of dwellings and households used as illustration in~\ref{indoc_theoretical}), thus \textit{also making in practice DPP being another sample-free method}.

Regarding the \textit{method}, the sample-based methods rely on a fit-and-generate scheme to reweight micro samples of dwellings composing households.  
Sample-free methods generate A and B, then iteratively create links between A and B. 
Our method also relies on a solve-and-generate scheme, where solving analytically the entire system (including the discrete version) makes the generation step direct. 
Different methods lead to different \textit{allocations of errors}: sample-based methods introduce errors in the weights of persons (when they control household), household (when they control person) or split the errors when they control both levels at a time (\ref{indoc_stateart_reweighting}). 
Sample-free methods introduce errors depending on their algorithm (\ref{indoc_stateart_samplefree}).
DPP introduces biases in different locations according to relaxation parameters defined by the user. We claim this originality is a benefit of our method, because it enables the usage of the same framework and algorithm in different scenarios were preserving the sizes and/or frequencies and/or degrees and/or pairing are of importance or not.

Concerning the \textit{application scope}, our method can be applied to any type of entities A and B to be connected together. We illustrated here the application on dwellings and households, which was not tackled yet in the literature.
The sample-based methods are limited to the cases where the composition relationship is already known (in practice, households and persons collected during the census - see~\ref{indoc_stateart_pbsample}). 
The sample-free methods were designed for the specific case of households and persons, but can likely be easily extended to any other kind of $0:n$ relationships with minor changes in the methodology. 
While we claim the genericity of our approach, it also might make this approach less relevant for specific cases. 
The existing sample-free methods developed for the creation of households made of persons (\ref{indoc_stateart_samplefree}) do not only create households-person links, but also create \textit{consistent} households made of persons having socio coherent demographic characteristics (such as for spouses ages, etc.). Our method focus on the sole consistency between the sources' and targets' characteristics of each link (dyadic approach in social network analysis), but does not constraints anything about the other dependencies. As a consequence, our solution is generic, but is not suitable for the specific case of creating \textit{coherent} groups of entities composed inside another entity.
% As usual, there is no free lunch: a more generic algorithm leads to worse results than an algorithm designed for a specific case. 

%As a consequence, our method is more generic and can be applied to any type of entities relevant for transportation studies (household / car, dwelling / household), energy (dwelling / appliances, dwelling / household), economics (firm / worker) or social studies (husband / wife). However, this method is creating links based solely upon the characteristics of the two entities of interest, without considering the others (approach denoted dyadic in social network analysis (TODO)). As a consequence, there is no guarantee of consistency of the various entities linked together to the same entity. We emphasised this limitation with the \textit{pairing} concept in the name of the method, Direct Probabilistic Pairing.

%Regarding the synthetic population to generate, we do not propose an approach specific to the creation of households made of persons, but we tackle instead the general question of 
%generating populations made of two types of entities with links between them depending to their characteristics. 
%This method as no strong requirement such as the sample-based methods (TODO link section) on the data, such as having a common identifier shared between the samples of the two populations.

% position it in the state of the art

Provided these many facets and differences, it is difficult to assign DPP to the current categories of "sample-based" and "sample-free" synthetic reconstruction methods. 
Like sample-free methods, we explicitly create links between two types of entities and accept the related input parameters. Like sample-based methods, we use a solve-and-generate pattern in order to fix inconsistencies prior to generation. 
A way to see the DPP method is to see it as \textit{an extension of the fitting approach used so far on sample-based SR methods} (with the frequencies $f_i$ and $f_j$ being the two multi-way tables fit in these methods) which \textit{manages pairing like in the sample-free SR methods} but with an analytical rather than algorithmic solving. 

\subsection{Solver}

We see the core of our proposal as the theoretical framework we highlighted. 
In the same way Beckman only introduced IPF as one tool to solve the reweighing problem he had identified \cite{beckman1996creating}, we only consider the solver we introduced as a tool necessary for dealing with the problem which might be replaced if a better tool is found. This first version suffers many limitations. 
This solver is not able to explore trade-offs; for instance the generated population will contain the exact count of agents $\hat{n}_A=n_A$, or of $\hat{n}_B=n_B$, or both, but will not propose an intermediate solution with the error being split between $\hat{n}_A$ and $\hat{n}_B$. Additional research should explore how to enhance this process, either by the analytic solving of the problem understood as the minimization of the aggregate summed error, using an iterative solving of this problem, or (inspired by combinatorial optimization methods) using any state-of-the-art optimization method like genetic algorithms to solve the various variables.

\subsection{Research directions}

A main originality of our proposal is to fit the entire system of frequencies, degrees and links all together, prior to generation. Another innovation is to formulate the integer version of the fitted problem and also solve it immediately in a consistent way.
As a consequence, the generation stage is free of errors. 
We claim this enables to master errors and keep free of allocating them where preferred, instead of discovering the errors after generation. 

Our initial framing of the pairing problem and our theoretical formulation
induce limits which we intend to question in future research. 
We only considered the creation of populations made of two types of entities after reweighing of micro samples. Can we link entities of one type only, for instance to create social links between entities? Can we chain several generations with DPP in order to create multi-level synthetic populations like dwellings linked with appliances, dwellings with households, households made of persons?

%We did not introduced any spatial population here. Yet if spatial area codes are available in the micro samples, these codes will be found in the result as well; they will either be there with a statistical dependancy given by the micro sample, or constrained if they are part of the attributes normed by the classes. As a consequence, 

% a landscape made of diverse approaches
%The generation of synthetic populations is a set of methods and tools which offer to generate synthetic populations of various types, with various constraints and based on different types of input data. This method thus completes the state of the art on an aspect which was not yet covered. 

We consider the Generation of Synthetic Populations as a set of methods and tools which tackle different problems depending to the initial data available, the expectations of the model and simulation experiment, and the constraints the user wishes to enforce. 
This DPP proposal stands a solution complementary of the state of the art, and not in complete opposition with past methods.

%We designed this solution for it to be generic. Will only be the case by applying it on many different cases. In order to facilitate this process, we opened the corresponding software as an opensource code TODO, documented with examples. We developed it within the ecosystem of R \cite{Wickham2018} which is compliant with many other statistical packages and features, interfacing of many data formats, and is more and more spread among humanities practitionners. 

We designed this method so it is generic. In order to challenge and demonstrate this genericity, this algorithm should be applied to different variations of the pairing problem such as the creation of horizontal links (such as social structure) instead of compositions ones, and to different applications including other types of entities.
The extrapolation of this theoretical framework to the creation of groups instead of pairs is feasible but would require further investigation.
%
%Extrapolate the theoretical framework to three or more types to be linked all together. Pairing probabilties are today two dimenstional to reflect the fact we create links. The creation of hyperlinks, that is edges with more than two vertices, would theoretically possible. It would render the parmaeters moaire complex. But it might offer the possibility to generate more complex situations TODO. 
%
%We also might introduce in the solving algorithm a way to overrepresent small probabilities in order to avoid not representing cases.

\clearpage

\section*{Acknowledgements}

This study was funded by the EDF energy utility and the EIFER European Institute For Energy Research. The study also benefited from the academic support of the GENStar project funded by the French National Research Agency and led by IRD.

The authors would like to thank Aude Sturma, Frederic Amblard, and Paul Chapron for earlier exchanges on this research topic; 
the ANR GenStar project team, notably Kevin Chapuis, for stimulating exchanges on the generation of synthetic populations.
We also would like to thanks Ashreeta Prasanna, Maria Sipowicz and Stephan Lehmers for their advices on the presentation of this research. 

%\section{Bibliography}

\bibliography{../commons/biblio_bibtex/library.bib}

\begin{thebibliography}{10}
\expandafter\ifx\csname url\endcsname\relax
  \def\url#1{\texttt{#1}}\fi
\expandafter\ifx\csname urlprefix\endcsname\relax\def\urlprefix{URL }\fi
\expandafter\ifx\csname href\endcsname\relax
  \def\href#1#2{#2} \def\path#1{#1}\fi

\bibitem{orcutt1957}
G.~H. Orcutt, \href{http://www.jstor.org/stable/1928528?origin=crossref}{{A New
  Type of Socio-Economic System}}, The Review of Economics and Statistics
  39~(2) (1957) 116.
\newblock \href {http://dx.doi.org/10.2307/1928528}
  {\path{doi:10.2307/1928528}}.
\newline\urlprefix\url{http://www.jstor.org/stable/1928528?origin=crossref}

\bibitem{Merz1991}
J.~Merz, {Microsimulation - A survey of principles, developments and
  applications}, International Journal of Forecasting 7~(1) (1991) 77--104.
\newblock \href {http://dx.doi.org/10.1016/0169-2070(91)90035-T}
  {\path{doi:10.1016/0169-2070(91)90035-T}}.

\bibitem{baroni_2007}
E.~Baroni, M.~Richiardi, {Orcutt's Vision, 50 years on}, Tech. rep.,
  LABORatorio R. Revelli, Centre for Employment Studies (2007).

\bibitem{Lovelace2016}
R.~Lovelace, M.~Dumont,
  \href{https://spatial-microsim-book.robinlovelace.net/}{{Spatial
  microsimulation with R.}}, CRC Press, 2016.
\newline\urlprefix\url{https://spatial-microsim-book.robinlovelace.net/}

\bibitem{heppenstall2011agent}
A.~J. Heppenstall, A.~T. Crooks, L.~M. See, M.~Batty, {Agent-based models of
  geographical systems}, Springer Science {\&} Business Media, 2011.

\bibitem{samuel_thiriot:bib_sma_simulation:gilbert_1999_1}
N.~Gilbert, K.~Troitzsch, {Simulation for the Social Scientists}, Open
  University Press., 1999.

\bibitem{phan_2007}
D.~Phan, F.~Amblard, {Agent-based Modelling}, The Bardwell Press, Oxford, 2007.

\bibitem{Smajgl_20014}
A.~Smajgl, O.~Barreteau, {Empirical Agent-Based Modelling - Challenges and
  Solutions}, Springer-Verlag New York, 2014.

\bibitem{Waddell2002a}
P.~Waddell, {UrbanSim - Modeling urban development for land use,
  transportation, and environmental planning}, Journal of the American Planning
  Association 68~(August) (2002) 297--314.
\newblock \href {http://dx.doi.org/10.1080/01944360208976274}
  {\path{doi:10.1080/01944360208976274}}.

\bibitem{Balmer2006}
M.~Balmer, K.~Axhausen, K.~Nagel, {Agent-based demand-modeling framework for
  large-scale microsimulations}, Transport. Res. Rec. J. Transpo (2006)
  125--134.

\bibitem{Salvini2005}
P.~Salvini, E.~Miller, {ILUTE: an operational prototype of a comprehensive
  microsimulation model of urban systems}, Networks Spatial Econ. 5~(2) (2005)
  217----234.

\bibitem{Amouroux2014}
E.~Amouroux, T.~Huraux, F.~Semp{\'{e}}, N.~Sabouret, Y.~Haradji, {SMACH:
  Agent-Based Simulation Investigation on Human Activities and Household
  Electrical Consumption}, Communications in Computer and Information Science
  449~(October).
\newblock \href {http://dx.doi.org/10.1007/978-3-662-44440-5}
  {\path{doi:10.1007/978-3-662-44440-5}}.

\bibitem{thiriot2011referral}
S.~Thiriot, Z.~Lewkovicz, P.~Caillou, J.-D. Kant, {Referral hiring and labor
  markets: a computational study}, in: Emergent Results of Artificial
  Economics, Springer, 2011, pp. 15--25.

\bibitem{Batty2007}
M.~Batty,
  \href{http://www.springerlink.com/index/j3863x4mm7gu8645.pdf}{{Planning
  support systems: progress, predictions, and speculations on the shape of
  things to come}}, Tech. rep., University College London -- Centre for
  Advanced Spatial Analysis (2007).
\newblock \href {http://arxiv.org/abs/0706.0024} {\path{arXiv:0706.0024}},
  \href {http://dx.doi.org/10.1103/PhysRevE.78.016110}
  {\path{doi:10.1103/PhysRevE.78.016110}}.
\newline\urlprefix\url{http://www.springerlink.com/index/j3863x4mm7gu8645.pdf}

\bibitem{beckman1996creating}
R.~J. Beckman, K.~A. Baggerly, M.~D. McKay, {Creating synthetic baseline
  populations}, Transportation Research Part A: Policy and Practice 30~(6)
  (1996) 415--429.

\bibitem{muller2011hierarchical}
K.~M{\"{u}}ller, K.~W. Axhausen, {Hierarchical IPF: Generating a synthetic
  population for Switzerland}, Eidgen{\"{o}}ssische Technische Hochschule
  Z{\"{u}}rich, IVT, 2011.

\bibitem{samuel_thiriot:bib_sma_simulation:deming_1940_1}
W.~E. Deming, F.~F. Stephan, {On a least squares adjustment of a sampled
  frequency table when the expected marginal totals are known}, The Annals of
  Mathematical Statistics 11~(4) (1940) 427--444.

\bibitem{bowman2004comparison}
J.~L. Bowman,
  \href{http://www.jbowman.net/papers/2004.Bowman.Comparison{\_}of{\_}PopSyns.pdf}{{A
  comparison of population synthesizers used in microsimulation models of
  activity and travel demand}}, Tech. rep. (2004).
\newline\urlprefix\url{http://www.jbowman.net/papers/2004.Bowman.Comparison{\_}of{\_}PopSyns.pdf}

\bibitem{Guo2007}
J.~Guo, C.~Bhat,
  \href{http://trrjournalonline.trb.org/doi/10.3141/2014-12}{{Population
  Synthesis for Microsimulating Travel Behavior}}, Transportation Research
  Record: Journal of the Transportation Research Board 2014 (2007) 92--101.
\newblock \href {http://dx.doi.org/10.3141/2014-12}
  {\path{doi:10.3141/2014-12}}.
\newline\urlprefix\url{http://trrjournalonline.trb.org/doi/10.3141/2014-12}

\bibitem{Ye2009}
X.~Ye, K.~Konduri, R.~M. Pendyala, B.~Sana, P.~Waddell, {A methodology to match
  distributions of both household and person attributes in the generation of
  synthetic populations}, in: 88th Annual Meeting of the Transportation
  Research Board, Washington, DC., 2009.

\bibitem{muller2010population}
K.~M{\"{u}}ller, K.~W. Axhausen, {Population synthesis for microsimulation:
  State of the art}, in: 90th Annual Meeting of the Transportation Research
  Board, ETH Z{\"{u}}rich, Institut f{\"{u}}r Verkehrsplanung,
  Transporttechnik, Strassen-und Eisenbahnbau (IVT), 2010.

\bibitem{lovelace2015evaluating}
R.~Lovelace, M.~Birkin, D.~Ballas, E.~van Leeuwen,
  \href{http://jasss.soc.surrey.ac.uk/18/2/21.html}{{Evaluating the performance
  of Iterative Proportional Fitting for spatial microsimulation: new tests for
  an established technique}}, Journal of Artificial Societies and Social
  Simulation 18~(2) (2015) 21.
\newblock \href {http://dx.doi.org/10.18564/jasss.2768}
  {\path{doi:10.18564/jasss.2768}}.
\newline\urlprefix\url{http://jasss.soc.surrey.ac.uk/18/2/21.html}

\bibitem{Ma2015}
L.~Ma, S.~Srinivasan, {Synthetic population generation with multilevel
  controls: A fitness-based synthesis approach and validations}, Computer-Aided
  Civil and Infrastructure Engineering 30~(2) (2015) 135--150.
\newblock \href {http://dx.doi.org/10.1111/mice.12085}
  {\path{doi:10.1111/mice.12085}}.

\bibitem{pritchard2012advances}
D.~R. Pritchard, E.~J. Miller, {Advances in population synthesis: Fitting many
  attributes per agent and fitting to household and person margins
  simultaneously}, Transportation 39~(3) (2012) 685--704.
\newblock \href {http://dx.doi.org/10.1007/s11116-011-9367-4}
  {\path{doi:10.1007/s11116-011-9367-4}}.

\bibitem{Arentze2007}
T.~Arentze, H.~Timmermans, F.~Hofman,
  \href{http://trrjournalonline.trb.org/doi/10.3141/2014-11}{{Creating
  Synthetic Household Populations: Problems and Approach}}, Transportation
  Research Record: Journal of the Transportation Research Board 2014 (2007)
  85--91.
\newblock \href {http://dx.doi.org/10.3141/2014-11}
  {\path{doi:10.3141/2014-11}}.
\newline\urlprefix\url{http://trrjournalonline.trb.org/doi/10.3141/2014-11}

\bibitem{bar2009estimating}
H.~Bar-Gera, K.~C. Konduri, B.~Sana, X.~Ye, R.~M. Pendyala, {Estimating survey
  weights with multiple constraints using entropy optimization methods}, in:
  Transportation Research Board 88th Annual Meeting, no. 09-1354, 2009.

\bibitem{auld2010efficient}
J.~A. Auld, A.~Mohammadian, {Efficient methodology for generating synthetic
  populations with multiple control levels}, Transportation Research Record:
  Journal of the Transportation Research Board 2175~(1) (2010) 138--147.

\bibitem{Lovelace2013}
R.~Lovelace, D.~Ballas, {‘Truncate, replicate, sample': A method for creating
  integer weights for spatial microsimulation}, Computers, Environment and
  Urban Systems 41 (2013) 1--11.
\newblock \href {http://dx.doi.org/10.1016/j.compenvurbsys.2013.03.004.White}
  {\path{doi:10.1016/j.compenvurbsys.2013.03.004.White}}.

\bibitem{Smith2017}
A.~P. Smith, R.~Lovelace, M.~Birkin, {Population synthesis with quasirandom
  integer sampling}, Jasss 20~(4).
\newblock \href {http://dx.doi.org/10.18564/jasss.3550}
  {\path{doi:10.18564/jasss.3550}}.

\bibitem{samuel_thiriot:bib_sma_simulation:williamson_1998_1}
P.~Williamson, M.~Birkin, P.~H. Rees, Others, {The estimation of population
  microdata by using data from small area statistics and samples of anonymised
  records}, Environment and Planning A 30 (1998) 785--816.

\bibitem{birkin2006synthetic}
M.~Birkin, A.~Turner, B.~Wu, {A synthetic demographic model of the UK
  population: Methods, progress and problems}, in: Regional Science Association
  International British and Irish Section, 36th Annual Conference, Citeseer,
  2006.

\bibitem{voas2000evaluation}
D.~Voas, P.~Williamson, {An evaluation of the combinatorial optimisation
  approach to the creation of synthetic microdata}, International Journal of
  Population Geography 6~(5) (2000) 349--366.

\bibitem{huang2001comparison}
Z.~Huang, P.~Williamson, {A comparison of synthetic reconstruction and
  combinatorial optimisation approaches to the creation of small-area
  microdata}, Tech. rep. (2001).

\bibitem{ryan2009population}
J.~Ryan, H.~Maoh, P.~Kanaroglou, {Population synthesis: Comparing the major
  techniques using a small, complete population of firms}, Geographical
  Analysis 41~(2) (2009) 181--203.

\bibitem{smith2009improving}
D.~M. Smith, G.~P. Clarke, K.~Harland, {Improving the synthetic data generation
  process in spatial microsimulation models}, Environment and planning. A
  41~(5) (2009) 1251.

\bibitem{samuel_thiriot:bib_sma_simulation:gargiulo_2010_1}
F.~Gargiulo, S.~Ternes, S.~Huet, G.~Deffuant, {An iterative approach for
  generating statistically realistic populations of households}, PLoS One 5~(1)
  (2010) e8828.

\bibitem{barthelemy2013synthetic}
J.~Barthelemy, P.~L. Toint, {Synthetic population generation without a sample},
  Transportation Science 47~(2) (2013) 266--279.

\bibitem{birkin1988synthesis}
M.~Birkin, M.~Clarke, {SYNTHESIS--a synthetic spatial information system for
  urban and regional analysis: methods and examples}, Environment and planning
  A 20~(12) (1988) 1645--1671.

\bibitem{Huynh2016}
N.~Huynh, J.~Barth{\'{e}}lemy, P.~Perez, {A heuristic combinatorial
  optimisation approach to synthesising a population for agent-based modelling
  purposes}, Journal of Artificial Societies and Social Simulation 19~(4).
\newblock \href {http://dx.doi.org/10.18564/jasss.3198}
  {\path{doi:10.18564/jasss.3198}}.

\bibitem{samuel_thiriot:bib_perso:thiriot_2008_3}
S.~Thiriot, J.-D. Kant, {Generate country-scale networks of interaction from
  scattered statistics}, in: The Fifth Conference of the European Social
  Simulation Association, Brescia, Italy, 2008.

\bibitem{lewkovicz2011detailed}
Z.~Lewkovicz, S.~Thiriot, P.~Caillou,
  \href{https://hal.inria.fr/inria-00579620/document}{{How detailed should
  social networks be for labor market's models?}}, in: SNAMAS@AISB 2011, 2011.
\newline\urlprefix\url{https://hal.inria.fr/inria-00579620/document}

\bibitem{samuel_thiriot:bib_psycho:wasserman_1994_2}
S.~Wasserman, K.~Faust, {Social network analysis, methods and applications},
  Cambridge: Cambridge University Press, 1994, Ch. Social Net, pp. 3--25.

\bibitem{Voas2001}
D.~Voas, P.~Williamson, {Evaluating goodness-of-fit measures for synthetic
  microdata}, Geographical and Environmental Modelling 5~(2) (2001) 177--200.
\newblock \href {http://dx.doi.org/10.1080/13615930120086078}
  {\path{doi:10.1080/13615930120086078}}.

\bibitem{read1993freeman}
C.~B. Read, {Freeman—Tukey chi-squared goodness-of-fit statistics},
  Statistics {\&} probability letters 18~(4) (1993) 271--278.

\bibitem{guo2007population}
J.~Y. Guo, C.~R. Bhat, {Population synthesis for microsimulating travel
  behavior}, Transportation Research Record: Journal of the Transportation
  Research Board 2014~(1) (2007) 92--101.

\bibitem{Farooq2013}
B.~Farooq, M.~Bierlaire, R.~Hurtubia, G.~Fl{\"{o}}tter{\"{o}}d, {Simulation
  based Population Synthesis}, Transportation Research Part B: Methodological
  58 (2013) 243--263.

\bibitem{Zhu2014}
Y.~Zhu, J.~Ferreira,
  \href{http://trrjournalonline.trb.org/doi/10.3141/2429-18}{{Synthetic
  Population Generation at Disaggregated Spatial Scales for Land Use and
  Transportation Microsimulation}}, Transportation Research Record: Journal of
  the Transportation Research Board 2429 (2014) 168--177.
\newblock \href {http://dx.doi.org/10.3141/2429-18}
  {\path{doi:10.3141/2429-18}}.
\newline\urlprefix\url{http://trrjournalonline.trb.org/doi/10.3141/2429-18}

\bibitem{RCoreTeam2018}
{R Core Team}, \href{https://www.r-project.org/}{{R: A Language and Environment
  for Statistical Computing}}, R Foundation for Statistical Computing, Vienna,
  Austria (2018).
\newline\urlprefix\url{https://www.r-project.org/}

\bibitem{Wickham2018}
H.~Wickham, R.~Fran{\c{c}}ois, L.~Henry, K.~M{\"{u}}ller,
  \href{https://cran.r-project.org/package=dplyr}{{dplyr: A Grammar of Data
  Manipulation}} (2018).
\newline\urlprefix\url{https://cran.r-project.org/package=dplyr}

\bibitem{insee_ponderations_2018}
INSEE,
  \href{https://www.insee.fr/fr/statistiques/fichier/2383177/fiche-ponderation.pdf}{{Fiche
  th{\'{e}}matique - les pond{\'{e}}rations}}, Tech. rep., INSEE (2018).
\newline\urlprefix\url{https://www.insee.fr/fr/statistiques/fichier/2383177/fiche-ponderation.pdf}

\bibitem{Besbeas2014}
P.~Besbeas, B.~J. Morgan, {Goodness-of-fit of integrated population models
  using calibrated simulation}, Methods in Ecology and Evolution 5~(12) (2014)
  1373--1382.
\newblock \href {http://dx.doi.org/10.1111/2041-210X.12279}
  {\path{doi:10.1111/2041-210X.12279}}.

\bibitem{insee_2018_logements}
INSEE,
  \href{https://www.insee.fr/fr/statistiques/fichier/2383177/fiche-logements.pdf}{{Recensement
  de la population - Les Logements}}, Tech. rep., INSEE (2018).
\newline\urlprefix\url{https://www.insee.fr/fr/statistiques/fichier/2383177/fiche-logements.pdf}

\bibitem{insee_2018_menages}
INSEE,
  \href{https://www.insee.fr/fr/statistiques/fichier/2383177/fiche-menages-familles.pdf}{{Recensement
  de la population - M{\'{e}}nages et familles}}, Tech. rep., INSEE (2018).
\newline\urlprefix\url{https://www.insee.fr/fr/statistiques/fichier/2383177/fiche-menages-familles.pdf}

\end{thebibliography}

\newpage 

\section{Annex}

% more images !

\begin{table}[pth]
	\centering
	% latex table generated in R 3.4.4 by xtable 1.8-3 package
% Tue Oct  2 08:27:49 2018
\begingroup\footnotesize
\begin{tabular}{rrrrrrrr}
  \hline
 & ACHL & CATL & CHFL & HLML & IPONDL & NBPI & SURF \\ 
  \hline
1 & 211 &   1 &   3 &   2 & 3.71 &   5 &   5 \\ 
  2 & 211 &   1 &   3 &   2 & 3.71 &   5 &   5 \\ 
  3 & 211 &   1 &   3 &   2 & 3.71 &   3 &   5 \\ 
  4 & 211 &   1 &   3 &   2 & 3.71 &   3 &   5 \\ 
  5 & 314 &   1 &   2 &   2 & 0.98 &   4 &   6 \\ 
  6 & 311 &   1 &   3 &   2 & 0.98 &   2 &   2 \\ 
  7 & 212 &   1 &   3 &   2 & 3.23 &   1 &   2 \\ 
  8 & 311 &   1 &   3 &   2 & 3.23 &   2 &   3 \\ 
  9 & 112 &   1 &   2 &   2 & 3.32 &   2 &   2 \\ 
  10 & 311 &   1 &   3 &   2 & 3.16 &   3 &   4 \\ 
  11 & 211 &   1 &   3 &   2 & 3.23 &   1 &   1 \\ 
  12 & 212 &   1 &   2 &   2 & 3.41 &   2 &   1 \\ 
  13 & 112 &   1 &   1 &   2 & 3.23 &   2 &   3 \\ 
  14 & 212 &   1 &   3 &   2 & 0.98 &   3 &   3 \\ 
  15 & 312 &   1 &   3 &   2 & 3.16 &   3 &   4 \\ 
  16 & 211 &   1 &   1 &   2 & 3.32 &   6 &   7 \\ 
  17 & 112 &   1 &   2 &   1 & 3.23 &   3 &   6 \\ 
  18 & 212 &   1 &   3 &   2 & 3.23 &   2 &   2 \\ 
  19 & 311 &   1 &   3 &   2 & 0.98 &   2 &   3 \\ 
   \hline
\end{tabular}
\endgroup
	
	\caption{Excerpt of the sample of dwellings provided by INSEE. 
		The sample contains variables ACHL (year of achievement), CATL (occupancy status), 
		CHFL (type of heater), HLML (social housing), NBPI (count of rooms) and SURF (surface).
		The column IPONDL contains the weights of the sample.}\label{tab_sample_lille_dwellings}
\end{table}
\begin{table}[pth]
	\centering
	% latex table generated in R 3.4.4 by xtable 1.8-3 package
% Tue Oct  2 08:28:07 2018
\begingroup\footnotesize
\begin{tabular}{rrrrlrlr}
  \hline
 & AGEREV & AGEREVQ & COUPLE & EMPL & INPER & NA17 & IPONDI \\ 
  \hline
1 &  48 &  45 &   1 & 16 &   2 & OQ & 3.46 \\ 
  2 &  53 &  50 &   2 & ZZ &   1 & ZZ & 3.59 \\ 
  3 &  18 &  15 &   2 & ZZ &   1 & ZZ & 1.03 \\ 
  4 &  28 &  25 &   2 & 16 &   2 & GZ & 3.86 \\ 
  5 &  32 &  30 &   1 & 16 &   2 & JZ & 3.42 \\ 
  6 &  21 &  20 &   2 & ZZ &   1 & ZZ & 1.03 \\ 
  7 &  48 &  45 &   2 & 16 &   1 & JZ & 1.09 \\ 
  8 &  21 &  20 &   2 & ZZ &   3 & ZZ & 1.03 \\ 
  9 &  18 &  15 &   2 & ZZ &   1 & ZZ & 0.99 \\ 
  10 &  18 &  15 &   2 & ZZ &   1 & ZZ & 0.85 \\ 
  11 &  79 &  75 &   1 & ZZ &   2 & ZZ & 3.81 \\ 
  12 &  39 &  35 &   2 & 16 &   1 & OQ & 3.83 \\ 
  13 &  19 &  15 &   2 & ZZ &   1 & ZZ & 3.39 \\ 
  14 &  18 &  15 &   2 & 15 &   1 & MN & 1.03 \\ 
  15 &  24 &  20 &   1 & 16 &   3 & HZ & 3.51 \\ 
  16 &  63 &  60 &   1 & ZZ &   2 & ZZ & 1.09 \\ 
  17 &  38 &  35 &   1 & ZZ &   2 & ZZ & 3.38 \\ 
  18 &  75 &  75 &   2 & ZZ &   1 & ZZ & 1.19 \\ 
  19 &  92 &  90 &   1 & ZZ &   2 & ZZ & 3.61 \\ 
   \hline
\end{tabular}
\endgroup

	\caption{Excerpt of the sample of households provided by INSEE.
		AGEREV encodes the detailed age, AGEREVQ the age encoded in a quinquenal way, COUPLE the marital status,
		EMPL the employment status, INPER the count of persons in the household, NA17 (economical activity).
		IPONDI encodes the weight.}\label{tab_sample_lille_households}
\end{table}

\begin{sidewaystable}[ph]
	\renewcommand*\rot{\rotatebox{80}}
	\tiny
	\begin{adjustbox}{width=1\textwidth}
		\begin{tabular}{r|cccc|cccccccccc|c}
\multicolumn{5}{r|}{$\text{Cla}_i^A$}  & \rot{SURF=1,CATL=1} & \rot{SURF=1,CATL=2} & \rot{SURF=1,CATL=3} & \rot{SURF=1,CATL=4} & \rot{SURF=2,CATL=1} & \rot{SURF=2,CATL=2} & \rot{SURF=2,CATL=3} & \rot{SURF=2,CATL=4} & \rot{SURF=3,CATL=1}  & ...& \\
\hline
\multicolumn{5}{r|}{$f_i$}  & 0.1412 & 0.004396 & 0.002261 & 0.02554 & 0.1196 & 0.001501 & 0.001104 & 0.01605 & 0.2136 & ... & 1\\
\multicolumn{5}{r|}{$\tilde{d}_i$}  & 1.05 & 6.661E-16 & 6.661E-16 & 6.661E-16 & 1.05 & 6.661E-16 & 6.661E-16 & 6.661E-16 & 1.05 & ... & \\
\multicolumn{5}{r|}{$p_i$}  & 0.03916 & 0.03916 & 0.03916 & 0.03916 & 0.03319 & 0.03319 & 0.03319 & 0.03319 & 0.05926 & ... & 1\\

$\text{Cla}_j^B$ & $f_j$ & $\tilde{d}_j$ & $p_j$ & $j/i$ & 1&2&3&4&5&6&7&8&9 & ...&\\
\hline
INPER=1 & 0.5263 & 1 & 0.5263 & 1 & 0.03699 & 0.03699 & 0.03699 & 0.03699 & 0.02637 & 0.02637 & 0.02637 & 0.02637 & 0.03511 & ... & \\
INPER=2 & 0.2624 & 1 & 0.2624 & 2 & 0.001958 & 0.001958 & 0.001958 & 0.001958 & 0.005774 & 0.005774 & 0.005774 & 0.005774 & 0.01861 & ... & \\
INPER=3 & 0.09772 & 1 & 0.09773 & 3 & 0.0001402 & 0.0001402 & 0.0001402 & 0.0001402 & 0.0007214 & 0.0007214 & 0.0007214 & 0.0007214 & 0.00337 & ... & \\
INPER=4 & 0.0662 & 1 & 0.0662 & 4 & 4.969E-05 & 4.969E-05 & 4.969E-05 & 4.969E-05 & 0.0002275 & 0.0002275 & 0.0002275 & 0.0002275 & 0.00138 & ... & \\
INPER=5 & 0.03061 & 1 & 0.03059 & 5 & 2.429E-05 & 2.429E-05 & 2.429E-05 & 2.429E-05 & 5.813E-05 & 5.813E-05 & 5.813E-05 & 5.813E-05 & 0.0005401 & ... & \\
INPER=6 & 0.01119 & 1 & 0.0112 & 6 & 5.651E-17 & 5.651E-17 & 5.651E-17 & 5.651E-17 & 3.1E-05 & 3.1E-05 & 3.1E-05 & 3.1E-05 & 0.0001903 & ... & \\
INPER=7 & 0.003476 & 1 & 0.003476 & 7 & 5.651E-17 & 5.651E-17 & 5.651E-17 & 5.651E-17 & 4.8E-06 & 4.8E-06 & 4.8E-06 & 4.8E-06 & 2.873E-05 & ... & \\
INPER=8 & 0.001292 & 1 & 0.001291 & 8 & 5.651E-17 & 5.651E-17 & 5.651E-17 & 5.651E-17 & 2.427E-06 & 2.427E-06 & 2.427E-06 & 2.427E-06 & 1.698E-05 & ... & \\
INPER=9 & 0.0005671 & 1 & 0.0005671 & 9 & 5.651E-17 & 5.651E-17 & 5.651E-17 & 5.651E-17 & 5.651E-17 & 5.651E-17 & 5.651E-17 & 5.651E-17 & 6.599E-06 & ... & \\
INPER=10 & 0.0001078 & 1 & 0.0001078 & 10 & 5.651E-17 & 5.651E-17 & 5.651E-17 & 5.651E-17 & 5.651E-17 & 5.651E-17 & 5.651E-17 & 5.651E-17 & 5.651E-17 & ... & \\
INPER=12 & 6.155E-05 & 1 & 6.154E-05 & 11 & 5.651E-17 & 5.651E-17 & 5.651E-17 & 5.651E-17 & 5.651E-17 & 5.651E-17 & 5.651E-17 & 5.651E-17 & 5.651E-17 & ... & \\
INPER=14 & 3.104E-05 & 1 & 3.104E-05 & 12 & 5.651E-17 & 5.651E-17 & 5.651E-17 & 5.651E-17 & 5.651E-17 & 5.651E-17 & 5.651E-17 & 5.651E-17 & 5.651E-17 & ... & \\
\hline
 & 1 &  & 1 & &&&&&&&&&& & 1\\
\end{tabular}

	\end{adjustbox}
	\caption{Probabilistic view of the pairing problem of dwellings and households in Lille \textit{before resolution}. These values reflect initial user parameters and are not consistent.}\label{tab_lille_allfree_probaview_init}
\end{sidewaystable}

\begin{sidewaystable}
	\renewcommand*\rot{\rotatebox{80}}
	\tiny
	\begin{adjustbox}{width=1\textwidth} % 0.94
		\begin{tabular}{r|cccc|cccccccccc|c}
\multicolumn{5}{r|}{$\text{Cla}_i^A$}  & \rot{SURF=1,CATL=1} & \rot{SURF=1,CATL=2} & \rot{SURF=1,CATL=3} & \rot{SURF=1,CATL=4} & \rot{SURF=2,CATL=1} & \rot{SURF=2,CATL=2} & \rot{SURF=2,CATL=3} & \rot{SURF=2,CATL=4} & \rot{SURF=3,CATL=1}  & ...& \\
\hline
\multicolumn{5}{r|}{$\hat{f_i}$}  & 0.1412 & 0.004392 & 0.002262 & 0.02555 & 0.1196 & 0.0015 & 0.001108 & 0.01605 & 0.2136 & ... & 1\\
\multicolumn{5}{r|}{$\hat{\tilde{d}}_i$}  & 1.05 & 0 & 0 & 0 & 1.05 & 0 & 0 & 0 & 1.05 & ... & \\
\multicolumn{5}{r|}{$\hat{p}_i$}  & 0.1566 & 0 & 0 & 0 & 0.1327 & 0 & 0 & 0 & 0.237 & ... & 1\\

$\text{Cla}_j^B$ & $\hat{f}_j$ & $\hat{\tilde{d}}_j$ & $\hat{p}_j$ & $j/i$ & 1&2&3&4&5&6&7&8&9 & ...&\\
\hline
INPER=1 & 0.5263 & 1 & 0.5263 & 1 & 0.1479 & 0 & 0 & 0 & 0.1055 & 0 & 0 & 0 & 0.1404 & ... & \\
INPER=2 & 0.2624 & 1 & 0.2624 & 2 & 0.007836 & 0 & 0 & 0 & 0.0231 & 0 & 0 & 0 & 0.07445 & ... & \\
INPER=3 & 0.09772 & 1 & 0.09772 & 3 & 0.0005609 & 0 & 0 & 0 & 0.002886 & 0 & 0 & 0 & 0.01348 & ... & \\
INPER=4 & 0.0662 & 1 & 0.06619 & 4 & 0.0001951 & 0 & 0 & 0 & 0.0009105 & 0 & 0 & 0 & 0.00552 & ... & \\
INPER=5 & 0.03061 & 1 & 0.03061 & 5 & 9.755E-05 & 0 & 0 & 0 & 0.0002357 & 0 & 0 & 0 & 0.002162 & ... & \\
INPER=6 & 0.01119 & 1 & 0.01119 & 6 & 0 & 0 & 0 & 0 & 0.0001219 & 0 & 0 & 0 & 0.0007641 & ... & \\
INPER=7 & 0.003476 & 1 & 0.003479 & 7 & 0 & 0 & 0 & 0 & 1.626E-05 & 0 & 0 & 0 & 0.0001138 & ... & \\
INPER=8 & 0.001292 & 1 & 0.001293 & 8 & 0 & 0 & 0 & 0 & 8.129E-06 & 0 & 0 & 0 & 6.503E-05 & ... & \\
INPER=9 & 0.0005671 & 1 & 0.0005609 & 9 & 0 & 0 & 0 & 0 & 0 & 0 & 0 & 0 & 2.439E-05 & ... & \\
INPER=10 & 0.0001078 & 1 & 0.0001057 & 10 & 0 & 0 & 0 & 0 & 0 & 0 & 0 & 0 & 0 & ... & \\
INPER=12 & 6.155E-05 & 1 & 6.503E-05 & 11 & 0 & 0 & 0 & 0 & 0 & 0 & 0 & 0 & 0 & ... & \\
INPER=14 & 3.104E-05 & 1 & 3.252E-05 & 12 & 0 & 0 & 0 & 0 & 0 & 0 & 0 & 0 & 0 & ... & \\
\hline
 & 1 &  & 1 & &&&&&&&&&& & 1\\
\end{tabular}

	\end{adjustbox}
	\caption{Probabilistic view of the \textit{solved} pairing problem with \protect population sizes $n^A=130000$ and $n^B=120000$. Relaxation parameters are $\nu^A=\allowbreak\phi^A=\allowbreak\delta^A=\allowbreak\gamma=\allowbreak\delta^B=\allowbreak\phi^B=\allowbreak\nu^B=1$. The population generated sizes $\hat{n}^A=130000$ and $\hat{n}^B=123016$.
}
	\label{tab_lille_freeall_perspective_proba_solved}
\end{sidewaystable}

\begin{sidewaystable}
	\renewcommand*\rot{\rotatebox{80}}
	\tiny
	\begin{adjustbox}{width=1\textwidth}
		\begin{tabular}{r|cccc|cccccccccc|c}
\multicolumn{5}{r|}{$\text{Cla}_i^A$}  & \rot{SURF=1,CATL=1} & \rot{SURF=1,CATL=2} & \rot{SURF=1,CATL=3} & \rot{SURF=1,CATL=4} & \rot{SURF=2,CATL=1} & \rot{SURF=2,CATL=2} & \rot{SURF=2,CATL=3} & \rot{SURF=2,CATL=4} & \rot{SURF=3,CATL=1}  & ... & \\
\hline
\multicolumn{5}{r|}{$\hat{c_i}$}  & 18351 & 571 & 294 & 3321 & 15552 & 195 & 144 & 2087 & 27769 & ... & 130000\\
\multicolumn{5}{r|}{$\hat{\tilde{d}}_i$}  & 1.05 & 0 & 0 & 0 & 1.05 & 0 & 0 & 0 & 1.05 & ... & \\
\multicolumn{5}{r|}{$\hat{n}_i$}  & 19269 & 0 & 0 & 0 & 16330 & 0 & 0 & 0 & 29157 & ... & 123016\\

$\text{Cla}_j^B$ & $\hat{c}_j$ & $\hat{\tilde{d}}_j$ & $\hat{n}_j$ & $j/i$ & 1&2&3&4&5&6&7&8&9 & ...&\\
\hline
INPER=1 & 64744 & 1 & 64744 & 1 & 18200 & 0 & 0 & 0 & 12974 & 0 & 0 & 0 & 17276 & ... & \\
INPER=2 & 32284 & 1 & 32284 & 2 & 964 & 0 & 0 & 0 & 2842 & 0 & 0 & 0 & 9159 & ... & \\
INPER=3 & 12021 & 1 & 12021 & 3 & 69 & 0 & 0 & 0 & 355 & 0 & 0 & 0 & 1658 & ... & \\
INPER=4 & 8143 & 1 & 8143 & 4 & 24 & 0 & 0 & 0 & 112 & 0 & 0 & 0 & 679 & ... & \\
INPER=5 & 3766 & 1 & 3766 & 5 & 12 & 0 & 0 & 0 & 29 & 0 & 0 & 0 & 266 & ... & \\
INPER=6 & 1377 & 1 & 1377 & 6 & 0 & 0 & 0 & 0 & 15 & 0 & 0 & 0 & 94 & ... & \\
INPER=7 & 428 & 1 & 428 & 7 & 0 & 0 & 0 & 0 & 2 & 0 & 0 & 0 & 14 & ... & \\
INPER=8 & 159 & 1 & 159 & 8 & 0 & 0 & 0 & 0 & 1 & 0 & 0 & 0 & 8 & ... & \\
INPER=9 & 69 & 1 & 69 & 9 & 0 & 0 & 0 & 0 & 0 & 0 & 0 & 0 & 3 & ... & \\
INPER=10 & 13 & 1 & 13 & 10 & 0 & 0 & 0 & 0 & 0 & 0 & 0 & 0 & 0 & ... & \\
INPER=12 & 8 & 1 & 8 & 11 & 0 & 0 & 0 & 0 & 0 & 0 & 0 & 0 & 0 & ... & \\
INPER=14 & 4 & 1 & 4 & 12 & 0 & 0 & 0 & 0 & 0 & 0 & 0 & 0 & 0 & ... & \\
\hline
 & 123016 &  & 123016 & &&&&&&&&&& & 123016\\
\end{tabular}

	\end{adjustbox}
	\caption{Discrete view of the \textit{solved} pairing problem with \protect population sizes $n^A=130000$ and $n^B=120000$. Relaxation parameters are $\nu^A=\allowbreak\phi^A=\allowbreak\delta^A=\allowbreak\gamma=\allowbreak\delta^B=\allowbreak\phi^B=\allowbreak\nu^B=1$. The population generated sizes $\hat{n}^A=130000$ and $\hat{n}^B=123016$.
}
	\label{tab_lille_freeall_perspective_discrete}
\end{sidewaystable}

\subsection{Source and preparation of micro samples\label{annex_samples}}

Micro samples for dwellings and households come from the 2014 census information provided by the French national institute for statistics named INSEE.
The datasets are freely available on their website. We loaded them from CSV format into R \cite{RCoreTeam2018} using the sqldf package \cite{Besbeas2014} which enables the selection of the records of interest to our case.

The sample for population A is the weighted sample of dwellings. We download from the INSEE website the public dataset named "logements" (dwellings) \cite{insee_2018_logements}.of zone B (north of France) 
We retain only the elements relevant to our case: we keep elements of the city of Lille (variable 'COMMUNE' equal to 59350); we exclude the dwellings classified as being specific (CATL different of 2 and 4). 
The resulting dataset contains~\countA ~entities.
We present in table~\vref{tab_sample_lille_dwellings} a few lines of this sample.

The sample for population B is a weighted sample of households \cite{insee_2018_menages}. 
We download from the INSEE website the data sets named "logements" of zone B.
We retain only the elements relevant to the area of the city of Lille (variable 'CANTVILLE' = 5997). 
The original dataset is structured with several lines per representative household sharing the same household identifier. We focus on the only lines representing the head of household, that is records having `LPRM`=1.
The resulting dataset contains~\countB ~entities.

\end{document}